\documentclass[Journal,letterpaper,InsideFigs,NoLineNumbers]{ascelike-new}
\usepackage[utf8]{inputenc}
\usepackage[T1]{fontenc}
\usepackage[style=base,figurename=Fig.,labelfont=bf,labelsep=period]{caption}
\usepackage{amsfonts} 
\usepackage[colorlinks=true,citecolor=red,linkcolor=black]{hyperref}
\usepackage{mathrsfs} 

\usepackage{calc}
\usepackage{xcolor}
\usepackage{amsmath}
\usepackage{graphicx}
\usepackage{newtxtext,mathptmx} 
\usepackage{wrapfig}
\usepackage{color}\usepackage{xcolor}
\usepackage{bm}
\usepackage{placeins}
\usepackage{multirow,cuted}
\usepackage[list=true]{subcaption}
\usepackage{accents}
\usepackage[outdir=./]{epstopdf}
\usepackage{rotating}
\usepackage[flushleft]{threeparttable}
\usepackage{footnote}
\makesavenoteenv{table}
\makesavenoteenv{tabular}
\usepackage{ifthen}
\usepackage{natbib} 
\usepackage{tikz}
\usetikzlibrary{matrix} % for block alignment
\usetikzlibrary{arrows} % for arrow heads
\usetikzlibrary{calc} % for manimulation of coordinates
\usetikzlibrary{shapes}
\usetikzlibrary{fit}
\usetikzlibrary{3d}
\usetikzlibrary{positioning}
\usetikzlibrary{shadows}
\usetikzlibrary{external}
\tikzsetexternalprefix{figures/cache/tmp/}
\newcommand*{\WhiteShadow}[1]{%
	\begingroup
	\leavevmode
	\rlap{\kern0.1ex\raise-0.1ex\hbox{\color{white}#1}}%
	\hbox{#1}%
	\endgroup
}

\tikzstyle{block} = [draw,rectangle,thick,minimum height=2em,minimum width=2em]
\tikzstyle{sum} = [draw,circle,inner sep=0mm,minimum size=2mm]
\tikzstyle{connector} = [->,thick]
\tikzstyle{line} = [thick]
\tikzstyle{branch} = [circle,inner sep=0pt,minimum size=1mm,fill=black,draw=black]
\tikzstyle{guide} = []
\tikzset{>=latex}
\usepackage{pgfplots}
\pgfplotsset{compat=newest}

\makeatletter
\pgfmathdeclarefunction{invgauss}{2}{%
	\pgfmathln{#1}% <- might need parsing
	\pgfmathmultiply@{\pgfmathresult}{-2}%
	\pgfmathsqrt@{\pgfmathresult}%
	\let\@radius=\pgfmathresult%
	\pgfmathmultiply{6.28318531}{#2}% <- might need parsing
	\pgfmathdeg@{\pgfmathresult}%
	\pgfmathcos@{\pgfmathresult}%
	\pgfmathmultiply@{\pgfmathresult}{\@radius}%
}
\pgfmathdeclarefunction{randnormal}{0}{%
	\pgfmathrnd@
	\ifdim\pgfmathresult pt=0.0pt\relax%
	\def\pgfmathresult{0.00001}%
	\fi%
	\let\@tmp=\pgfmathresult%
	\pgfmathrnd@%
	\ifdim\pgfmathresult pt=0.0pt\relax%
	\def\pgfmathresult{0.00001}%
	\fi
	\pgfmathinvgauss@{\pgfmathresult}{\@tmp}%
}
\makeatother

\newcommand{\ubar}[1]{\text{\underline{${#1}$}}}%{\underaccent{\bar}{#1}}
\newcommand{\units}[1]{\,\mbox{#1}}
\newcommand\CDOT{\ensuremath{\cdot}}
\newcommand{\Bm}{{\mathbf B}}

\newcommand{\Cm}{{\mathbf C}}

\newcommand{\cb}{c\subscript{b}}

\newcommand{\C}{\mathcal{M}}

\newcommand{\D}{\pmb {\mathfrak{D}}}

\newcommand{\dd}{\mathbf{d}}

\newcommand{\G}{{\mathbf G }}

\newcommand{\I}{\mathcal{I}}

\newcommand{\kpre}{k\subscript{pre}}
\newcommand{\keq}{k\subscript{eq}}
\newcommand{\kpost}{k\subscript{post}}
\newcommand{\Km}{{\mathbf K}}

\newcommand{\Le}{{\mathcal{L}}}

\newcommand{\Mm}{{\mathbf M}}

\newcommand{\Mk}{{\mathcal{M}}_{{k}}}

\newcommand{\Ns}{N\subscript{s}}

\newcommand{\npow}{n\subscript{pow}}
\newcommand{\nm}{N_\mathrm{o}}
\newcommand{\Nvr}{N_\mathrm{vr}}

\newcommand{\Nr}{N_\mathrm{r}}

\newcommand{\pdf}{\mathrm{p}}

\newcommand{\q}{{\mathbf q}}

\newcommand{\Qy}{Q\subscript{y}}

\newcommand{\rr}{{\mathbf r}}

\newcommand{\rk}{r\subscript{k}}
\newcommand{\rd}{r\subscript{d}}
\newcommand{\rmd}{\mathrm{d}}

\newcommand{\ub}{x\subscript{b}}
\newcommand{\uu}{{\mathbf u }}

\newcommand{\ubd}{{\dot x}\subscript{b}}

\newcommand{\wt}{\mathcal{W}}

\newcommand{\xd}{x\subscript{d}}
\newcommand{\xy}{x\subscript{y}}

\newcommand{\X}{{\mathbf X}}

\newcommand{\y}{{\mathbf y}}

\newcommand{\ksieq}{\zeta\subscript{eq}}

\newcommand{\Sigm}{\boldsymbol{\Sigma}}

\newcommand{\Thetaa}{\boldsymbol{\Theta}}
\newcommand{\thetaa}{\boldsymbol{\theta}}

\renewcommand{\epsilon}{\text{\usefont{OML}{cmr}{m}{n}\symbol{15}}}
\newcommand{\epsb}{\pmb{\epsilon}}
\newcommand{\lbi}{\ubar\epsilon_i}%{T\subscript{lb,{\it i}}}
\newcommand{\ubi}{\overline\epsilon_i}%{T\subscript{ub,{\it i}}}
\newcommand{\lbb}{\ubar\epsilon}%{T\subscript{lb,{\it i}}}
\newcommand{\ubb}{\overline\epsilon}%{T\subscript{ub,{\it i}}}
\newcommand{\zetaeq}{\zeta\subscript{eq}}
\newcommand{\ie}{{\emph{i.e.}}, }
\newcommand{\eg}{{\emph{e.g.}}, }

\newcommand{\subb}{_{\mathrm{b}}}
\newcommand{\subg}{_{\mathrm{g}}}
\newcommand{\subs}{_{\mathrm{s}}}
\newcommand{\fref}[1]{Figure~\ref{#1}}

\newcommand{\transpose}{^{\mathrm{T}}}
\newcommand{\superscript}[1]{\ensuremath{^{\textrm{#1}}}}
\newcommand{\subscript}[1]{\ensuremath{_{\textrm{#1}}}}

\newcommand{\sgn}{\operatorname{sgn}}
\newcommand{\Tstrut}{\rule{0pt}{2.6ex}}       % top strut
\newcommand{\Bstrut}{\rule[-1.2ex]{0pt}{0pt}} % bottom strut

\DeclareMathAlphabet{\mathcal}{OMS}{cmsy}{m}{n}

\pgfplotsset{
	colormap={whitered}{color(0cm)=(blue!15);  color(0.1cm)=(orange!60!red); }
}
\tikzset{
	declare function={gaussx(\xx,\mm,\sm)=1/(\sm*sqrt(2*pi))*exp(-(\xx-\mm)^2/(2*\sm^2));},
	declare function={bivarxy(\xx,\yy,\ma,\sa,\mb,\sb)= 1/(2*pi*\sa*\sb) * exp((-((\xx-\ma)^2/\sa^2 + (\yy-\mb)^2/\sb^2))/2);},
	declare function={bivarxy_rho(\xx,\yy,\ma,\sa,\mb,\sb,\rh)= 1/(2*pi*\sa*\sb*sqrt(1-\rh^2)) * exp(
		(-((\xx-\ma)^2/\sa^2 + (\yy-\mb)^2/\sb^2 - 2*\rh*(\xx-\ma)*(\yy-\mb)/\sa/\sb))/2/(1-\rh^2));},
	declare function={range1=1.4;},
	declare function={range2=1.4;},
	declare function={marginalscaling=1;},
	declare function={xzscale=\pgfkeysvalueof{/pgfplots/zmax}/2/range1;},
	declare function={yzscale=\pgfkeysvalueof{/pgfplots/zmax}/2/range2;},
	yzplane/.style={canvas is yz plane at x=#1},yzloc/.style={shift={#1},scale=0.7,xscale=1.000}, %note: the yscale must be in the node
	xzplane/.style={canvas is xz plane at y=#1},xzloc/.style={shift={#1},scale=0.7,xscale=0.625}, %note: the yscale must be in the node attributes for some reason
}
\newcommand\FWERepsbar{0.6}
\newcommand\FDRepsbarone{0.4705}
\newcommand\FDRepsbartwo{0.6111}
\tikzset{
	declare function={lambda1FWERepsbar=gaussx(\FWERepsbar,mu2,sigma2)*marginalscaling;},
	declare function={lambda2FWERepsbar=gaussx(\FWERepsbar,mu2,sigma2)*marginalscaling;},
	declare function={lambda1FDRepsbarone=gaussx(\FDRepsbarone,mu1,sigma1)*marginalscaling;},
	declare function={lambda1FDRepsbartwo=gaussx(\FDRepsbartwo,mu1,sigma1)*marginalscaling;},
	declare function={lambda2FDRepsbarone=gaussx(\FDRepsbarone,mu2,sigma2)*marginalscaling;},
	declare function={lambda2FDRepsbartwo=gaussx(\FDRepsbartwo,mu2,sigma2)*marginalscaling;},
}

\NameTag{De \textit{et al.}, \today}
\begin{document}

% You will need to make the title all-caps
\title{Model Falsification for Predicting Dynamical Responses of Uncertain Structural Systems }

\author[1]{Subhayan De, Member, ASCE} 
\author[2]{Tianhao Yu} 
\author[3]{Patrick T.~Brewick, Member, ASCE} 
\author[4]{Erik A.~Johnson*, Member, ASCE}
\author[5]{Steven F.~Wojtkiewicz, Member, ASCE} 

\affil[1]{{Department of Mechanical Engineering}, {Northern Arizona University, Flagstaff}, {{AZ 86011}, {USA}}}
\affil[2]{{Previously at Sonny Astani Department of Civil and Environmental Engineering}, {University of Southern California, Los Angeles}, {USA}. 
Currently at College of Civil Engineering, Hefei University of Technology, Hefei 230009, P.~R.~China} 
% yutianhao@hfut.edu.cn, chasing1990yth@outlook.com 
\affil[3]{{Department of Civil and Environmental Engineering and Earth Sciences}, {University of Notre Dame, Notre Dame}, {{IN 46556}, {USA}}}
\affil[4]{{Sonny Astani Department of Civil and Environmental Engineering}, {University of Southern California, Los Angeles}, {USA}}
\affil[5]{{Department of Civil and Environmental Engineering}, {Clarkson University,
		Potsdam}, {{NY 13699}, {USA}}}
\maketitle

% Please include an abstract:
\begin{abstract}
Accurate prediction of dynamical response of structural system depends on the correct modeling of that system. However, modeling becomes increasingly challenging when there are many candidate models available to describe the system behavior. 
% Even with suitable models for different components of a dynamic system, parameter variations can still lead to significant uncertainty in the predicted response. 
% In recent years, passive control devices, such as base isolation devices or tuned-mass-dampers are often used with the structure to improve its response when subjected to earthquake or wind loads. 
% Moreover, selecting among these candidates does not eliminate the challenge, as the chosen representations may still involve parameter variations that affect accuracy. 
% While different model classes can be used to represent the behavior of these control devices, uncertainties can be present even for the parameters of these model classes.
%passive control devices are often fitted to structures to alleviate 
While different model classes can be used to represent the behavior of different components of the dynamic system, uncertainties can be present even for the parameters of these model classes. 
The plausibility of each input-output model class of the structures with uncertain {\color{black}components} can be determined by a Bayesian approach from measured dynamic responses to one or more input records; predictions of the structural system response to alternate input records can then be made. % in a manner that is robust to model class uncertainty.  
However, this approach may require many model simulations, even though most of those model classes are quite implausible. 
An approach is proposed herein to use a bound, computed from the false discovery rate, on the likelihood of measured data to falsify models considering uncertainties in the passive control devices that do not reproduce the measured data to sufficient accuracy.  Response prediction is then performed using the unfalsified models in an approximate Bayesian sense by assigning weights, computed from the likelihoods, only to the unfalsified models. When predicting the response to one or more new input scenarios, this approach incurs a fraction of the computational cost of the standard Bayesian approach because response simulations are no longer required for the models that have been falsified.  %The robustness of the predicted responses are also tested for varying noise levels and different initial model sets. 
The proposed approach for response prediction is illustrated using three structural examples: an earthquake-excited four--degree-of-freedom building model with a hysteretic isolation layer; a 1623--degree-of-freedom three-dimensional building model, with tuned mass dampers attached to its roof, subjected to wind loads; and a full-scale four-story base-isolated building tested on world's largest shake table in Japan's E-Defense lab. The results exhibit very accurate response predictions and significant computational savings, thereby illustrating the potential of the proposed method. 
\end{abstract} 

% \keywords{Response prediction, passive control devices, model falsification, false discovery rate.} 

\section{Introduction}

The attempt to describe a physical system through mathematical models is often driven by the need to predict its behavior, such as is necessary in control design, reliability estimation, health monitoring and lifetime prognosis.
Typically, appropriate models are chosen by computing their responses to one or more input scenarios and comparing model responses with the corresponding measurements; these models are then used to predict responses to alternate inputs that could not be explored physically (\eg responses to other historical or synthetic earthquake records).
However, there is always uncertainty in such modeling because of measurement noise, finite response durations, a limited set of candidate models, and so forth; thus, a probabilistic framework is necessary to quantify the plausibility of the candidate models.

{\color{black}Uncertain dynamic systems, where nonlinearities are often spatially localized yet critically influence global dynamics, present significant challenges for accurate modeling and response prediction. Examples include buildings with nonlinear base isolators \citep{kelly1990base,nagarajaiah1991nonlinear}, rubber bearings \citep{kelly1993earthquake}, sliding bearings \citep{mostaghel1983response,yang1990response}, magnetorheological dampers \citep{ramallo2002smart,yoshioka2002smart}, tuned mass damper (TMD) \citep{warburton1982optimum,sadek1997method,hoang2008optimal,kareem1995performance,yang2022vibration}, distributed mass damper systems \citep{yamaguchi1993fundamental,abe1994dynamic,igusa1994vibration,fu2010distributed}, bridges with frictional joints \cite{ali1995modeling}, structures interacting with soil \citep{lou2011structure}, structures with bolted or riveted joints exhibiting microslip and hysteresis \citep{ruderman2014modeling}, fluid-structure interaction in systems like pipelines and valves \citep{hou2012numerical,de2022prediction}, and so on. 
These systems are often affected by a combination of localized nonlinearities, parametric uncertainty, and measurement noise, and exhibit strong sensitivity to unmodeled phenomena \citep{gattulli2004nonlinear,alexander2009exploring,chatzi2010experimental,De2017selection}. However, constructing a single model that captures all relevant dynamics is rarely feasible, especially when experimental data is sparse or noisy. As a result, engineers typically rely on an ensemble of approximate models, each embodying different physical assumptions or idealizations.} %Exhaustive Bayesian model selection becomes computationally impractical in such cases, especially when dealing with high-dimensional models or large candidate sets. Instead, the model falsification framework provides a statistically rigorous and computationally scalable alternative by focusing on the early rejection of implausible models, thereby narrowing the space to a reduced set of unfalsified models for response prediction. This paradigm is particularly suited to high-dimensional, heterogeneous, and partially observed systems, as commonly encountered in both civil and mechanical engineering practice. 
In this paper, different structures with passive control devices, such as base isolation or tuned mass dampers, are used for response prediction in the presence of uncertainty.

Using prior and posterior distributions of model parameters, prediction has been performed in a Bayesian framework \citep{beck1998,muto2008bayesian,beck2013prior}, %in a manner that is robust to modeling uncertainty (\eg Beck and Katafygiotis \cite{beck1998}, Muto and Beck \cite{muto2008bayesian}, Beck and Taflanidis \cite{beck2013prior}), 
though it is often assumed that the \textit{true} model class is in the candidate pool, which may not always be the case.
Some approximate Bayesian model selection methods exist --- \eg Laplace's approximation or the approximations described in \cite{wasserman2000bayesian} --- but these require estimation of marginal likelihood (also known as model evidence) using either maximum \textit{a posteriori} or maximum likelihood estimates of the parameters, which can also be computationally expensive to determine for complex realistic systems.
However, the model simulations that are required for model selection and subsequent prediction can incur significant computational cost: the computation time to run a single model simulation multiplied by the number of models (which may be large to ensure fully explored posterior parameter space) multiplied by the number of input scenarios (both those for which one has measurements and those for which subsequent response predictions are desired).

To reduce this computational burden of dynamical response prediction of uncertain systems, the study herein proposes to use the computational tool called \textit{model falsification} to reject the invalid models of the {\color{black}uncertain components of the dynamic systems} that cannot sufficiently explain the measurement data \citep{Popper,BrugarolasSafonov2004}; the remaining unfalsified models can then be used for prediction, significantly reducing the number of model simulations for subsequent response prediction to alternate input records. 
{Applications of model falsification include structural identification \citep{gouletDyn,gouletUncert,goulet2010etal,gouletMonitor,De2017falsification}, the condition assessment of bridges \citep{cao2019enhancing} and buildings \citep{reuland2019measurement}, leak detection in pipe networks \citep{gouletPipes,moser2018leak}, occupant tracking \citep{drira2019model}, and excavation \citep{wang2020comparative}. \cite{pai2022methodology} discussed strategies to select methods such as Bayesian model updating and (residual) error-domain model falsification to interpret any monitoring data. 
Note that error-domain model falsification can be shown to be similar to Bayesian inference with a modified likelihood \citep{pai2017comparing,pai2018comparing}.  
\cite{de2019hybrid} combined model falsification and model selection in a Bayesian setting to address some of the shortcomings of each of these methods and applied this framework to structural systems. Recently, \cite{dasgupta2024model} showed model falsification can be viewed as an approximate Bayesian computation \citep{beaumont2002approximate}.} 
Among various model falsification strategies \citep{De2017falsification}, a likelihood-bound method is used herein, rejecting models that predict low probabilities of 
observing the measurement data. To compute the likelihood bound, a false discovery rate (FDR) \citep{benjamini1995controlling} criterion is chosen herein. FDR, which is defined as the average number of incorrect measurement rejections of a valid model response normalized by the total number of measurement rejections, is useful in rejecting many invalid models of a dynamical system with many measured data points \citep{De2017falsification}.
After likelihood-bound model falsification, subsequent response predictions are performed herein according to Bayes' theorem by assigning weights to the unfalsified models (excluding the falsified models to reduce computation cost). %, thereby providing a robustness to the uncertainty in modeling.

This approach provides a computationally inexpensive tool for response prediction while providing a sanity check on the initial candidate model and model class pool.
This approach is illustrated by three numerical examples. The first example uses a three-story superstructure on a hysteretic base isolation layer; different linear and nonlinear model classes are used to model the isolation layer. After the application of model falsification models from the remaining model classes are used to predict the structure response under seismic excitation.
The second example employs a complex 1623 degree-of-freedom model of a building with three roof-mounted nonlinear tuned mass dampers (TMD), each modeled with several linear and nonlinear model classes, subjected to wind load. Again, the unfalsified models are used to predict the response of the structure under a different wind excitation.
The third example uses measurements from experiments in which a full-scale base-isolated four-story building mounted on world's largest shake table at Japan's E-Defense lab was subjected to random base excitations. Four model classes for the superstructure are considered for model falsification. Unfalsified models from two of these classes are used next for response prediction.
The results from these examples show that the proposed method using the unfalsified models accurately predicts responses to alternate input scenarios.

\section{Methodology}

A \emph{model class} is a set of equations with uncertain parameters that attempts to describe the input-output behavior of a physical system and a \emph{model} is a particular parameterization of those equations (others may call these a \emph{model} and a choice of \emph{parameters}, respectively).
Let $\mathscr{M}=\{\C_1, \C_2, \dots\}$ be the set of different model classes considered to describe a particular system.
A model, within some model class $\Mk$, is specified by the value of a parameter vector $\thetaa^{(k)}$ (the superscript is subsequently omitted for notational simplicity). 
A model's $\nm$ outputs (due to some input record) and the corresponding actual response measurements are assembled in the $N_\mathrm{o}\times1$ vectors ${\mathbf{h}}(\thetaa)$ and $\mathbf{d}$, respectively, in a stacked form; \eg $\dd=\left[\y^\mathrm{T\!}(0\Delta t) ~~ \y^\mathrm{T\!}(1\Delta t) ~~ \cdots ~~ \right]^\mathrm{T}$ if the measurements $\y(t)$ are sampled at time interval $\Delta t$.
The residual error vector, defined as
\begin{equation}
\epsb=\mathbf{h}(\thetaa)-\mathbf{d},
\end{equation}
contains the differences between measurements and model predictions and must be used to evaluate the model suitability because the \textit{true} response of the system is unknown.
These residuals $\epsb$ are modeled as continuous random variables with marginal probability density functions {$\pdf_{E_{i}}(e_{i}|\thetaa)$}
%These residuals may be modeled as (typically, continuous) random variables with marginal probability density functions {$\pdf_{E_{i}}(e_{i}|\thetaa)$}, 
where {$E_{i}$} is the random variable denoting the $i$\textsuperscript{th} residual error, $e_i$ is a possible value of random variable $E_i$, and $\epsilon_i$ is the actual residual error (realization of $E_i$) for a particular model. 

\subsection{Likelihood-bound Model Falsification}
The errors $\epsb$ are expected to be small for models that predict the system response reasonably well. Various approaches can be used to provide criteria for accepting or rejecting a model.
%Since the residual error is the quantity based on which the model falsification is performed it can directly be compared to some bounds on it. However, 
For exploratory studies, \cite{De2017falsification} showed that a bound based on a \emph{likelihood function} is useful for rejecting most of the invalid models while keeping the valid ones. The \emph{likelihood function} $\Le(\thetaa;\D)$  is defined as the probability of observing the measurement data $\D=\{\dd\}$ given the model $\thetaa$ from a model class; \ie
\begin{equation}
\Le(\thetaa;\D)
=\pdf_\mathbf{E}(\mathbf{h}(\thetaa)-\dd|\thetaa)
\end{equation}
The common assumption is that the residual errors are jointly Gaussian distributed
\begin{equation}
\Le(\thetaa;\D)=\frac{1}{(2\pi)^{{\nm}/{2}}|\Sigm|^{{1}/{2}}}{ \exp\left(-\frac{1}{2}\epsb\transpose\Sigm^{-1}\epsb \right)  }{}
\end{equation}
where $\Sigm$ is the chosen (or assumed) covariance structure of the residuals $\epsb$.
A likelihood-bound model falsification accepts a model if its likelihood exceeds a particular threshold 
\citep{De2017falsification}:
\begin{equation} \label{eq:intra1}
%\lambda\subscript{low}<\Le(\boldsymbol{\theta};d)\leq\lambda\subscript{up} 
\Le(\thetaa;\D)>\ubar \Le \quad\Rightarrow\quad \text{accept $\thetaa$} 
\end{equation}
where $\ubar\Le$ is a likelihood lower bound defined based on some error criterion. 
(The likelihoods and the likelihood bound are often very small numbers, so their calculations are performed in log scale to avoid numerical errors.)
While several error criteria have been explored by the authors to define likelihood bounds \citep{De2017falsification}, this study calculates the likelihood bounds from multiple-hypotheses-testing ranges on the residual errors $[\lbi,\ubi]$, computed using the false discovery rate (FDR) introduced by {\cite{benjamini1995controlling}}
\begin{equation}\label{eq:lbL2}
\ubar\Le\
%=\min_{\lbbb \,\le\, \mathbf{e} \,\le\, \ubbb} \; \pdf_E(\mathbf{e}|\thetaa)
=\prod_{i=1}^{\nm}\min_{\lbb_i \,\le\, e_i \,\le\, \ubb_i} \; \pdf_{E_i}(e_i|\thetaa) 
%=\pdf(\epsb^{*}|\thetaa)
\end{equation} 
%{\color{purple!70!blue}(EAJ: Something is not right with $\ubi$ and $\lbi$. SD: I agree. However, the same commands produce different results when not used inside $\min$ or $\int$. EAJ: changing the definition of \textbackslash ubar to use \textbackslash text\{...\} instead of \textbackslash mbox\{...\} fixes the problem.)}

\subsection{Use of False Discovery Rate (FDR)}

The FDR is defined as the expected fraction of measurement rejections that are incorrect. For example, if a model is falsified based on rejecting $\Nr$ of the $\nm$ measurements, but $\Nvr$ of those $\Nr$ rejections were incorrect (\ie the model did accurately predict those $\Nvr$ responses but measurement noise or modeling error caused the rejections), then
\begin{equation}
{\text{FDR}}=
\begin{cases}
\mathbb{E}\left[\left.({\Nvr}/{\Nr})\right\vert \Nr>0\right]\mathbb{P}(\Nr>0) \qquad \text{for }R\neq 0\\
0 \qquad\qquad\qquad\qquad\qquad\qquad\quad~~~\! \text{for }R=0
\end{cases}
\end{equation}
On average, FDR control ensures that the fraction of incorrect rejections is below some predefined significance value $\alpha$, and provides better 
statistical power (\ie the probability of rejecting a model when it is invalid)
than other conventional methods such as family-wise error rate control (FWER) % while allowing some false positive results
\citep{bouaziz2012multiple}.
Hence, FDR control should perform better in falsifying invalid models with many data points \citep{De2017falsification}.
The Benjamini-Hotchberg (BH) procedure for controlling FDR at $\alpha$ suggests first sorting the residual errors according to their $p$-values.\footnote{The $p$-values for two-sided distributions can be defined as:
	\begin{align*}\label{eq:pvalues}
	p_i &= 2 \, \min \left\{ \mathbb{P}\left(E_{i} \leq \epsilon_{i} |\thetaa \right), \mathbb{P}\left(E_{i} \geq \epsilon_{i} |\thetaa \right) \right\}, \qquad i = 1,\dots,\nm
	\\\nonumber
	\mbox{}&= 2 \, \min \left\{ \int_{-\infty}^{\epsilon_{i}} \pdf(e_i|\thetaa) \rmd e_i, \int_{\epsilon_{i}}^{\infty} \pdf(e_i|\thetaa) \rmd e_i \right\}
	\end{align*}
}, \ie 
\begin{equation}\label{eq:porder}
0 \leq p_1 \leq p_2 \leq \dots \leq p_{\nm} \leq 1
\end{equation}
After the sorting, the significance level for each residual error $\epsilon_i$ is chosen as
\begin{equation}\label{eq:bh2}
\bar\alpha_{i} =
\frac{i}{\nm} \alpha,
\quad \quad
i=1,\dots,\nm
\end{equation}
where the target identification probability is {$\phi=1-\alpha$}. Typically $\phi$ is chosen as 0.95 or 0.90 in hypothesis testing. %which comes from hypothesis testing (typically $\phi=0.95$, or 0.90). 
The error bounds $[\lbi,\ubi]$ are then determined by
\begin{equation}\label{eq:lim_eval}
\begin{split}
\frac{1}{2}{\bar{\alpha}}_i &= \mathbb{P}(E_i \leq \lbi|\thetaa) = \mathbb{P}(E_i \geq \ubi|\thetaa)\\
&= \int_{-\infty}^{\lbi}\pdf_{E_{i}}(e_{i}|\thetaa)\rmd e_{i}
= \int_{\ubi}^{\infty}\pdf_{E_{i}}(e_{i}|\thetaa)\rmd e_{i}\\
%&\qquad \qquad \quad i=1,\dots,\nm\\
\end{split}
\end{equation} 
% {\color{purple!70!blue}(EAJ: $\ubi$ and $\lbi$ are of different sizes inside the integration. SD: They seem to be okay outside the integration. No idea why.)}

\begin{figure}
	\centering
	\includegraphics[scale=1.5]{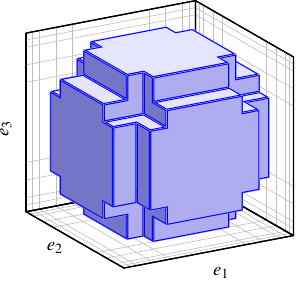}
	\caption{Using false discovery rate models with their residual error falling inside the shaded box are unfalsified.}
	\label{fig:fdr}
\end{figure}
\begin{figure}%{r}{7.5cm}%[htb!]
%	\vspace{-5mm}
	\centering
	\includegraphics[scale=1.2]{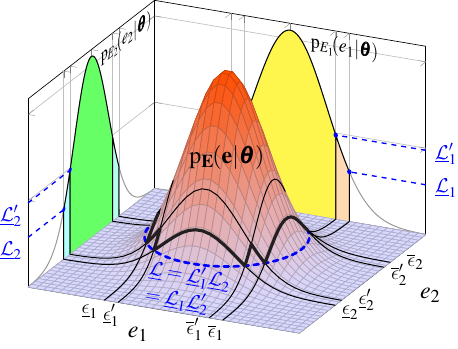} 
	\caption{Likelihood bounds with two measurements.
		(${\bar\alpha}_1=\alpha/2$ determines $\ubar\epsilon_j$ and $\overline{\epsilon}_j$; ${\bar\alpha}_2=\alpha$ determines $\ubar\epsilon_j^{\prime}$ and $\overline{\epsilon}_j^{\prime}$.
		%($\lbb_j$ and $\ubb_j$ are decided using ${\bar\alpha}_1=\alpha/2$ and $\lbb_j^{\prime}$ and $\ubb_j^{\prime}$ are decided using ${\bar\alpha}_2=\alpha$.) 
		Uncorrelated joint densities are depicted but this need not be assumed.) \label{fig:hyptest2}}%\vspace{-10mm}
\end{figure}

To better visualize the effect of the FDR, consider three measurements and their corresponding residual errors $e_1,e_{2},$ and $e_3$. \fref{fig:fdr} shows the combination of these residual errors for which the model will be accepted.
% where the shaded complicated box represents all the residual errors from models that will be accepted.
These error bounds $[\lbi,\ubi]$ are then used in \eqref{eq:lbL2} to compute $\ubar\Le$. For example, \fref{fig:hyptest2} shows the use of FDR to evaluate $\ubar\Le$ in a two-measurement case.
\pgfmathdeclarefunction{gauss}{2}{%
	\pgfmathparse{1/(#2*sqrt(2*pi))*exp(-((x-#1)^2)/(2*#2^2))}%
}%

\subsection{Response Prediction}

The \textit{posterior model probability} $\pdf(\thetaa|\D,\C)$ for model class $\C$ is computed via Bayes' Theorem
\begin{equation}\label{eq:posteriors}
%\begin{split}
\pdf{(\thetaa|\D,\C)}=\frac{\Le(\thetaa;\D)\pdf(\thetaa|\C)}{\int\Le(\thetaa;\D)\pdf(\thetaa|\C)\rmd\thetaa}
%&=\frac{1}{\mathcal{Z}}\left[{\Le(\thetaa;\D)\pdf(\thetaa|\C)}\right]\\
%&=\frac{\pdf(\D|\thetaa^{(k)},\Mk)\pdf(\thetaa^{(k)}|\Mk)}{\pdf(\D|\Mk)}\\
%\end{split}
\end{equation}
using model likelihood $\Le(\thetaa;\D)=\pdf(\D|\thetaa,\C)$ and the \textit{prior model probability} $\pdf(\thetaa|\C)$ (which is constant if, prior to data collection and analysis, all models are assumed equally likely, or non-constant based on the modeler's expert judgement or other knowledge of the distribution of $\thetaa$).
Then, following \cite{beck2013prior}, the model class' parameters and subsequent response can be estimated %in a manner that is robust to the modeling uncertainty 
by using the theorem of total probability, which is an average prediction weighted by the models' posterior probabilities.  The parameter estimation, then, is
\begin{equation}\label{eq:param_est}
\begin{split}
\widehat{\thetaa}
&= \mathbb{E}[\thetaa|\D,\C]= \int
%_{\Dth}
\thetaa \; \pdf(\thetaa|\D,\C)\; \mathrm{d}\thetaa\\
%\approx \sum_{i=1}^{\Ns} {\wt}_i \, \thetaa_i
\end{split}
\end{equation}

To predict some quantity of interest $\q(\thetaa|\C,\I)$, which is a response to input record $\I$ (which may be the same input record used to perform the model selection or, more likely, some alternate input for which responses are also desired), the posterior model probabilities $\pdf(\thetaa|\D,\C)$ are again used in a theorem of total probability:
\begin{equation}\label{eq:rob_pred}
\begin{split}
\widehat{\q}(\C,\I)
&= \mathbb{E}[\q(\thetaa|\C,\I)|\D]= \int
%_{\Dth}
\q(\thetaa|\C,\I) \, \pdf(\thetaa|\D,\C) \, \rmd \thetaa\\
%\approx \sum_{i} {\wt}_i \: \q(\thetaa_i|\C,\I)
\end{split}
\end{equation}
While {robustness} to model class uncertainty \citep{beck2013prior} could be evaluated by incorporating models from multiple model classes into \eqref{eq:rob_pred} --- \ie compute $\widehat{\q}(\I)$ by integrating over model classes $\C$  --- this aspect of the proposed approach is not evaluated herein.

\subsection{Model Confidence and Post-falsification Response Prediction}

The falsification proposed in \cite{De2017falsification} can be extended to quantify post-falsification model confidence, and to use unfalsified models to provide parameter and response estimates. % that are robust to the modeling uncertainty. 
In \eqref{eq:posteriors}, the denominator $\pdf(\D|\C)$ --- known as the likelihood %
%$\Le(\C)$
of, or the evidence %
%$\mathcal{E}_{\C}$
for, model class $\C$ --- is the same normalization factor for all models $\thetaa$ in model class $\C$, so it need not be explicitly computed and the numerators can be used in a relative sense.
%that is simply the integral or sum of the numerator over all models $\thetaa$ in model class $\C$.  Thus, the posterior model weight is proportional to the likelihood and the prior model weight.
%$\Dthu$ denotes the set of unfalsified models.  
%Omitting the model class $\C$ for notational simplicity, the 
Suitable post-falsification weights for a sampling-based Bayesian method are then
\begin{equation}\label{eq:weightsBayesian}
\begin{split}
{\wt}_i &= \frac{ {\widehat\wt}_i }{ \displaystyle\sum_{j} {\widehat\wt}_j },\quad{\widehat\wt}_i 
= \Le(\thetaa_i;\D) \pdf(\thetaa_i|\C)\\
\end{split} %= \left\{ \begin{array}{ll} \Le(\thetaa_i;\D) \pdf(\thetaa_i|\C),\mbox{} & \thetaa_i\mbox{ is unfalsified} \\ 0 & \thetaa_i\mbox{ is falsified} \\ \end{array} \right.
\end{equation}
where, for the $\Ns$ models in model class $\C$, the weights $\wt_i$ are normalized so that their sum is unity.
%\comment{It may be noted that \eqref{eq:weightsBayesian} is approximate in that it uses only the finite number of models, and avoids the computationally expensive evaluation of the denominator in the exact Bayesian approach \eqref{eq:posteriors}.}{EAJ: do we need this? or is it obvious?}

As an alternative, consider using non-zero weights only for the unfalsified models:
\begin{equation}\label{eq:weights}
\begin{split}
{\wt}_i &= \frac{ {\widehat\wt}_i }{ \displaystyle\sum_{j} {\widehat\wt}_j },\quad{\widehat\wt}_i 
= \left\{ \begin{array}{ll} \Le(\thetaa_i;\D) \pdf(\thetaa_i|\C),\mbox{} & \thetaa_i\mbox{ is unfalsified} \\ 0 & \thetaa_i\mbox{ is falsified} \\ \end{array} \right.\\
\end{split}
\end{equation}
Using these weights --- and the notation that 
$\Thetaa_\mathrm{f} = \{\thetaa_i: \Le(\thetaa_i;\D) \le \ubar \Le\}$ is the set of the $N_\mathrm{f}$ falsified models from model class $\C$ and $\Thetaa_\mathrm{u} = \{\thetaa_i: \Le(\thetaa_i;\D) > \ubar \Le\}$ is the set of the $N_\mathrm{u}$ unfalsified models (where $N_\mathrm{u}+N_\mathrm{f}=\Ns$)
---
a parameter estimate %that is robust to the uncertainty in modeling 
can be computed with%
\begin{equation}\label{eq:param_est2}
\begin{split}
\widehat{\thetaa}
%= \mathbb{E}[\thetaa|\D,\C]
%= \int_{\Dth} \thetaa \; \pdf(\thetaa|\D,\C)\; \mathrm{d}\thetaa
\approx \sum_{i=1}^{\Ns} {\wt}_i \, \thetaa_i
&= \sum_{\thetaa_i\in\Thetaa_\mathrm{u}} {\wt}_i \, \thetaa_i
+ \sum_{\thetaa_i\in\Thetaa_\mathrm{f}} {\wt}_i \, \thetaa_i\approx \sum_{\thetaa_i\in\Thetaa_\mathrm{u}} {\wt}_i \, \thetaa_i\\
\end{split}
\end{equation}
and the corresponding response prediction is
\begin{equation}\label{eq:rob_pred2}
\begin{split}
\widehat{\q}(\C,\I)
%= \mathbb{E}[\q(\thetaa|\C)|\D]
%= \int_{\Dth} \q(\thetaa|\C) \, \pdf(\thetaa|\D,\C) \, \rmd \thetaa
&\approx
%\sum_{i=1}^{\Ns} {\wt}_i \: \q(\thetaa_i|\C,\I) =
\sum_{\thetaa_i\in\Thetaa_\mathrm{u}} {\wt}_i \: \q(\thetaa_i|\C,\I)
+ \sum_{\thetaa_i\in\Thetaa_\mathrm{f}} {\wt}_i \: \q(\thetaa_i|\C,\I)\approx \sum_{\thetaa_i\in\Thetaa_\mathrm{u}} {\wt}_i \: \q(\thetaa_i|\C,\I)\\
\end{split}
\end{equation}

The computational cost is dominated by the number of model simulations required; the approximation on the right side of \eqref{eq:rob_pred2} uses only the $N_\mathrm{u}$ unfalsified models, whereas a standard sampling-based Bayesian approach would always use all $\Ns>N_\mathrm{u}$ models. Thus, for each additional input scenario $\I$, the proposed approach has computational cost savings of $N_\mathrm{f}/\Ns$: this will asymptotically approach 100\% when many of the models are falsified and, at worst when very few models are falsified, will approach 0\% and the proposed method reverts to a standard sampling-based Bayesian approach.

The accuracies of the approximations in
\eqref{eq:param_est2} and \eqref{eq:rob_pred2}
depend, of course, on the number of models --- the number in the high likelihood regions and, for the proposed approach, the number unfalsified --- which should be sufficiently large (either through a sufficient number of initial candidate models so that they are representative of the whole model class or, for the falsification approach, through a sufficiently large value of $\phi$) for accurate predictions. 
Alternately, an iterative strategy could be implemented
in which the number of models evaluated within a model class is iteratively increased (\eg doubled) until the fraction of models that are falsified converges (within some tolerance). % where $\Ns$ number of models is first used for every model class and it is increased by a factor of two at each iteration until the change in percent of unfalsified models becomes smaller than a certain threshold.
Similarly, if it is found that only a very few models dominate (\ie only a few ${\wt}_i$ are much larger than the rest), then a similar iterative process could be pursued (\eg until the largest ${\wt}_i$ is below some threshold). 
%(EAJ: Should cite Agni's work on the number of models needed to reliably falsify a model class. SD: However, I did not use that number here. EAJ: While not used herein, there are works )
Note that related works have also explored the number of models necessary to reliably falsify a model class \citep{dasgupta2024model}. 

\section{Numerical Illustrations}
Three numerical examples are used next to illustrate the proposed method for response prediction. %that is robust to uncertainty in modeling. 
In the examples, a relative root-mean-square (RMS) error is defined as 
\begin{equation}
\mathrm{RMS~~error} = \frac{\lVert u_{\mathrm{true}}(t)-u_{\mathrm{est}}(t)\rVert_2 \;}{\; \lVert u_{\mathrm{true}}(t)\rVert_2}
\end{equation}
where $u_{\mathrm{true}}$ is the true value of some quantity of interest, $u_{\mathrm{est}}$ is its estimate, and the two norm is defined as $\lVert u(t) \rVert_2 = \left[{\frac{1}{t_\mathrm{f}}\int_{0}^{t_\mathrm{f}}u^2(t)\mathrm{d}t}\right]^{1/2}$.

\subsection{Example I: Base-Isolated Building (4~\!DOF)}
\begin{figure}[htb!]
	\centering
	\includegraphics[scale =1.5]{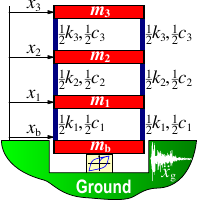}
	\caption{Base-isolated building model}\label{fig:4dof:a}
\end{figure}
Consider the base-isolated building model shown in \fref{fig:4dof:a}, with a hysteretic base isolation system \citep{De2017falsification}. 
Accurate modeling of the hysteretic isolation elements is necessary for useful system response simulation and control design.
The superstructure masses are $m_1=m_2=m_3=300\units{Mg}$; the corresponding story stiffnesses are ${k}_1={k}_2={k}_3=40\units{MN/m}$.
Rayleigh superstructure damping is introduced with 3\% damping in the first two superstructure modes. The base mass is $m\subb=500\units{Mg}$, making the structure weight $W=g\!\cdot\!1400\units{Mg}\approx13.729\units{MN}$.

The equations of motion of the superstructure, if it were fixed base, are given by
\begin{equation}
\Mm\subs\ddot{\X}\subs+\Cm\subs\dot\X\subs+\Km\subs\X\subs=-\Mm\subs\mathbf{1}\ddot x\subscript{g}
\end{equation}
where $\Mm\subs$ and $\Km\subs$ are the $3\times3$ mass and stiffness matrices, respectively, 
% \begin{equation}
$\Mm\subs =\left[\begin{array}{c c c}
m_1 & 0 & 0 \\
0&m_2&0\\
0&0&m_3
\end{array} \right]\!$, $\Km\subs=\left[ \begin{array}{ccc}
{k}_1+{k}_2 &-{k}_2&0\\
-{k}_2&{k}_2+{k}_3&-{k}_3\\
0&-{k}_3&{k}_3
\end{array} \right]\!$, 
% \end{equation}
% with $m_1=m_2=m_3=300$~Mg; $\Km\subs$ is the $3\times3$ stiffness matrix 
% \begin{equation}
% \Km\subs=\left[ \begin{array}{ccc}
% {k}_1+{k}_2 &-{k}_2&0\\
% -{k}_2&{k}_2+{k}_3&-{k}_3\\
% 0&-{k}_3&{k}_3
% \end{array} \right]\!
% \end{equation}
% with ${k}_1={k}_2={k}_3=40$~MN/m; 
$\mathbf{1}_{3\times 1}$ is a column vector of ones, and $\X\subs=\left[x_1\quad x_2\quad x_3\right]\transpose$ contains the floor displacements relative to the ground. %Rayleigh damping with 3\% damping for the first two modes are assumed.
Along with the isolation layer the equations of motion of the full system become 
\begin{align}
&\Mm\subs \ddot{\X}\subs+\Cm\subs \dot{\X}\subs +\Km\subs \X\subs=-\Mm\subs\mathbf{1} \ddot x\subg+\Cm\subs\mathbf{1} \dot{x}\subb+\Km\subs \mathbf{1} x\subb\\\nonumber
&m\subb\ddot{x}\subb+\mathbf{1}\transpose\Cm\subs\mathbf{1} \dot x\subb  + \mathbf{1}\transpose\Km\subs \mathbf{1} x\subb+f\subb=-m\subb\ddot x\subg+\mathbf{1}\transpose\Cm\subs\dot{\X}\subs +\mathbf{1}\transpose\Km\subs \X\subs
\end{align}
% where $m\subb=500$ Mg is the base mass. 
where $f\subb$ is the force exerted by the hysteretic isolation layers, which is described by 
different model classes as 
% are to describe the restoring force in the isolation layer are 
discussed next.

\subsubsection{Candidate Model Classes for Base-Isolation Layer}
Two nonlinear and four linear model classes are candidates to represent the behavior of the isolation layer: a bilinear model, a smoother and more realistic \citep{Nagarajaiah2000} Bouc-Wen hysteresis model \citep{wen1976method}, and linear models specified by AASHTO (American Association of State Highway and Transportation Officials), JPWRI (Japanese Public Works Research Institute), Caltrans (California Department of Transportation) and a modified AASHTO.

\begin{figure}
	\centering
	\includegraphics[scale=0.25]{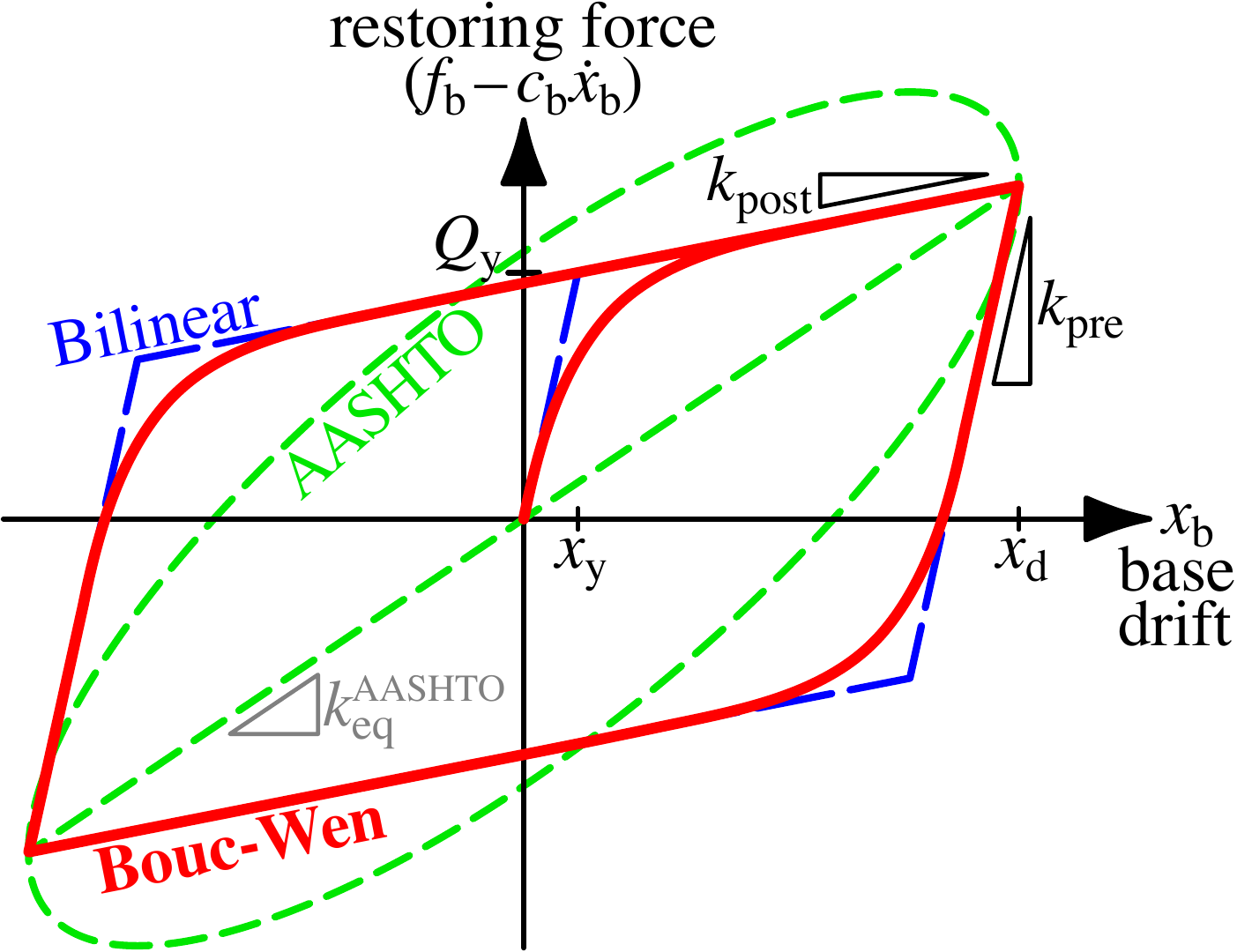}
	\caption{Several model classes to represent the force-displacement loops.}\label{fig:4dof:b}
\end{figure}

In these model classes, $\kpre$, $\kpost$, $\Qy$, and $\rd$ are the pre-yield and post-yield stiffnesses, yield force, and the ratio of design displacement $\xd$ to yield displacement $\xy=\Qy/\kpre$, respectively, as shown in \fref{fig:4dof:b}.The non-elastic restoring force in the nonlinear model classes is represented by $q_\mathrm{y} z$, where $q_\mathrm{y}=\Qy\left(1-\rk\right)$, the hardness ratio is $\rk=\kpost/\kpre$, and $z$ is an evolutionary variable given by% \cite{ma2004parameter}
\begin{equation} \label{eq:bouc}
\dot z = A \ubd - \beta \ubd|z|^{\npow}-\gamma z|\ubd||z|^{\npow-1}
\end{equation}
Herein, $A = 2\beta = 2\gamma = \kpre/\Qy$ is used so that
the loading and unloading stiffnesses are the same \citep{ramallo2002smart}. 
Further, $\npow=1$ is assumed for the Bouc-Wen models and $\npow=100$ is used to represent the bilinear models. In addition to the hysteretic restoring force, a linear viscous damping force $\cb$ is also assumed. 

The linear model classes approximate the hysteretic isolation behavior by defining a linear stiffness and damping with roughly equivalent per-cycle energy dissipation.
For these models the isolator force can be written as
\begin{equation}
\begin{split}
f\subscript{b}&=\left[\cb+c\subscript{eq}\right]\ubd+\keq\ub
%\\&
=\left[\cb+2\zetaeq\sqrt{\keq m}\right]\ubd+\keq\ub
\end{split}
\end{equation}
where $\ksieq$ is the equivalent damping ratio  and $\keq$ equivalent stiffness.
Linear models of AASHTO (American Association of State Highway and Transportation Officials) and JPWRI (Japanese Public Works Research Institute) uses the following expressions for $\ksieq$ and $\keq$
\begin{equation}\label{eq:AASHTO_JPWRI}
\begin{split}
\ksieq&=\frac{2(1-\rk)(1-\rho^{-1})}{\pi[1+\rk(\rho-1)]},~~
%%%\qquad\mbox{and}\qquad
\keq=\frac{\kpre}{\rho}{[1+\rk(\rho-1)]},~~\rho=\begin{cases}
    \rd,~~~~~\quad \text{AASHTO}\\
    0.7\rd,\quad \text{JPWRI} 
\end{cases}\\ 
% \rho&=\rd,~~~~~\quad \text{AASHTO}\\
% \rho&=0.7\rd,\quad \text{JPWRI}\\
%&\rd=\xd/\xy\\
\end{split}
\end{equation}
where $\rd=\xd/\xy$ is known as the shear ductility ratio \citep{kawashima1992manual,hwang1996equivalent}. A modified AASHTO model 
\citep{hwang1996equivalent} using \eqref{eq:AASHTO_JPWRI} but with correction factors $\rd^{0.58}/(6-10\rk)$ and $\left[1-0.737{(\rd-1)}{/\rd^2}\right]^{-2}$ for $\ksieq$ and $\keq$, respectively, is also used as a linear model class herein.
The fourth linear model class is specified by Caltrans as \citep{hwang1994practical}
\begin{equation}
\begin{split}
&\ksieq=0.0587(\rd-1)^{0.371}\\
&\keq={\kpre}{\{1+\ln[1+0.13(\rd-1)^{1.137}]\}^{-2}}\\
\end{split}
\end{equation}
A comparison of typical hysteresis loops of some of these model classes is shown in \fref{fig:4dof:b}.
Further details of these linear and nonlinear model classes are provided by \cite{De2017falsification,De2017selection}. 
% {\color{purple!70!blue}(EAJ: Italicize et al. SD: natbib does not italicize it.)}

\subsubsection{Measurement Data}
The building model is subjected to ground acceleration $\ddot x\subscript{g}(t)$ and simulated to generate nonlinear dynamic base absolute accelerations $\ddot x_\mathrm{b}^\mathrm{a}(t)$ using the Bouc-Wen model as the \textit{true} isolation layer model.  The $i$\textsuperscript{th} element of data vector $\mathbf{d}$ is, then, $d_i = \ddot x_\mathrm{b}^\mathrm{a}([i-1]\Delta t) + v_i$ where $\Delta t = 0.05\units{s}$, $N_\mathrm{o}=600$, and measurement noise $v_i$ is independent zero-mean Gaussian with a standard deviation that is 20\% that of the noise-free response. Herein, the ``measured'' responses are generated when $\ddot x\subscript{g}$ is the N-S \emph{El Centro}, CA (Imperial Valley Irrigation District substation) earthquake record during the 1940 Imperial Valley earthquake, sampled at 50\units{Hz}, with a peak acceleration of $3.42\units{m/s\textsuperscript{2}}$.
The true Bouc-Wen model parameter values and the prior distributions of the model classes' parameters are listed in Table~\ref{tab2prior2}.
\begin{table}[t]
	%\small
		\caption{True Bouc-Wen model parameters and prior distributions of the model class parameters. (Std.~Dev.~denotes standard deviation.)}
	\label{tab2prior2}
	\centering
	\begin{threeparttable}
		\begin{tabular}{@{} l l l l l l l l l l@{}} 
			\hline
			\Tstrut
			\multirow{2}{*}{Parameter} & \multirow{2}{*}{True value}&\multicolumn{3}{c}{Prior Distribution} &&
			\Bstrut\\
			\cline{3-5}  \Tstrut & & Type & Mean & Std.~Dev. && \Bstrut\\
			\hline
			\Tstrut
			$\kpost$ [MN/m] & $4.0$ & Lognormal &$4.5$  & $0.25$ && \\
			%\hline
			$\cb$ [kN\CDOT s/m] & $20$ & Lognormal &  $20$  & $4$ && \\
			%\hline
			$\rk$ & 0.1667&Uniform & $0.1600$ & $0.0058$ && \\
			%\hline
			$\rd$ & n/a\tnote{\textdagger}&Uniform & $2.5$ & $0.2887$ &&  \\
			%\hline
			$\Qy$ (\%$W$) & 5.00$^*$&Uniform & $4.75$ & $0.2887$ && \Bstrut\\
			%(in \%) & &\\
			\hline
		\end{tabular}
		\begin{tablenotes}
			\item[\textdagger] {\footnotesize{The nonlinear models do not require $\rd$.}  }
			\item[*] {\footnotesize{The linear models do not require $\Qy$.}}
			%\item[\textdaggerdbl] {\footnotesize{Maximum Likelihood estimate of the Bilinear model parameters is same as $\widehat{\thetaa}$ up to four decimal points.}  }
		\end{tablenotes}
	\end{threeparttable}
%	\vspace{-12pt}
\end{table}
\subsubsection{Results}
For each model class $\C$, $\Ns=2000$ models randomly generated from the prior distribution $\pdf(\thetaa|\C)$ are used for falsification. Each residual $\epsilon_i$ is assumed to be Gaussian distributed $\mathcal{N}(0,\sigma^2)$, where the residual standard deviation $\sigma$ is assumed to be  15\% of the standard deviation of the measured absolute base acceleration. 

\begin{table}[htb!]
		\caption{Falsification results for Example I.} \label{tab2fals}
	\centering
	\begin{tabular}{ l  |c c c  c c c c} 
		\hline
		\Tstrut  Model Class & \% Unfalsified\Bstrut \\ 
		%& \multicolumn{2}{|c|}{8 Hz sampling rate} & \multicolumn{2}{|c|}{20 Hz sampling rate} \Tstrut\Bstrut\\ \cline{2-5} \Tstrut
		%	\Tstrut	& \multicolumn{3}{c|}{Error-bound} & \multicolumn{4}{c}{Likelihood-bound}\Bstrut\\
		%		& Bonferroni & \v{S}id\'ak & BH &  Bonferroni & \v{S}id\'ak & BH & CPM \Tstrut\Bstrut\\
		\hline
		\Tstrut Bouc-Wen &  89.8\\
		%		Cubic Polynomial &  96.25  \\
		%		CALTRANS   & 0 \\
		%		 mod.~AASHTO  & 0 \\
		%		Bouc-Wen & 32.2& 32& 29.75& 100&100 & 89.45 & 16.7 \\
		Bilinear	& ~~5.1\\
		AASHTO & ~~0.0\\
		JPWRI & ~~0.0\\
		modified AASHTO & ~~0.0\\
		Caltrans & ~~0.0\Bstrut\\
		%		Bilinear  & 5.1\Bstrut\\
		\hline
	\end{tabular}
	
\end{table}

\begin{figure}[htb!]
	\centering
	\begin{subfigure}[htb!]{0.48\textwidth}
%		\begin{tikzpicture}
%		\node (fig1) at (0,0)
%		{\includegraphics[scale=0.31]{figures/pred_bw.pdf}};
%		\node (fig2) at (0.1,-1.6)
%		{\includegraphics[scale=0.06]{figures/pred_zoomed_BW.pdf}};  
%		\draw (-0.1,-0.8) -- (0.3,-0.8) -- (0.3,0.4) -- (-0.1,0.4) -- (-0.1,-0.8);
%		\draw (-0.1,-0.8) -- (-0.525,-1.05);
%		\draw (0.3,-0.8) -- (0.78,-1.05);
%		\end{tikzpicture}
		        \centering
		   \includegraphics[scale =1]{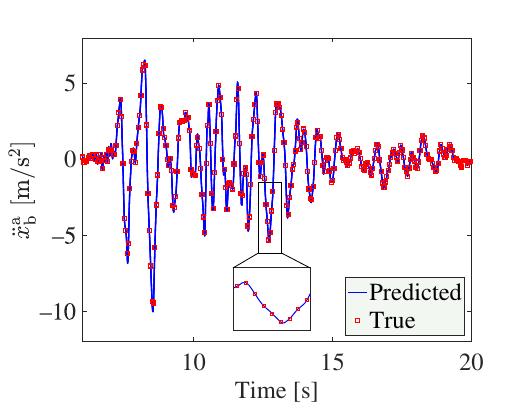}
		\caption{Bouc-Wen model class}\label{fig:kobe:a}
	\end{subfigure}%
	\hfill
	\begin{subfigure}[htb!]{0.48\textwidth}
		        \centering
%		\begin{tikzpicture}
%		\node (fig1) at (0,0)
%		{\includegraphics[scale=0.31]{figures/pred_bilin.pdf}};
%		\node (fig2) at (0.1,-1.6)
%		{\includegraphics[scale=0.06]{figures/pred_zoomed_bilin.pdf}};  
%		\draw (-0.1,-0.8) -- (0.3,-0.8) -- (0.3,0.4) -- (-0.1,0.4) -- (-0.1,-0.8);
%		\draw (-0.1,-0.8) -- (-0.525,-1.05);
%		\draw (0.3,-0.8) -- (0.78,-1.05);
%		
%		\end{tikzpicture}
		        \includegraphics[scale=1]{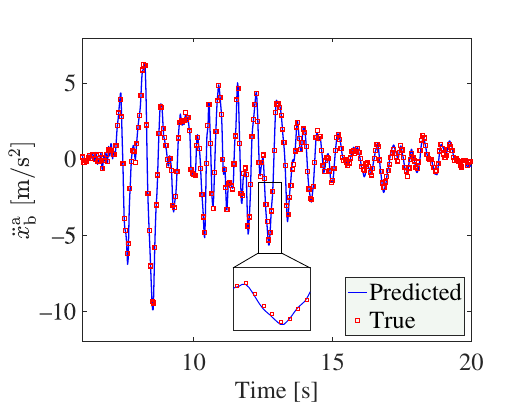}
		\caption{Bilinear model class}\label{fig:kobe:b}
	\end{subfigure}

%	\vspace{-12pt}
\caption{True absolute base accelerations of the 4\,DOF building subjected to the Kobe earthquake, and those predicted from models likelihood-bound-unfalsified using the El Centro data.} \label{fig:kobe}
\vspace{-12pt}
\end{figure}

\begin{table}[t]
	%\small
		\caption{True Bouc-Wen model parameters and estimates of the Bouc-Wen and Bilinear model parameters. (ML denotes maximum likelihood parameter estimates; $\widehat{\thetaa}$ are parameter estimates.)}
	\label{tab2results}
	\centering
	\begin{threeparttable}
		\begin{tabular}{@{} l l l l l l l l l l@{}} 
			\hline
			\Tstrut
			\multirow{2}{*}{Parameter} & \multirow{2}{*}{True value}&& \multicolumn{2}{c}{Bouc-Wen} && Bilinear%\tnote{\textdaggerdbl}
			\Bstrut\\
			\cline{4-5} \cline{7-7} \Tstrut &  && ML & $~~~~\widehat{\thetaa}$ && ML, $\widehat{\thetaa}$\Bstrut\\
			\hline
			\Tstrut
			$\kpost$ [MN/m] & $4.0$ & & 4.0733 & 4.0609 && 3.8701\\
			%\hline
			$\cb$ [kN\CDOT s/m] & $20$ & & 23.3901 & 22.5401 && 19.5726\\
			%\hline
			$\rk$ & 0.1667&& 0.1687 & 0.1681 && 0.1634\\
			%\hline
			%	$\rd$ & n/a\tnote{\textdagger}&& n/a\tnote{\textdagger} & n/a\tnote{\textdagger} && n/a\tnote{\textdagger} \\
			%\hline
			$\Qy$ (\%$W$) & 5.00$^*$&& 4.9315 & 4.9450 && 4.3468\Bstrut\\
			%(in \%) & &\\
			\hline
		\end{tabular}
		%	\begin{tablenotes}
		%	\item[\textdagger] {\footnotesize{The nonlinear models do not require $\rd$.}  }
		%	\item[*] {\footnotesize{The linear models do not require $\Qy$.}}
		%\item[\textdaggerdbl] {\footnotesize{Maximum Likelihood estimate of the Bilinear model parameters is same as $\widehat{\thetaa}$ up to four decimal points.}  }
		%		\end{tablenotes}
	\end{threeparttable}
%	\vspace{-12pt}
\end{table}
Using the likelihood-bound falsification (with target identification probability $\phi=0.95$), every candidate model of each of the linear model classes is falsified; 89.8\% of the candidate Bouc-Wen models and 5.1\% of the bilinear models are unfalsified, as shown in Table \ref{tab2fals}. Hence, even if the Bouc-Wen model class were not considered, the method correctly chooses some bilinear models that reproduce the system responses reasonably well (see \fref{fig:kobe}). However, if the initial candidate set contains only the four linear model classes, the standard Bayesian approach would fail to identify that all of the candidate model classes are indeed poor descriptions of the system. In contrast, the proposed method can identify this situation by falsifying all linear models since it provides a check on the initial candidate model class set.

The estimated parameters, using both maximum likelihood estimation and estimation with \eqref{eq:param_est2} using the unfalsified Bouc-Wen models, are shown in Table \ref{tab2results} and are very close to the corresponding true values.

Finally, the models unfalsified based on the El Centro response data are used to predict the isolated structure response to the 1995 \textit{Kobe} earthquake (N-S record at the Japanese Meteorological Agency in Kobe, Japan, during the 1995 Hy\={o}go-ken Nanbu earthquake, sampled at 50\units{Hz}, with a peak acceleration of 8.18\units{m/s\textsuperscript{2}}).  
%Using the FDR/BH error-bound falsification 
%%(but computing likelihoods for the unfalsified models), 
%(assigning equal weights in \eqref{eq:rob_pred} to all unfalsified models),
%the relative RMS error predicting the Kobe-induced absolute base acceleration of the building is 9.7467\%; 
For the unfalsified Bouc-Wen models using the FDR/BH likelihood-bound falsification with weights assigned according to \eqref{eq:weights} used in \eqref{eq:rob_pred2}, the relative RMS error in predicting the absolute base acceleration is 0.8639\%; the actual and predicted responses are shown in \fref{fig:kobe}.  The corresponding error using only the parameter estimate $\widehat{\thetaa}$
% (rather than the robust prediction)
is a similar 0.8788\%, and is 0.9752\% using the maximum likelihood parameter estimate $\arg\max_i\{\Le(\thetaa_i)\}$. %{\color{red}(EAJ: So the reviewer may ask: ``0.86\% vs. 0.88\%? big deal.'' How do we modify to anticipate this?)} {\color{red}(SD: Are max lik estimates needed? If we remove the max lik estimates from Table 1 then we will have space for Bilinear parameter estimates.)}
Hence, using either \eqref{eq:param_est2} or \eqref{eq:rob_pred2} with the weights $\mathcal{W}_i$ of the unfalsified models provides very accurate response predictions; estimation using \eqref{eq:rob_pred2} is slightly better than the direct use of the estimated parameters \eqref{eq:param_est2}. 
Using the unfalsified bilinear models, prediction \eqref{eq:rob_pred2} gives a 10.1957\% error, which is larger because the parameter estimates from the bilinear models significantly underestimate $\kpost$ and $\Qy$; the bilinear models' maximum likelihood and estimates using \eqref{eq:param_est2}, tabulated together in the rightmost column of Table \ref{tab2prior2}, are the same to four decimal digits because few bilinear models remain unfalsified and one of them, with a high likelihood, dominates the others.  By eliminating the falsified models, predicting the responses to the Kobe earthquake requires only about 1900 model simulations (some bilinear and most Bouc-Wen), whereas the standard Bayesian approach would have simulated all 12,000 models, so the proposed approach would be about six times faster.

\begin{figure}%[scale=0.4\textwidth]
	%  \vspace{-25pt}
	\begin{center}
		\begin{subfigure}[t]{0.48\textwidth}
			\centering
			\includegraphics[scale=0.98]{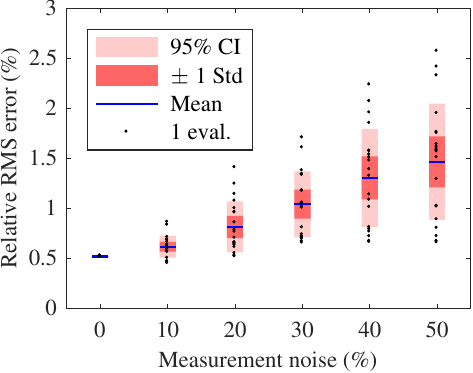}
			\caption{Prediction error vs.~noise level for one candidate model set}\label{fig:4doferror:a}
		\end{subfigure}%
		\hfill
		\begin{subfigure}[t]{0.48\textwidth}
			\centering
			\includegraphics[scale=0.98]{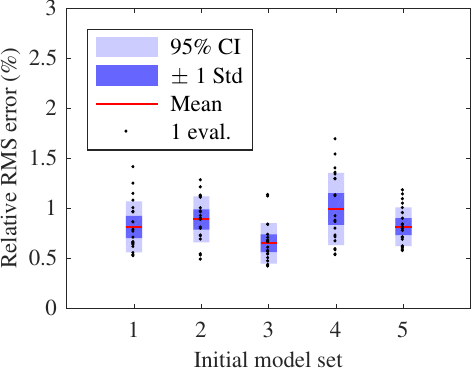}
			\caption{Prediction error for different candidate model sets (20\% measurement noise)}\label{fig:4doferror:b}
		\end{subfigure}
		%    \\[0.5ex]
		%    \includegraphics[scale=1.5]{figures/four_dof_carts.pdf}
	\end{center}
%	\vspace{-22pt}
	\caption{Relative RMS error in predicting the 4\,DOF building's absolute base acceleration response to the 1995 Kobe earthquake with varying measurement noise level and for different candidate model sets. (CI = confidence interval, Std = standard deviation.)} \label{fig:4doferror}
%	\vspace{-12pt}
\end{figure}

The primary sources of model uncertainty are the measurement noise and the choice of candidate models; these are both explored next. % to evaluate the predictions' robustness to the model uncertainty. %The robustness of the proposed method is tested next by varying the measurement noise level and the initial model set for falsification.
The falsification is first repeated for the $\Ns=2000$ Bouc-Wen models with measurement noise levels ranging from 0\% to 50\% of the RMS of the actual absolute base acceleration (but using likelihood bounds that assume a constant 20\% error residual standard deviation) when subjected to the El Centro earthquake.  For each measurement noise level, the resulting unfalsified Bouc-Wen models are used to predict the responses to the Kobe earthquake.  As the measurement noise level increases, \fref{fig:4doferror:a} shows that the relative RMS prediction error remains quite small, and is less than 1.5\% average error even for 50\% measurement noise (note that these errors would all decrease if $\Ns$ increases), verifying that the predictions are relatively {robust} to measurement noise level.
Next, the errors in the predicted Kobe responses are shown in \fref{fig:4doferror:b} for five randomly-chosen sets of $\Ns=2000$ Bouc-Wen candidate models, using the 20\% measurement noise, demonstrating that the predictions are relatively {robust} to the set of candidate models. 
% {\color{purple!70!blue}(EAJ: Why is ``robust" in red? SD: In the initial comments, we were criticized for using the term ``robust." So, I wanted to make sure we are using this here in a non-controversial sense.)}

\subsection{Example II: Complex Wind-Excited Building (1623 DOF)}

\begin{figure}%{r}{5.5cm}
	\vspace{-3pt}
	%\begin{figure}
	%\begin{minipage}{0.375\textwidth}
	%\parbox{.35\linewidth}{
	\centering
	\includegraphics[scale=0.75]{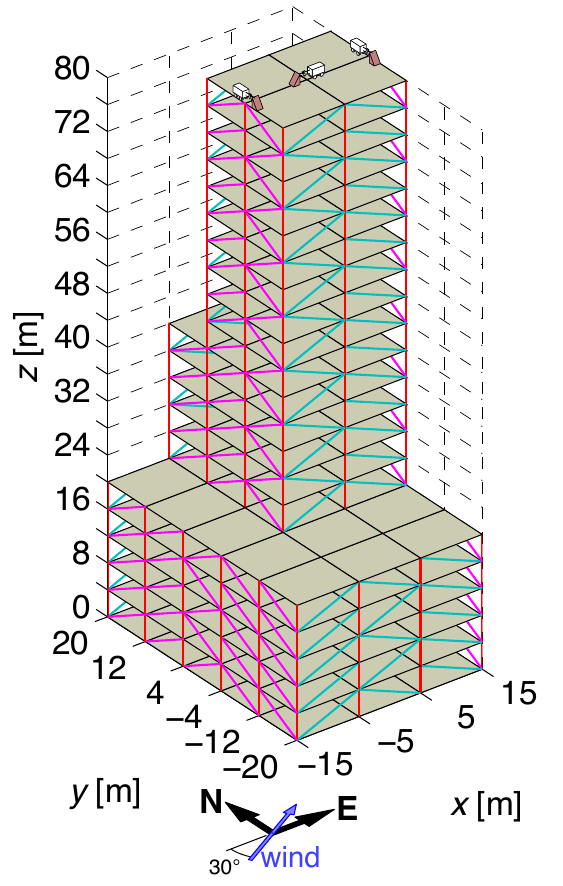}
	\caption{1623 DOF model of a building with TMDs on its roof, subjected to wind load.} \label{fig:building2}
	\vspace{-5mm}
\end{figure}

A complex 20-story moment-resisting frame building model, adapted from \cite{WojtkiewiczJohnson2013_sensitivity_timedomain}, with a height of 80\units{m}, is shown in \fref{fig:building2}.
Cross braces provide additional
stiffness for lateral bending, torsion, and in-plane floor stiffness.
The structural model without the passive control devices has 1620 DOFs, with its fundamental modes in the $y$-direction at 0.5718\units{Hz}, $x$-direction at 0.5893\units{Hz} and torsional at 0.9363\units{Hz}.
 Two TMDs are placed in the $y$-direction (each 0.55\% of building mass) and one TMD in the $x$-direction (1.1\% of building mass). The building is subjected to wind excitation (oriented toward the east-northeast, at a 30$^{\circ}$ angle from the $x$ axis as shown in \fref{fig:building2}), which is one realization of a narrowband filtered Gaussian white noise process (most of the excitation energy is in the range of 0.35--1.5\units{Hz}, exciting primarily the fundamental mode in the east-west $x$-direction),  vertically shaped proportional to the height to the 0.3 power \citep{holmes1996along}
 and, for simplicity, assumed to be fully correlated at all heights along the building.
 The three TMDs, in the \textit{true} model, exert power-law damping forces
 \begin{equation}\label{eq:tmd}
 \begin{split}
 f\subscript{TMD}^x&=200 \text{\units{kN\CDOT(s/m)$^{0.8}$}} |\dot u|^{0.8}\sgn\dot u +30 \text{\units{kN\CDOT(s/m)}} \,\dot u\\
 f\subscript{TMD}^y&=100 \text{\units{kN\CDOT(s/m)$^{0.8}$}} |\dot u|^{0.8}\sgn\dot u +15 \text{\units{kN\CDOT(s/m)}} \,\dot u %0.5\overline{c}\subscript{NL}\left(1+\xi\subscript{NL}\right)|\dot u|^{\beta\subscript{NL}}\sgn\dot u+c\subscript{L}\,\dot u
 \end{split}
 \end{equation}
 where $\dot u$ is the velocity of a TMD relative to its roof connection.
 The values in \eqref{eq:tmd} are chosen so that the effects of the nonlinearities are pronounced in the structure's roof accelerations. The ($N_\mathrm{o}=1200$ element) measurement data vector contains sampled $x$- and $y$-direction roof-center acceleration time histories --- $[\ddot u_x^\mathrm{roof}(0\Delta t) ~~ \ddot u_y^\mathrm{roof}(0\Delta t) ~~ \ddot u_x^\mathrm{roof}(1\Delta t) ~~ \ddot u_y^\mathrm{roof}(1\Delta t) ~~ \cdots]^\mathrm{T}$ with $\Delta t = 0.05\units{s}$ --- plus additive Gaussian pulse-process measurement noise with a standard deviation that is 30\% that of that noise-free vector.
 \begin{table}[!htb]
 	%	\small
 	\caption{Nonlinear damping model classes, where $u$ is a TMD displacement relative to its roof attachment point, and $W\subscript{tmd}$ is the weight of the corresponding TMD. }
 	\label{tab:tmd_models}
 	\centering
 	\begin{threeparttable}
 		\begin{tabular}{@{} l l l @{}} 
 			\hline
 			\Tstrut
 			Model class $\Mk$ & Damping force & Parameters\Bstrut\\
 			\hline
 			\Tstrut
 			$\C_1$ Linear & $f\subscript{lin}=c\subscript{1}\,\dot u$ & $c\subscript{1}$ [kN\CDOT s/m] \\
 			\hline 
 			%\hline 
 			\multirow{2}{*}{$\C_2$ Cubic polynomial} & \multirow{ 2}{*}{$f\subscript{cub}={c}\subscript{3}\,\dot u^3+c\subscript{1}\,\dot u$} &${c}\subscript{3}$ [kN\CDOT(s/m)\textsuperscript{3}] \\
 			 & & $c\subscript{1}$ [kN\CDOT s/m] \\
 			\hline 
 			\multirow{ 2}{*}{$\C_3$ Bouc-Wen} &$f\subscript{H}=q_{\mathrm{y}} z+\kpost u$ $^\mathsection$ & $\rk=\kpost/\kpre$\\ 
 			& $\npow=1$; ~$\kpre$ fixed & $\Qy$ [\%$W\subscript{tmd}$]\\
 						\hline
 		\end{tabular}
 	\begin{tablenotes}
 	\item[$\mathsection$]{\footnotesize{ $z$ is defined in \eqref{eq:bouc}.}}
 \end{tablenotes}
\end{threeparttable}
\end{table}

\begin{table}[!htb]
	\caption{Prior distributions for uncertain parameters of different model classes used to describe the TMD damping forces in the structure ($W\subscript{tmd}$ is the weight of the corresponding TMD; Std.~Dev.~is the standard deviation).} 
    		\label{tab:priors}
\centering
\begin{threeparttable}
%		\small
\centering
\begin{tabular}{@{}l |  c l l l @{}c l l l@{}} 
%\hline
\hline \Tstrut
\multirow{2}{*}{\parbox{\widthof{Model}}{Model Class}} & \multirow{2}{*}{Parameter}&  \multirow{2}{*}{Distribution}& \multicolumn{2}{c}{$x$ TMD} && \multicolumn{2}{c}{$y$ TMDs} & \\\cline{4-5}\cline{7-8}
\Tstrut &   & & Mean & Std.~Dev. && Mean & Std.~Dev. & 
\\\hline\Tstrut $\C_1$ Linear  & $c_1$& Normal & $250$  & $75.0$ && 150& 45.0 & 
			\\\hline \Tstrut
\multirow{2}{*}{\parbox{\widthof{polynomial}}{ $\C_2$ Cubic\\polynomial}} & $c_3$ &Lognormal & $50.0$ & $22.5$  && 25.0&6.0&\\
			& $c_1$&Normal & $250$ & $75.0$ && 150&45.0&\Bstrut
						\\\hline\Tstrut \multirow{2}{*}{$\C_3$ Bouc-Wen} & $\rk$ &Uniform & 0.1667 & 0.05 && 0.1667 & 0.0144 & \\
			& $\Qy$ &Normal & 7.5 & 0.5 && 7.5 & 0.5 &\Bstrut
			\\\hline 
		\end{tabular}
	\begin{tablenotes}
	\item[$\mathsection$]{\footnotesize{ $z$ is defined in \eqref{eq:bouc}.}}
\end{tablenotes}
	\end{threeparttable}
%	\vspace{-12pt}
\end{table}

	% \newcolumntype{P}[1]{>{\centering\arraybackslash}p{#1}}
\begin{table}[htb!]
		\caption{Fractions of models unfalsified within each model class.} \label{tab:fals_res}
	\centering 
	\begin{tabular}{ c  c| c c c} %P{\widthof{Bouc-Wen}} P{\widthof{Bouc-Wen}} P{\widthof{Bouc-Wen}} }%
		\hline
		\Tstrut  {} & & \multicolumn{3}{c}{$x$ TMD}\Bstrut \\ 
		{} & &  linear & cubic & Bouc-Wen\Bstrut\\
		%& \multicolumn{2}{|c|}{8 Hz sampling rate} & \multicolumn{2}{|c|}{20 Hz sampling rate} \Tstrut\Bstrut\\ \cline{2-5} \Tstrut
		%	\Tstrut	& \multicolumn{3}{c|}{Error-bound} & \multicolumn{4}{c}{Likelihood-bound}\Bstrut\\
		%		& Bonferroni & \v{S}id\'ak & BH &  Bonferroni & \v{S}id\'ak & BH & CPM \Tstrut\Bstrut\\
		\hline
		\Tstrut \multirow{3}{*}{$y$ TMDs} & linear & 48.6\% & 41.2\% & 0.0\% \\
		%		Cubic Polynomial &  96.25  \\
		%		CALTRANS   & 0 \\
		%		 mod.~AASHTO  & 0 \\
		%		Bouc-Wen & 32.2& 32& 29.75& 100&100 & 89.45 & 16.7 \\
		& cubic	& 43.9\% & 45.6\% & 0.0\% \\
		& Bouc-Wen & ~~0.0\% & ~~0.0\% & 0.0\%\Bstrut\\
		%		Bilinear  & 5.1\Bstrut\\
		\hline
	\end{tabular}
	
\end{table}

\begin{table}[htb!]
	\caption{Estimated parameters of two model classes.} \label{tab:est_lin}
	
	\centering
	\begin{tabular}{ c  |c c c  c c c c} 
		\hline
		\Tstrut  Model Class & $\widehat{c}^{~x}_1$ $[$kN\CDOT s/m$]$ & $\widehat{c}^{~y}_1$ $[$kN\CDOT s/m$]$ & $\widehat{c}^{~x}_3$ $[$kN\CDOT (s/m)$^3]$ & $\widehat{c}^{~y}_3$ $[$kN\CDOT (s/m)$^3]$\Bstrut \\ 
				\hline
		\Tstrut $x$-linear, $y$-linear & 399.1 & 154.3 & n/a & n/a \\
				$x$-cubic, $y$-cubic	& 410.5 & 140.3 & 30.1 & 31.1\Bstrut\\
		%		Bilinear  & 5.1\Bstrut\\
		\hline
	\end{tabular}

\end{table}

\begin{figure}
	\centering
	\begin{subfigure}[htb!]{0.48\textwidth}
		        \centering
%		\begin{tikzpicture}
%		\node (fig1) at (0,0)
%		{\includegraphics[scale=0.31]{figures/pred_linear_vFINAL.pdf}};
%		\node (fig2) at (0.3,-1.6)
%		{\includegraphics[scale=0.06]{figures/pred_linear__zoomed_vFINAL.pdf}};  
%		\draw (0.45,-0.1) -- (0.8,-0.1) -- (0.8,0.9) -- (0.45,0.9) -- (0.45,-0.1);
%		\draw (0.45,-0.1) -- (-0.325,-1.05);
%		\draw (0.8,-0.1) -- (0.98,-1.05);
%		
%		\end{tikzpicture}
		   \includegraphics[scale =1]{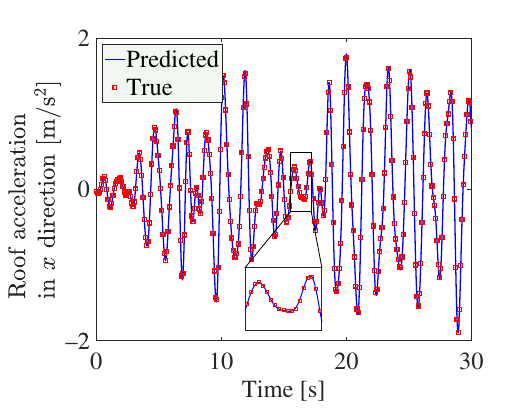}
		\caption{Linear damping forces in both directions}\label{fig:rob_pred2:a}
	\end{subfigure}%
	\hfill
	\begin{subfigure}[htb!]{0.48\textwidth}
		        \centering
%		\begin{tikzpicture}
%		\node (fig1) at (0,0)
%		{\includegraphics[scale=0.31]{figures/pred_cubic_vFINAL.pdf}};
%		\node (fig2) at (0.3,-1.6)
%		{\includegraphics[scale=0.06]{figures/pred_cubic__zoomed_vFINAL.pdf}};  
%		\draw (0.45,-0.1) -- (0.8,-0.1) -- (0.8,0.9) -- (0.45,0.9) -- (0.45,-0.1);
%		\draw (0.45,-0.1) -- (-0.325,-1.05);
%		\draw (0.8,-0.1) -- (0.98,-1.05);
%		
%		\end{tikzpicture}
		        \includegraphics[scale=1]{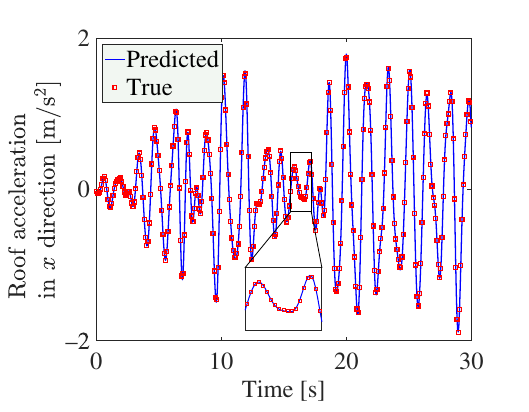}
		\caption{Cubic damping forces in both directions}\label{fig:rob_pred2:b}
	\end{subfigure}
%	\vspace{-12pt}
\caption{Predicted and true absolute roof acceleration of the 1623 DOF building subjected to a different realization of the wind excitation.} \label{fig:rob_pred2}
%\vspace{-12pt}
\end{figure}
\subsubsection{Candidate Model Classes}
The three model classes for TMD damping forces and their corresponding uncertain parameters are shown in Table \ref{tab:tmd_models}. 
For each TMD, the first model class has linear viscous damping with coefficient $c_1$, the second adds a cubic damping term with coefficient $c_3$, and the third is a Bouc-Wen hysteresis with parameters as the hardness ration $\rk=\kpost/\kpre$ with a fixed $\kpre$ and yield force $\Qy$. 
The $3\times3=9$ candidate model classes are formed by combining these TMD force models for the $x$- and $y$-direction TMDs. 
The \textit{true} TMD damping force model is (intentionally) omitted from the candidate model classes, but two of these TMD damping force models --- linear and cubic polynomial --- cause TMD behaviors similar to the true one. The priors for these model classes are assumed according to Table \ref{tab:priors}.
\subsubsection{Results}
For each model class $\C$, a set of $\Ns=2000$ models is randomly generated from the prior distribution $\pdf(\thetaa|\C)$. Each residual $\epsilon_i$ is assumed normally distributed $\mathcal{N}(0,\sigma^2)$, where the residual standard deviation $\sigma$ is assumed to be 0.08\units{m/s\textsuperscript{2}}, which is about 15\% of the standard deviation of the measured data. The fractions of models unfalsified within each candidate model class using a target identification probability $\phi=0.95$ are shown in Table \ref{tab:fals_res}.
All Bouc-Wen models are falsified because the more boxy shapes of their hysteresis loops, for the priors chosen here, are very different from the elliptical shapes of the other damping models.
The parameters estimated using the unfalsified models of each of the two model classes with the most unfalsified models --- linear in both directions and cubic polynomial in both directions --- are shown in Table \ref{tab:est_lin}.

Then, response prediction is performed, using these unfalsified models with weights assigned according to \eqref{eq:weights} and used in \eqref{eq:rob_pred2}, for response due to a different realization of the stochastic wind excitation; the results for both model classes are shown in \fref{fig:rob_pred2}. The relative RMS errors in predicting the roof acceleration in the $x$-direction are 1.7995\% and 1.8081\% for response prediction using the linear-in-both-directions and cubic-polynomial-in-both-directions models, respectively. Hence, the response prediction using the proposed method provides very good accuracy for this example.
Further, out of all models from the nine model classes, over 80\% were falsified, so the number of model simulations required for the prediction step reduces by about a factor of 5.

\begin{figure}[htb!]%[scale=0.4\textwidth]
	%  \vspace{-25pt}
	\begin{center}
		\begin{subfigure}[t]{0.48\textwidth}
			\centering
			\includegraphics[scale=0.98]{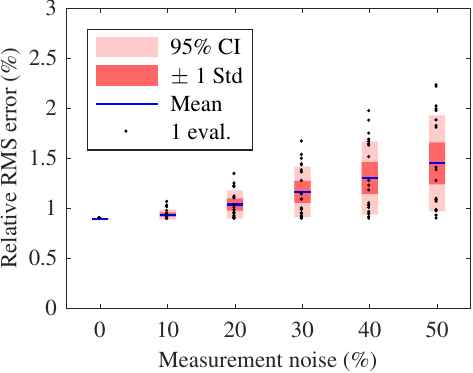}
			\caption{Prediction error vs.~noise level for one candidate model set}\label{fig:1623doferror:a}
		\end{subfigure}%
		\hfill
		\begin{subfigure}[t]{0.48\textwidth}
			\centering
			\includegraphics[scale=0.98]{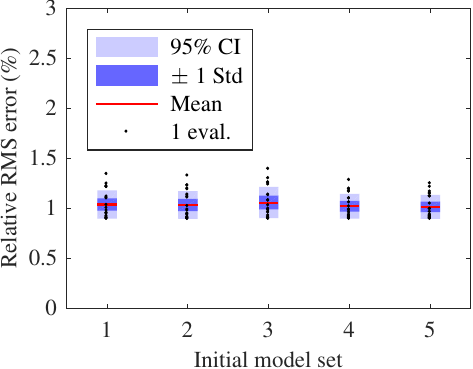}
			\caption{Prediction error for different candidate model sets (20\% measurement noise).}\label{fig:1623doferror:b}
		\end{subfigure}
		%    \\[0.5ex]
		%    \includegraphics[scale=1.5]{figures/four_dof_carts.pdf}
	\end{center}
%	\vspace{-22pt}
	\caption{Relative RMS error in predicting the 1623-DOF building's $x$-direction roof acceleration response to a different realization of wind excitation with varying measurement noise level and for different candidate model sets. (CI = confidence interval, Std = standard deviation)} \label{fig:1623doferror}
%	\vspace{-12pt}
\end{figure}

To evaluate the {robustness} of the Kobe response predictions, the relative RMS prediction errors are computed for six measurement noise levels and for five sets of candidate models in which the TMDs in both directions are cubic polynomials; the likelihood bounds for falsification are chosen assuming 20\% measurement noise. 
The relative RMS errors in predicting the $x$-direction roof acceleration are shown in \fref{fig:1623doferror:a} for various measurement noise levels, indicating mean errors as small as 1.5\% even when the measurement noise level is 50\%.  With 20\% measurement noise but varying the randomly-chosen candidate model set, \fref{fig:1623doferror:b} shows that the average prediction error changes only modestly over different sets of candidate models. %Hence, in this example, the predictions using the proposed method are also robust to both the modeling uncertainty induced by measurement noise and randomness in candidate model sets.

\subsection{Example III: A Full-Scale Base-Isolated Four-Story Building}

\begin{figure}[htb!]
	\centering
	\includegraphics[width=0.45\textwidth]{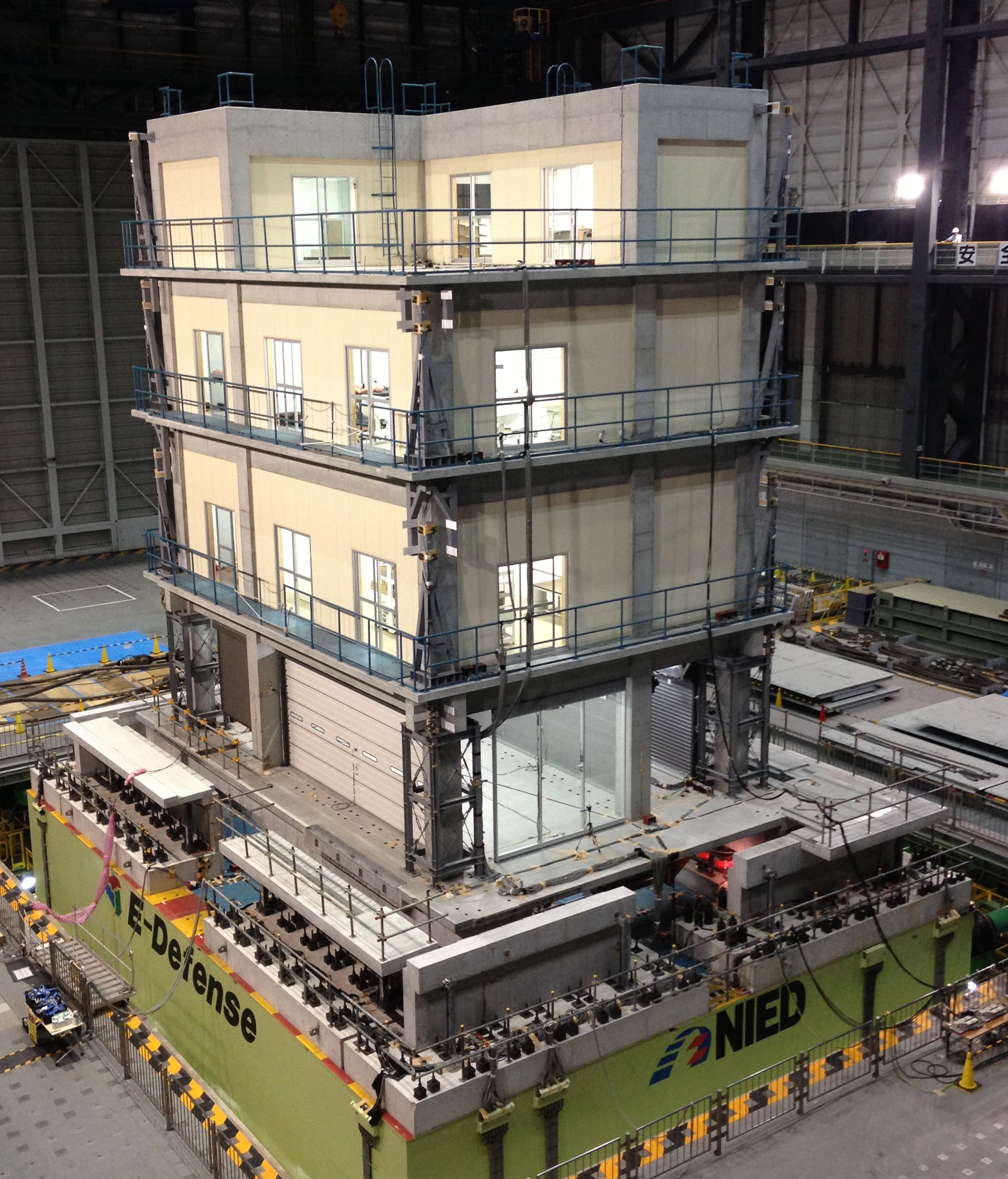}
	\caption{The experimental set-up.}\label{fig:test}
\end{figure}
\begin{figure}[htb!]
	\centering
	\includegraphics[scale=0.75]{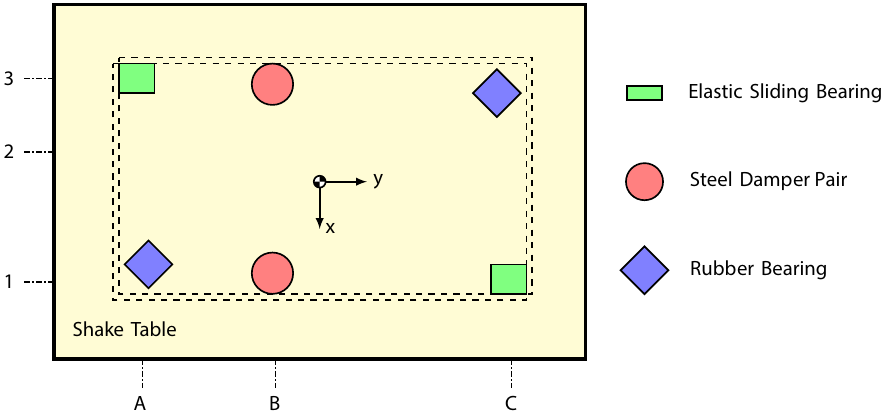}
	\caption{Isolation device placements in the base layer. (Note that $1-3$ and $\mathrm{A}-\mathrm{C}$ are used to identify the column locations.) %{\color{purple!70!blue}(EAJ: Steel Damper Pair; SD: I don't have the source file, only pdf. PB, do you have it?)}
    } 
	\label{fig:devices}
\end{figure}

A base-isolated test structure mounted on the world's largest six degree-of-freedom shake table   at Japan's E-Defense lab was tested in March 2013 and again in August 2013 (see \fref{fig:test}) \citep{EDef,Tianhaopaper}. In this study, measurements from tests performed on 8 August 2013 are used.
The structure consists of a four-story, asymmetric, moment frame with a setback and coupled transverse-torsional motion. The superstructure has a mass of 686 tons and has dimensions 14m$\times$10m$\times$15m. 
The isolation layer is composed of two rubber bearings, two elastic sliding bearings, and two pairs of passive U-shaped steel yielding dampers. 
A schematic of the location of these devices is shown in \fref{fig:devices}.
The building was subjected to random excitation along different table axes, \ie   in the $x$, $y$ and $z$ directions, as well as scaled versions of historical and synthetic earthquake ground motions.

\subsubsection{Measurement Sensors}
\begin{figure}[htb!]
	\centering
	\includegraphics[scale=1.15]{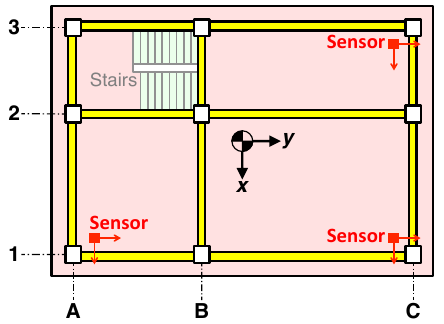}
	\caption{Typical accelerometer placements for the bottom three floors.}
	\label{fig:sensors}
\end{figure}
Tri-directional accelerometers recording responses in the $x$, $y$, and $z$ directions were used at the three corner locations on each floor except the roof as shown in \fref{fig:sensors}. On the roof, the accelerometers were placed only on two corners (locations 1-A and 3-C in \fref{fig:sensors}) because of the fourth story's setback. This leads to 42 channels of accelerations. Sensors were also placed at four more locations on the shake table. A sampling rate of 1 kHz with a low-pass filtering using a 50 Hz cut-off frequency were used to record the responses. 
Force and displacement transducers were also used below each of the isolation devices. 
Tests 010 and 012, in which random excitations were applied to all shake table  degrees-of-freedoms, are used herein. Further, Test 014, which is a scaled version of 2011 Mw9.0 Tohoku-Oki earthquake (K-NET Furukawa record) \citep{BrewickEDefenseLinearID}, is also used to characterize the nonlinearity in the base isolator devices. 

\subsubsection{Case (a): Uncertain Superstructure with Linear Isolation Layer}

\textit{Candidate Model Classes}\\
\begin{figure}[htb!]
	\centering
	\includegraphics[width=0.4\textwidth]{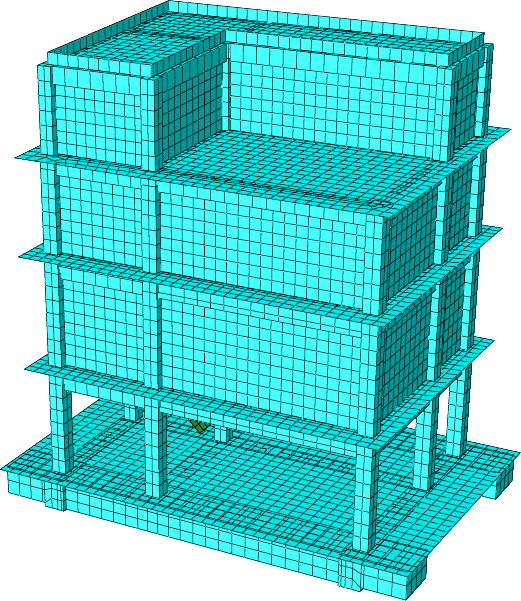}
	\caption{The finite element model has ~85,000 degree-of-freedom.}\label{fig:fem}
\end{figure}
\indent A finite element model with about 85,000 degrees-of-freedom has been developed from the design drawings as shown in \fref{fig:fem} \citep{Tianhaopaper}. Solid elements are used for the beams, columns, and shear walls. The steel reinforced bars are modeled using truss elements. Shell elements are used for the floor slabs and the nonstructural walls (autoclaved lightweight concrete plates). 
For the tests considered in this paper, the measurements from the isolation devices show a linear relationship between restoring force and device displacements. Therefore, they are modeled using bi-directional springs. 
The equations of motion of this finite element model thus become 
\begin{equation}
\Mm \ddot{\uu}+\Cm(\thetaa)\dot{\uu}+\Km(\thetaa)\uu = \Bm\Km\subscript{b}\uu\subscript{b} - \Mm\rr\ddot{\uu}\subscript{t}
\end{equation}
where $\Mm$, $\Cm$, and $\Km$ are the mass, damping, and stiffness matrices of the building model; $\thetaa$ is the uncertain parameter vector; $\uu$ is the displacement vector of the building with respect to the shake table; $\Bm$ is the influence matrix for the linear base-isolation device restoring force; $\Km\subscript{b}$ is linear base-isolation device stiffness; $\uu\subscript{b}=\G \uu$ is the base-isolation displacements; $\rr$ consists of zeros and ones based on the table motion's influence on that degree-of-freedom; and $\ddot{\uu}\subscript{t}$ is the table acceleration.
In this paper, the parameter vector $\thetaa$ is assumed to consist of elastic moduli of the superstructure components. Four different model classes are defined based on the choice of parameter set $\thetaa$ as shown in Table \ref{tab:ex3_prior}. 
The first model class $\C_1$ assumes that all beams and columns in the building have the same elastic modulus. The second model class $\C_2$ assumes all beams have the same elastic moduli, but different from the moduli of all columns. 
The third model class $\C_3$ also considers the beams on first floor different than those on other floors. The fourth model class $\C_4$ additionally assumes that the moduli of the columns on the 4\superscript{th} floor is different from that of the columns on other floors. The prior distributions of these parameters are assumed Gaussian with means and standard deviations as shown in Table \ref{tab:ex3_prior}.
\begin{table}[htb!]
		\caption{ Different model classes with their parameters, mean and standard deviation (Std.~dev.) of their prior  distribution, and estimated parameters  (The suffixes \textit{Beam} and \textit{Col} represent beams and columns in the building, respectively, and the numbers represent the floor numbers).}%\vspace{-15pt}
	\label{tab:ex3_prior}
	\begin{center}
		
		\begin{tabular}{l |l l l l c c c } 
			\hline
			\Tstrut	Model  & \multirow{2}{*}{Parameter} & Mean & Std.~dev. &  Estimated \\
			Class $\Mk$ & & [GPa] & [GPa] & [GPa] \Bstrut\\
			\hline
			\Tstrut	$\C_1$ & $E\subscript{Beam,Col}$ & 27  & 2.5 & ---  \Bstrut\\
			\hline
			\Tstrut	\multirow{2}{*}{$\C_2$} & $E\subscript{Beam}$ & 27  &2.5 & \multirow{2}{*}{---}  \\
			& $E\subscript{Col}$ & 23  & 2.5    \Bstrut\\
			%			\hline
			%			\Tstrut	\multirow{3}{*}{$\C_3$} & $E\subscript{Beam,1}$ & 27 GPa & 2.5 GPa  \\
			%			& $E\subscript{Beam,2,3,4}$ & 25 GPa & 2.5 GPa &  &  \\
			%			& $E\subscript{Col}$ & 23 GPa & 2.5 GPa  \Bstrut\\
			\hline
			\Tstrut	\multirow{4}{*}{$\C_3$} & $E\subscript{Beam,1}$ & 27  & 2.5  & 31.9976 \\
			& $E\subscript{Beam,2,3,4}^x$ & 27  & 2.5 & 24.4356  \\
			& $E\subscript{Beam,2,3,4}^y$ & 23  & 2.5 & 18.4283  \\
			& $E\subscript{Col}$ & 23  & 2.5 & 19.4563  \Bstrut\\
			\hline
			\Tstrut	\multirow{5}{*}{$\C_4$} & $E\subscript{Beam,1}$ & 27  & 2.5 & 28.7508 \\
			& $E\subscript{Beam,2,3,4}^x$ & 27  & 2.5 & 28.6286  \\
			& $E\subscript{Beam,2,3,4}^y$ & 23  & 2.5  & 16.0084  \\
			& $E\subscript{Col}^x$ & 23  & 2.5 & 21.1343  \\
			& $E\subscript{Col}^y$ & 24  & 2.5 & 23.9262 \Bstrut\\
			\hline
		\end{tabular}
	\end{center}
	%\vspace{-20pt}
\end{table}

\begin{figure}
	\centering
	\begin{subfigure}[htb!]{0.32\textwidth}
	\centering
		\includegraphics[scale=0.5]{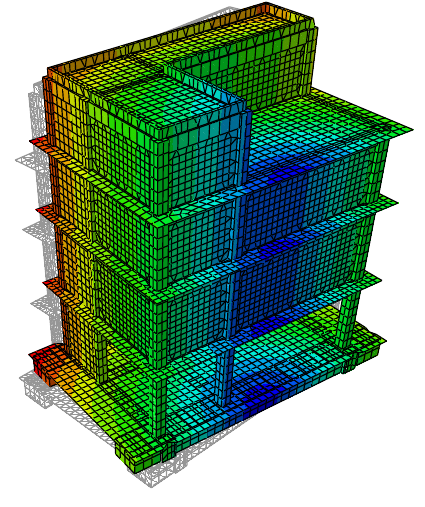}
		\caption{1$^\text{st}$ mode shape}
	\end{subfigure}
	\begin{subfigure}[htb!]{0.32\textwidth}
	\centering
		\includegraphics[scale=0.5]{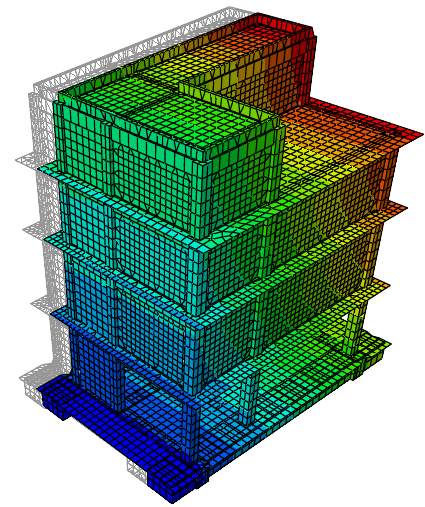}
		\caption{2$^\text{nd}$ mode shape}
	\end{subfigure}
	\begin{subfigure}[htb!]{0.32\textwidth}
	\centering
		\includegraphics[scale=0.5]{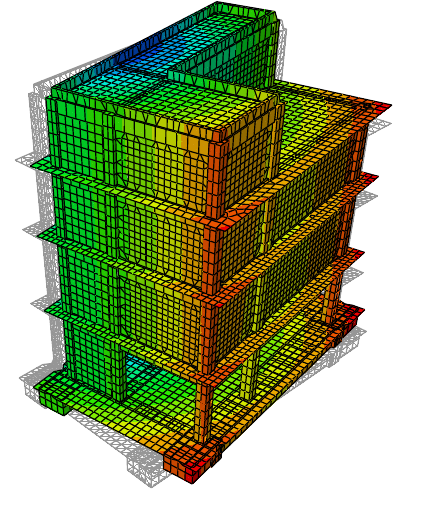}
		\caption{3$^\text{rd}$ mode shape}
	\end{subfigure}
	\begin{subfigure}[htb!]{0.32\textwidth}
	\centering
		\includegraphics[scale=0.5]{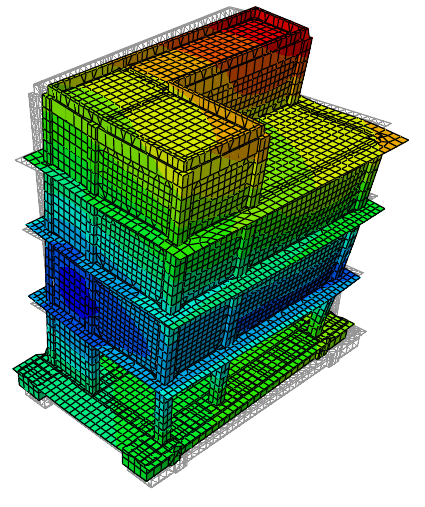}
		\caption{4$^\text{th}$ mode shape}
	\end{subfigure}
	\begin{subfigure}[htb!]{0.32\textwidth}
	\centering
		\includegraphics[scale=0.5]{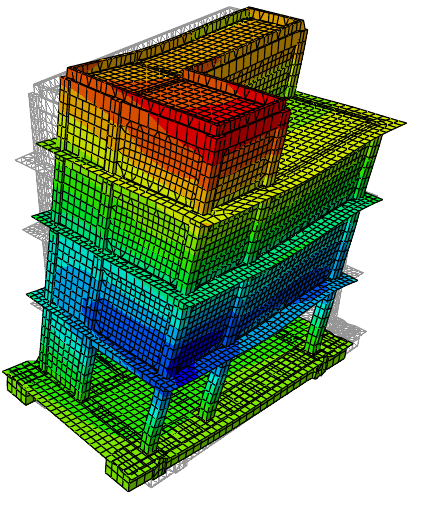}
		\caption{5$^\text{th}$ mode shape}
	\end{subfigure}
	\begin{subfigure}[htb!]{0.32\textwidth}
	\centering
		\includegraphics[scale=0.5]{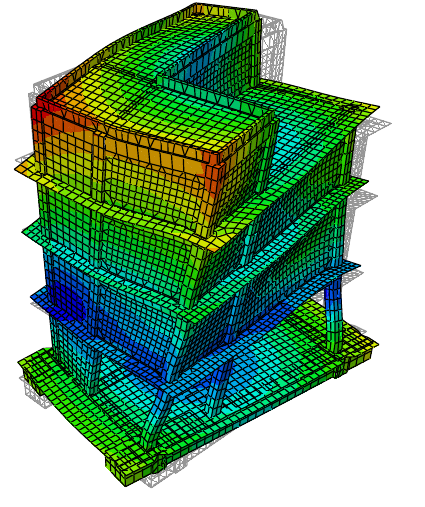}
		\caption{6$^\text{th}$ mode shape}
	\end{subfigure}
	\caption{First six mode shapes of the building (undeformed shape is shown in gray).}
	\label{fig:modeshapes}
\end{figure}

\begin{table}[htb!]
		\caption{First six identified natural frequencies using 12 recorded inputs are used in this example.} \label{tab:freq}
	\centering
	\begin{tabular}{ c  |c c c  c c c c} 
		\hline
		\Tstrut  \multirow{2}{*}{Mode} & Natural frequency\\ %& $\widehat{c}^{~y}_1$ $[$kN\CDOT s/m$]$ & $\widehat{c}^{~x}_3$ $[$kN\CDOT (s/m)$^3]$ & $\widehat{c}^{~y}_3$ $[$kN\CDOT (s/m)$^3]$\Bstrut \\ 
		%& \multicolumn{2}{|c|}{8 Hz sampling rate} & \multicolumn{2}{|c|}{20 Hz sampling rate} \Tstrut\Bstrut\\ \cline{2-5} \Tstrut
		&  $\omega_i$ (Hz)\\
		%	\Tstrut	& \multicolumn{3}{c|}{Error-bound} & \multicolumn{4}{c}{Likelihood-bound}\Bstrut\\
		%		& Bonferroni & \v{S}id\'ak & BH &  Bonferroni & \v{S}id\'ak & BH & CPM \Tstrut\Bstrut\\
		\hline
		\Tstrut 1\superscript{st} &  0.6853\\
		%		Cubic Polynomial &  96.25  \\
		%		CALTRANS   & 0 \\
		%		 mod.~AASHTO  & 0 \\
		%		Bouc-Wen & 32.2& 32& 29.75& 100&100 & 89.45 & 16.7 \\
		2\superscript{nd}	&  	0.6975\\
		3\superscript{rd} & 	0.7095\\
		4\superscript{th} & 	4.7812\\
		5\superscript{th} & 	5.1749\\
		6\superscript{th} &	    6.1199
		\Bstrut\\
		%		Bilinear  & 5.1\Bstrut\\
		\hline
	\end{tabular}
\end{table}
\noindent\textit{Results}\\
\indent Using the low-intensity random excitation of Test 010, a subspace identification algorithm, namely,  \citep{vanoverschee1994n4sid} is used to estimate the first six natural frequencies and the mode shapes as in \cite{BrewickEDefenseLinearID} and in  \cite{Tianhaopaper}. The first six identified natural frequencies are shown in Table \ref{tab:freq}. \fref{fig:modeshapes} shows the first six mode shapes.
Model falsification is performed on 1000 models from each class using their prediction of first six natural frequencies and mode shapes with target identification probability $\phi=0.90$. 
{\color{black}Residual errors are computed by stacking errors in frequencies and six diagonal MAC (modal assurance criterion) values in a vector.} 
% {\color{purple}(SD: Tianhao, can you please comment here? TY: Residual errors are computed by stacking errors in frequencies and 6 MAC (modal assurance criterion) values in a vector.}
They are then used to compute the likelihood using a covariance matrix $\Sigm$ that is assumed as diagonal with a variance of %entries 
$0.02^2$ for natural frequencies, \ie approximately 3\% of the first natural frequency, and $0.25^2$ for MAC values; given the wider variation in MAC values between analytical and experimental mode shapes, a larger standard deviation is used for the MAC errors.
% the relative variance for the MAC values is larger given the greater uncertainty in estimating mode shapes. 
% {\color{purple!70!blue}(EAJ: There's no context regarding these values. SD: 0.02 is about 3\% of the first natural frequency. For MAC, we can refer to an identity matrix. Would this explanation be sufficient? EAJ: Put a parenthetical statement about 3\%, given wider variations of mode shapes, 25\%)}\\
% {\color{purple!70!blue}(EAJ: Clarify: acccel measurements are not used for falsification. only, frequencies and MAC values from N4SID used. SD: As far as I am aware. TY: Confirmed.)} \\ 
% {\color{purple!70!blue}(EAJ: Are the models simulated and then N4SID used to predict the frequencies and mode shapes, or an eigen analysis was done with mass and stiffness matrices? SD: The second one, as far as I know. TY: Confirmed.)}
\begin{figure}
	\centering
	\includegraphics[scale=0.5]{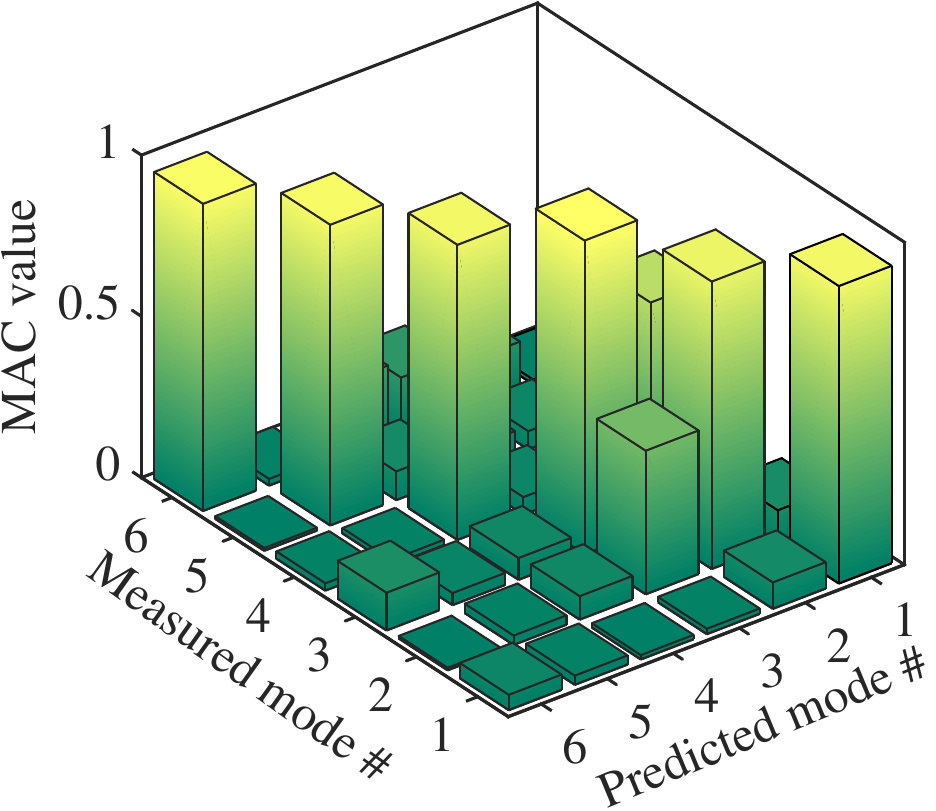}
	\caption{Typical MAC values of an unfalsified model from class 5.}
	\label{fig:mac_compare}
\end{figure}
\begin{table}[htb!]
		\caption{Falsification results for Example III (a).} \label{tab:ex3_fals}
	\centering
	\begin{tabular}{ l  |c c c  c c c c} 
		\hline
		\Tstrut  Model Class & \% Unfalsified\Bstrut \\ 
		%& \multicolumn{2}{|c|}{8 Hz sampling rate} & \multicolumn{2}{|c|}{20 Hz sampling rate} \Tstrut\Bstrut\\ \cline{2-5} \Tstrut
		%	\Tstrut	& \multicolumn{3}{c|}{Error-bound} & \multicolumn{4}{c}{Likelihood-bound}\Bstrut\\
		%		& Bonferroni & \v{S}id\'ak & BH &  Bonferroni & \v{S}id\'ak & BH & CPM \Tstrut\Bstrut\\
		\hline
		\Tstrut $\C_1$ &  0.0\\
		%		Cubic Polynomial &  96.25  \\
		%		CALTRANS   & 0 \\
		%		 mod.~AASHTO  & 0 \\
		%		Bouc-Wen & 32.2& 32& 29.75& 100&100 & 89.45 & 16.7 \\
		$\C_2$	& 0.0\\
		$\C_3$ & 3.7\\
		$\C_4$ & 4.0\Bstrut\\
		%\\
		%modified AASHTO & ~~0.0\\
		%Caltrans & ~~0.0
		%		Bilinear  & 5.1\Bstrut\\
		\hline
	\end{tabular}
	
\end{table}
%Model falsification is implemented with $\phi =0.90$ and $\Ns=1000$. 
The results are shown in Table \ref{tab:ex3_fals}. For the two unfalsified model classes the parameters are estimated using \eqref{eq:param_est2} as shown in Table \ref{tab:ex3_prior}. MAC values for an unfalsified model in class $\C_4$ is also shown in \fref{fig:mac_compare}. Further, the unfalsified models are used next to predict the acceleration responses of the structure when subjected to the scaled earthquake excitations in Test~012. These predictions are then compared with the actual test measurements. As shown in \fref{fig:ex3a_resppredM3} and \fref{fig:ex3a_resppredM4} predictions using the proposed method and unfalsified models from both $\C_3$ and $\C_4$ show good agreement with the experimental measurements. The relative RMS errors in different measurements are shown in Tables \ref{tab:ex3a_errorM3} and \ref{tab:ex3a_errorM4}, respectively. {Note that, the excitations in Test~012 had higher intensity compared to Test~010. This results in the assumed linear behavior of isolation layer devices not being adequate as evident from the Tables \ref{tab:ex3a_errorM3} and \ref{tab:ex3a_errorM4} though the computational cost is reduced by a factor of 52 by falsifying many models. However, the RMS errors are significantly reduced in the next section in which the nonlinear behavior of the base isolation layer devices is implemented. %{\color{purple!70!blue}(EAJ: )}
%Thus, if the validation results turn out to be desirable, some confidence could be given to the estimated parameters from model falsification as least when the magnitudes of the inputs are within those of Test~010 and Test~012.}

% \begin{table}[htb!]
% 			\caption{ Estimated parameters of two model classes.}%\vspace{-15pt}
% 	\label{tab:ex3_param}
% 	\begin{center}
		
% 		\begin{tabular}{l |l c l c c c} 
% 			\hline
% 			\Tstrut	Model  & {Parameter} & $\widehat{\thetaa}$ [GPa] \Bstrut\\
% 			\hline
% 			\Tstrut	\multirow{4}{*}{$\C_3$} & $E\subscript{Beam,1}$ & 31.9976   \Bstrut\\
% 			& $E\subscript{Beam,2,3,4}^x$ & 24.4356   \Bstrut\\
% 			& $E\subscript{Beam,2,3,4}^y$ & 18.4283   \Bstrut\\
% 			& $E\subscript{Col}$ & ~19.4563     \Bstrut\\
% 			\hline
% 			\Tstrut	\multirow{5}{*}{$\C_4$} & $E\subscript{Beam,1}$ & 28.7508  \Bstrut\\
% 			& $E\subscript{Beam,2,3,4}^x$ & 28.6286   \Bstrut\\
% 			& $E\subscript{Beam,2,3,4}^y$ & 16.0084    \Bstrut\\
% 			& $E\subscript{Col}^{1}$ & 21.1343   \Bstrut\\
% 			& $E\subscript{Col}^{2,3,4}$ & ~23.9262    \Bstrut\\
% 			\hline
% 		\end{tabular}
% 	\end{center}
% 	%\vspace{-20pt}
% \end{table}

\begin{figure}[htb!]
    \centering
    \begin{subfigure}[t]{0.5\textwidth}
        \centering
        \includegraphics[scale=0.3]{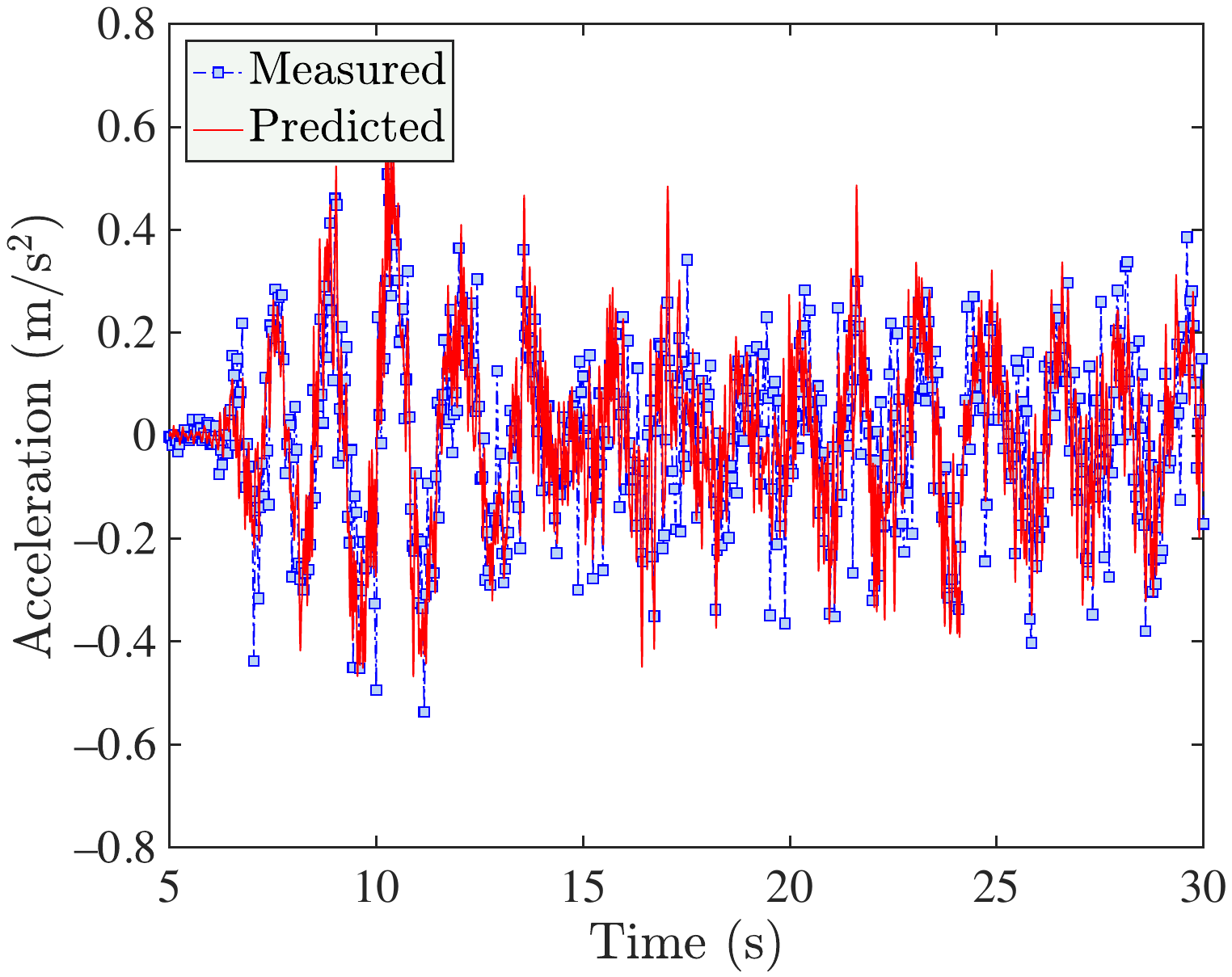}
        \caption{Base acceleration in $x$ direction.}
    \end{subfigure}%
    ~ 
    \begin{subfigure}[t]{0.5\textwidth}
        \centering
        \includegraphics[scale=0.3]{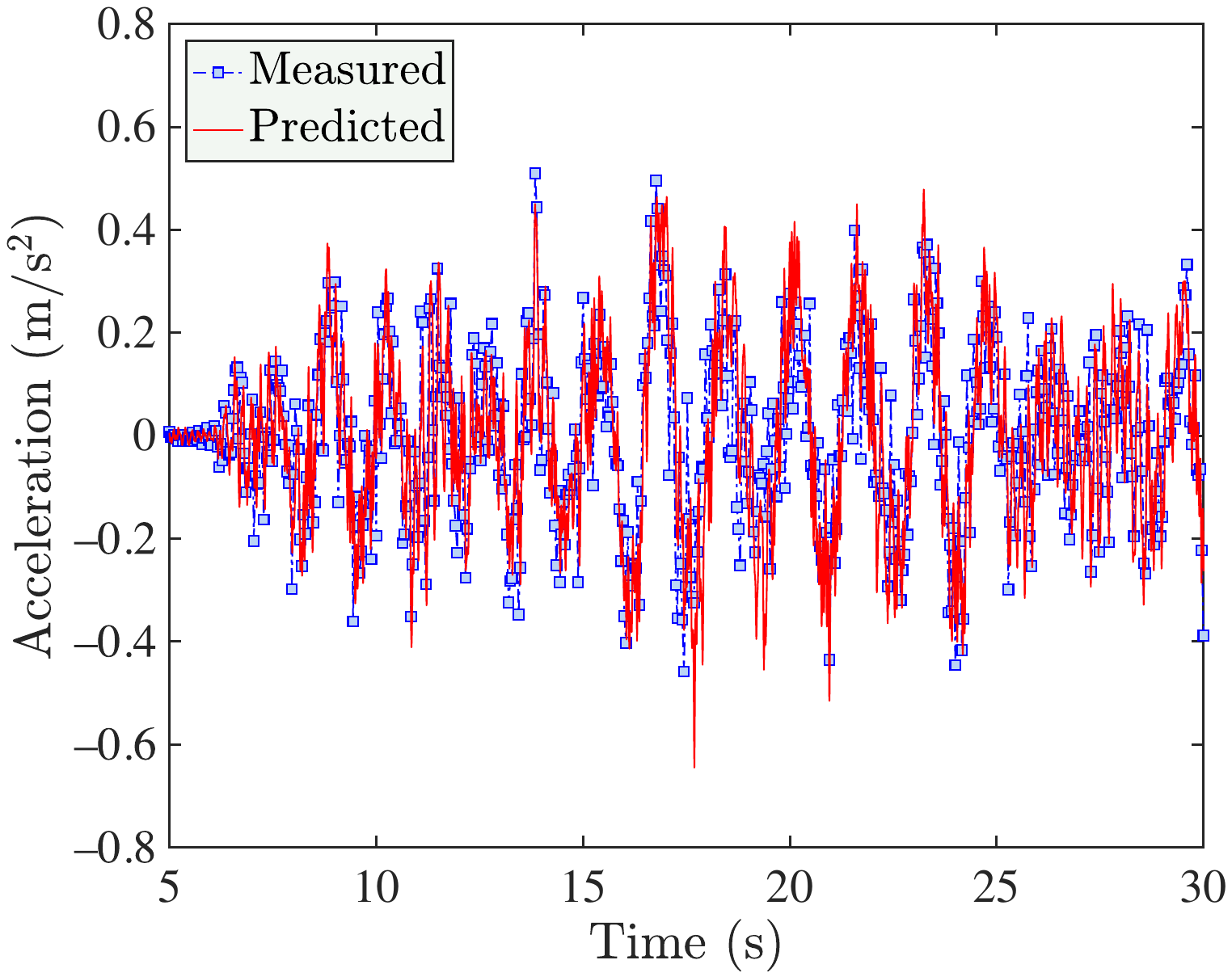}
        \caption{Base acceleration in $y$ direction. %{\color{purple!70!blue}(EAJ: Zoom in. SD: I can zoom in here. But, in case (b), I don't have the fig files.)}
        }
    \end{subfigure}
    \caption{Predicted base responses using unfalsified models from $\C_3$ are compared with measured during Test 012 using the sensor near 1-A (see \fref{fig:sensors}). For other floors, see \fref{fig:ex3a_resppredM3_allfloors} in Appendix \ref{sec:appendix1}.} \label{fig:ex3a_resppredM3}
\end{figure}

\begin{figure}[htb!]
    \centering
    \begin{subfigure}[t]{0.5\textwidth}
        \centering
        \includegraphics[scale=0.3]{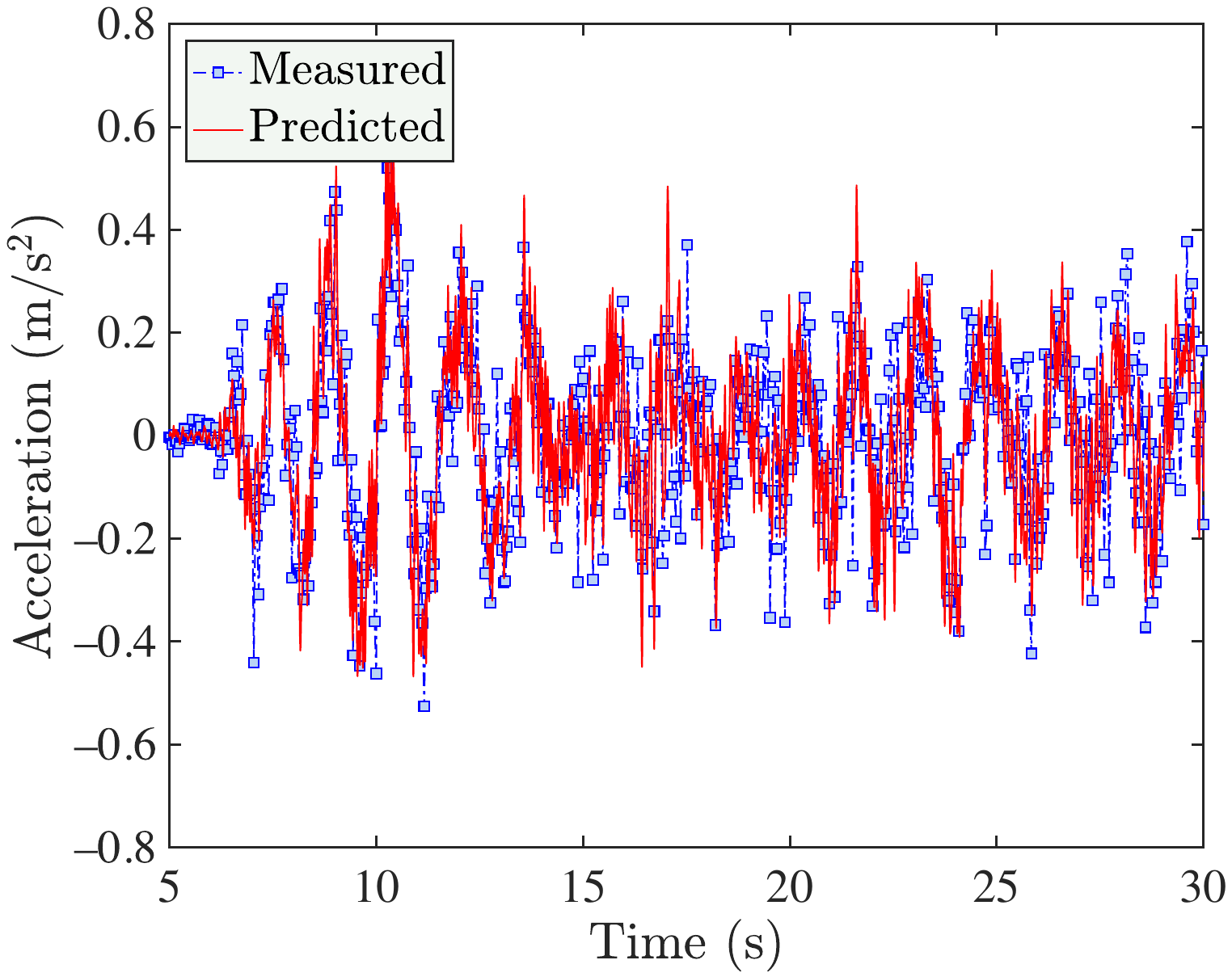}
        \caption{Base acceleration in $x$ direction.}
    \end{subfigure}%
    ~ 
    \begin{subfigure}[t]{0.5\textwidth}
        \centering
        \includegraphics[scale=0.3]{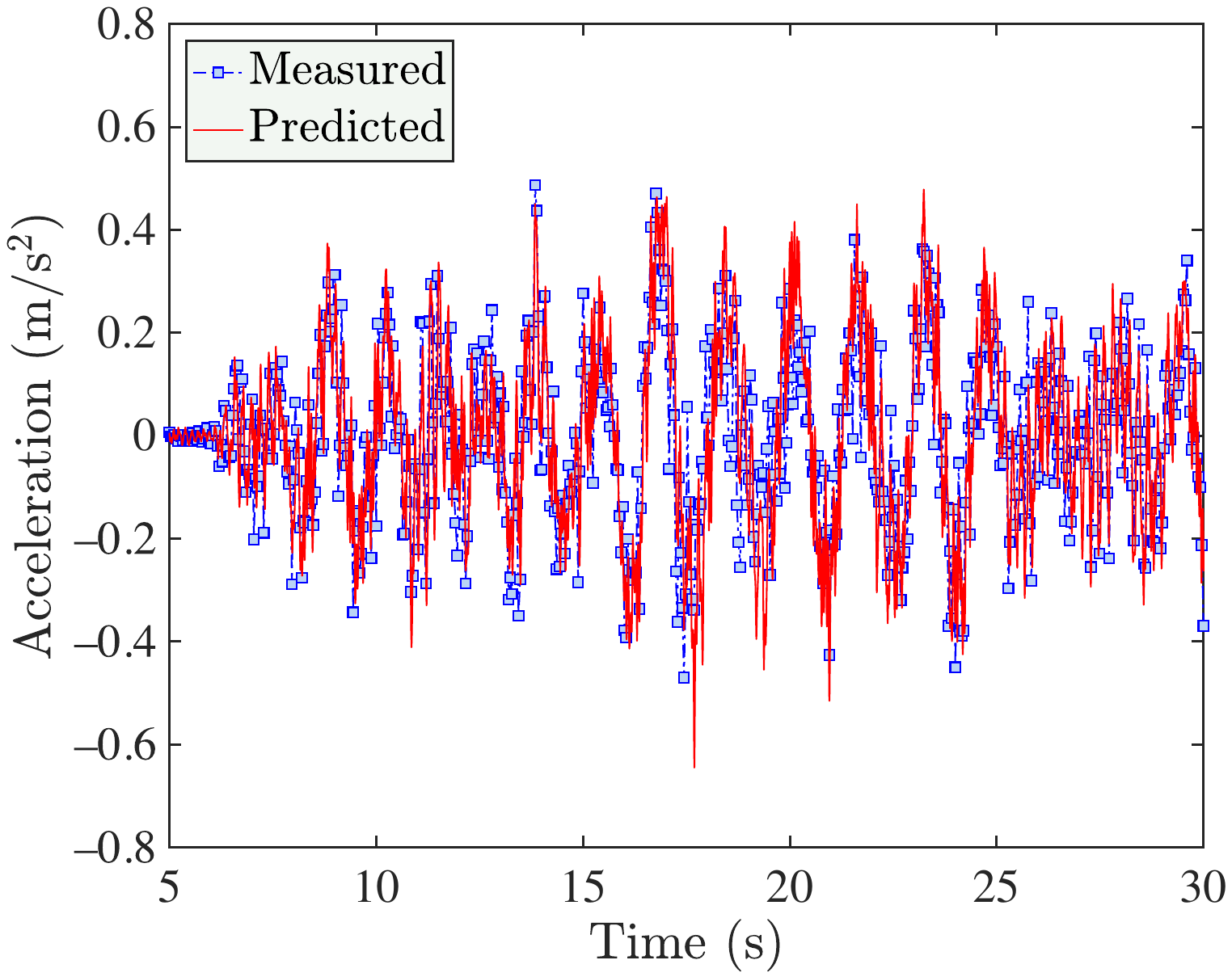}
        \caption{Base acceleration in $y$ direction.}
    \end{subfigure}
    \caption{Predicted base responses using unfalsified models from $\C_4$ are compared with measured during Test 012 using the sensor near 1-A (see \fref{fig:sensors}). For other floors, see \fref{fig:ex3a_resppredM4_allfloors} in Appendix \ref{sec:appendix1}.} \label{fig:ex3a_resppredM4}
\end{figure}

\begin{table}[htb!]
				\caption{Prediction errors using unfalsified models from model class $\C_3$ for Example III(a).}%\vspace{-15pt}
	\label{tab:ex3a_errorM3}
	\centering
% 		\begin{tabular}{c| c|c c c c c c} 
% 			\hline
% 			\Tstrut	\multirow{2}{*}{Floor} & \multirow{2}{*}{Sensor}  & \multirow{2}{*}{Direction} & \multicolumn{2}{c}{Error} \\
%             & & & $\C_3$ & $\C_4$ \Bstrut\\
% 			\hline
% 			\Tstrut	\multirow{6}{*}{Base} & \multirow{2}{*}{1-A} & $x$ & 0.1560 & 0.1559 \Bstrut\\
% 			& & $y$ & $0.0484$ & 0.0556  \Bstrut\\ \cline{2-5}
% \Tstrut		& \multirow{2}{*}{1-C} & $x$ & 0.0461 & 0.0493 \Bstrut\\
% 			& & $y$ & $0.0446$ & 0.0519  \Bstrut\\\cline{2-5}
% \Tstrut			& \multirow{2}{*}{3-C} & $x$ & 0.0438 & 0.0487 \Bstrut\\
% 			& & $y$ & $0.0100$  & 0.0091 \Bstrut\\
% 			\hline
% 		\end{tabular}
% ~~~~
\begin{tabular}{c|c c c c c c} 
			\hline
			\Tstrut	 \multirow{2}{*}{Sensor}  & \multirow{2}{*}{Direction} & \multicolumn{5}{c}{Prediction error} \Bstrut \\ \cline{3-7}
            \Tstrut & & Base & 1st floor & 2nd floor & 3rd floor & 4th floor \Bstrut\\
			\hline
			\Tstrut	 \multirow{2}{*}{1-A} & $x$ & 0.1560 & 0.1441 & 0.1507 & 0.1294 & 0.0753 \Bstrut\\
			& $y$ & 0.0484 & $0.0379$ & 0.0455 & 0.0286 & 0.0287 \Bstrut\\ \cline{2-7}
\Tstrut		\multirow{2}{*}{1-C} & $x$ & 0.0461 & 0.0176 & 0.0395 & 0.0366 & 0.0139 \Bstrut\\
			& $y$ & 0.0446 & $0.0318$ & 0.0523 & 0.0406 & 0.0077 \Bstrut\\\cline{2-7}
\Tstrut		\multirow{2}{*}{3-C} & $x$ & 0.0438 & 0.0165 & 0.0356 & 0.0334 & --- \Bstrut\\
			& $y$ & 0.0100 & $0.0207$ & 0.0124 & 0.0007 & --- \Bstrut\\
			\hline
		\end{tabular}%\\
		% \vspace{10pt}
% 		\begin{tabular}{c| c|c c c c c c} 
% 			\hline
% 			\Tstrut	Floor & Sensor  & Direction & Error \Bstrut\\
% 			\hline
% 			\Tstrut	\multirow{6}{*}{2nd} & \multirow{2}{*}{1-A} & $x$ & 0.1507  \Bstrut\\
% 			& & $y$ & $0.0455$   \Bstrut\\ \cline{2-4}
% \Tstrut		& \multirow{2}{*}{1-C} & $x$ & 0.0395  \Bstrut\\
% 			& & $y$ & $0.0523$   \Bstrut\\\cline{2-4}
% \Tstrut			& \multirow{2}{*}{3-C} & $x$ & 0.0356  \Bstrut\\
% 			& & $y$ & $0.0124$   \Bstrut\\
% 			\hline
% 		\end{tabular}
% 		~~~~
% 				\begin{tabular}{c| c|c c c c c c} 
% 			\hline
% 			\Tstrut	Floor & Sensor  & Direction & Error \Bstrut\\
% 			\hline
% 			\Tstrut	\multirow{6}{*}{3rd} & \multirow{2}{*}{1-A} & $x$ & 0.1294  \Bstrut\\
% 			& & $y$ & $0.0286$   \Bstrut\\ \cline{2-4}
% \Tstrut		& \multirow{2}{*}{1-C} & $x$ & 0.0366  \Bstrut\\
% 			& & $y$ & $0.0406$   \Bstrut\\\cline{2-4}
% \Tstrut			& \multirow{2}{*}{3-C} & $x$ & 0.0334  \Bstrut\\
% 			& & $y$ & $0.0007$   \Bstrut\\
% 			\hline
% 		\end{tabular}
% 		\\
% 		\vspace{10pt}
% 		\begin{tabular}{c| c|c c c c c c} 
% 			\hline
% 			\Tstrut	Floor & Sensor  & Direction & Error \Bstrut\\
% 			\hline
% 			\Tstrut	\multirow{4}{*}{4th} & \multirow{2}{*}{1-A} & $x$ & 0.0753  \Bstrut\\
% 			& & $y$ & $0.0287$   \Bstrut\\ \cline{2-4}
% \Tstrut		& \multirow{2}{*}{1-C} & $x$ & 0.0139  \Bstrut\\
% 			& & $y$ & $0.0077$   \Bstrut\\
% 			\hline
% 		\end{tabular}
	% \vspace{20pt}
\end{table}

\begin{table}[htb!]
				\caption{Prediction errors using unfalsified models from model class $\C_4$ for Example III(a).}%\vspace{-15pt}
	\label{tab:ex3a_errorM4}
	\centering 
\begin{tabular}{c|c c c c c c} 
			\hline
			\Tstrut	 \multirow{2}{*}{Sensor}  & \multirow{2}{*}{Direction} & \multicolumn{5}{c}{Prediction error} \Bstrut \\ \cline{3-7}
            \Tstrut & & Base & 1st floor & 2nd floor & 3rd floor & 4th floor \Bstrut\\
			\hline
			\Tstrut	 \multirow{2}{*}{1-A} & $x$ & 0.1559 & 0.1576 & 0.1601 & 0.1307 & 0.0901 \Bstrut\\
			& $y$ & 0.0556 & $0.0741$ & 0.0611 & 0.0375 & 0.0541 \Bstrut\\ \cline{2-7}
\Tstrut		\multirow{2}{*}{1-C} & $x$ & 0.0493 & 0.0196 & 0.0353 & 0.0420 & 0.0151 \Bstrut\\
			& $y$ & 0.0519 & $0.0670$ & 0.0678 & 0.0501 & 0.0158 \Bstrut\\\cline{2-7}
\Tstrut		\multirow{2}{*}{3-C} & $x$ & 0.0487 & 0.0189 & 0.0336 & 0.0389 & --- \Bstrut\\
			& $y$ & 0.0091 & $0.0337$ & 0.0166 & 0.0022 & --- \Bstrut\\
			\hline
		\end{tabular}

\end{table}

% \FloatBarrier 
\subsubsection{Case (b): Uncertain Base Layer}

\textit{Candidate Model Classes}

% \noindent 
\begin{itemize}
    \item \textbf{Rubber Bearings:} 
For earthquake excitations (Test 014), the rubber bearings still show linear relationship as seen in \cite{brewick:2020aa}. Hence, the restoring force in both $x$ and $y$ direction in the rubber bearings is given by
\begin{equation}\label{eq:linIII}
    f_{\mathrm{RB}} = k_{\mathrm{RB}}u_{\mathrm{RB}}
\end{equation}
where $f_{\mathrm{RB}}$ is the restoring force in the rubber bearings; $k_{\mathrm{RB}}$ is the rubber bearing stiffness coefficient; $u_{\mathrm{RB}}$ is the corresponding displacement of the rubber bearings. 
\item \textbf{Elastic Sliding Bearings:}
Similarly, for the elastic sliding bearings a linear relationship is considered at first. From Coulomb friction
\begin{equation}
    f_{\mathrm{ESB}} = \mu_\mathrm{ESB} W_\mathrm{ESB} \mathrm{sign}(\dot{u}_\mathrm{ESB})
\end{equation}
where $f_{\mathrm{RB}}$ is the restoring force in the rubber bearings; $\mu_\mathrm{ESB}$ is the coefficient of friction;  $W_\mathrm{ESB}$ is the weight on the sliding bearing; $\dot{u}_\mathrm{ESB}$ is the corresponding velocity of the elastic sliding bearings. 
To consider more possibilities, a Bouc-Wen relationship is also considered. Combining Bouc-Wen with Coulomb friction for sliding gives the restoring force as
\begin{equation}
   \left\{ \begin{array}{c}
         f_{\mathrm{ESB}}^x  \\
         f_{\mathrm{ESB}}^y
    \end{array} \right\} 
    = \mu_\mathrm{ESB} W_\mathrm{ESB}
     \left\{ \begin{array}{c}
         Z_{\mathrm{ESB}}^x  \\
         Z_{\mathrm{ESB}}^y
    \end{array} \right\}
\end{equation}
where $f_{\mathrm{ESB}}^x$ and $f_{\mathrm{ESB}}^y$ are the restoring forces in $x$ and $y$ directions, respectively;  $Z_{\mathrm{ESB}}^x$ and $Z_{\mathrm{ESB}}^y$
are the auxiliary variables of hysteresis in the $x$ and $y$ directions, respectively. The auxiliary variables $Z_{\mathrm{ESB}}^x$ and $Z_{\mathrm{ESB}}^y$ follow \citep{park1986random}
\begin{equation}\label{eq:hystIII}
    \begin{split}
        D^y\dot Z_{\mathrm{ESB}}^x &= A \dot u_{\mathrm{ESB}}^x - \beta\Big\lvert\dot u_{\mathrm{ESB}}^x Z_{\mathrm{ESB}}^x\Big\rvert Z_{\mathrm{ESB}}^x - \gamma \dot u_{\mathrm{ESB}}^x (Z_{\mathrm{ESB}}^x)^2
        -\beta \Big\lvert \dot u_{\mathrm{ESB}}^y Z_{\mathrm{ESB}}^y \Big\rvert Z_{\mathrm{ESB}}^x
        -\gamma \dot u_{\mathrm{ESB}}^yZ_{\mathrm{ESB}}^xZ_{\mathrm{ESB}}^y\\
        D^x\dot Z_{\mathrm{ESB}}^y &= A \dot u_{\mathrm{ESB}}^y - \beta\Big\lvert\dot u_{\mathrm{ESB}}^y Z_{\mathrm{ESB}}^y\Big\rvert Z_{\mathrm{ESB}}^y - \gamma \dot u_{\mathrm{ESB}}^y (Z_{\mathrm{ESB}}^y)^2
        -\beta \Big\lvert \dot u_{\mathrm{ESB}}^x Z_{\mathrm{ESB}}^x \Big\rvert Z_{\mathrm{ESB}}^y
        -\gamma \dot u_{\mathrm{ESB}}^x Z_{\mathrm{ESB}}^y Z_{\mathrm{ESB}}^x\\
    \end{split}
\end{equation}
 where $D^x$ and $D^y$ are yield displacements in $x$ and $y$ directions, respectively; the parameters $A$, $\beta$, and $\gamma$ control the shape of the hysteretic loops. Equation \eqref{eq:hystIII} is normalized by dividing both sides with $D^y$ and $D^x$, respectively. 

\item \textbf{Steel Damper Pairs:} For the steel damper pairs, the restoring forces are assumed linear at first, similar to Equation \eqref{eq:linIII}. Next, a hysteresis model is considered. In this model the restoring forces are given by
\begin{equation} 
    \begin{split}
         \left\{ \begin{array}{c}
         f_{\mathrm{SD}}^x  \\
         f_{\mathrm{SD}}^y
    \end{array} \right\} 
    = \alpha \Km_{\mathrm{SD}} \left\{ \begin{array}{c}
         u_{\mathrm{SD}}^x  \\
         u_{\mathrm{SD}}^y
    \end{array} \right\}  
    + (1-\alpha)\Km_\mathrm{SD} \left\{ \begin{array}{c}
         Z_{\mathrm{SD}}^x  \\
         Z_{\mathrm{SD}}^y
    \end{array} \right\} 
    \end{split}
\end{equation}
where $f_{\mathrm{ESB}}^x$ and $f_{\mathrm{ESB}}^y$ are the restoring forces in the $x$ and $y$ directions, respectively; $\Km_{\mathrm{SD}}$ is the stiffness matrix; $\alpha$ is the ratio of post-yield to pre-yield stiffness; the auxiliary variables of hysteresis $Z_{\mathrm{SD}}^x$ and $Z_{\mathrm{SD}}^y$ follow equations similar to \eqref{eq:hystIII}. However, $D^x = D^y = 1$ for steel dampers since $Z_{\mathrm{SD}}^x$ and $Z_{\mathrm{SD}}^y$ are not dimensionless as in \eqref{eq:hystIII}. %In the present study, $k_{\mathrm{RB}}$ for rubber bearings, $\beta$ and $\gamma$ for elastic sliding bearings, $k=k_{xx}=k_{yy}$, $k_{xy}$, $\alpha$, $\beta$, $\gamma$ for steel dampers are assumed to be uncertain. 
\end{itemize} 

Using different combinations of linear or hysteretic models for these passive control devices four candidate model classes are formed as shown in Table \ref{tab:exIII_mod_cls}.

\begin{table}[t]
	%\small
		\caption{Model class parameters for the base isolation devices. (Std.~Dev.~denotes standard deviation.)}
	\label{tab3prior}
	\centering
%	\begin{threeparttable}
		\begin{tabular}{@{} l l l l l l l l l l@{}} 
			\hline
			\Tstrut
			\multirow{2}{*}{Device} & \multirow{2}{*}{Relationship} & \multirow{2}{*}{Parameter} & \multicolumn{3}{c}{Prior Distribution} &&
			\Bstrut\\
			\cline{4-6}  \Tstrut & & & Type & Mean & Std.~Dev. && \Bstrut\\
			\hline
			\Tstrut
			Rubber Bearing & Linear & $k_{\mathrm{RB}}$ [kN/m] & Lognormal &$1100$  & $100$ && \Bstrut\\
			\hline \Tstrut 
			\multirow{3}{*}{Elastic Sliding Bearing} & Linear & $k_{\mathrm{ESB}}$ [kN/m] & Lognormal &  $1550$  & $100$ && \Bstrut \\
			%\hline
			\cline{2-8} \Tstrut 
			& \multirow{2}{*}{Hysteretic} & $\beta$ & Uniform & $0.25$ & $0.0289$ && \\
			%\hline
			& & $\gamma$ & Uniform & $0.35$ & $0.0289$ &&  \\
			\hline \Tstrut 
			\multirow{6}{*}{Steel Damper} & Linear
			& $k_{\mathrm{SD}}$ [kN/m] & Lognormal & $4100$ & $400$ & & \Bstrut\\
			%(in \%) & &\\
			\cline{2-8} \Tstrut 
			& \multirow{5}{*}{Hysteretic} & $k_{\mathrm{SD}}$ [kN/m] & Lognormal & 25 & 2 \\
			& & $k_{xy}$ [kN/m] & Lognormal & 0.5 & 0.025 \\
			& & $\alpha$ & Uniform & 0.65 & 0.0289\\
			& & $\beta$ & Uniform & 0.65 & 0.0289\\
			& & $\gamma$ & Uniform & $-0.150$ & 0.0144\\
			\hline
		\end{tabular}
%		\begin{tablenotes}
%			\item[\textdagger] {\footnotesize{The nonlinear models do not require $\rd$.}  }
%			\item[*] {\footnotesize{The linear models do not require $\Qy$.}}
			%\item[\textdaggerdbl] {\footnotesize{Maximum Likelihood estimate of the Bilinear model parameters is same as $\widehat{\thetaa}$ up to four decimal points.}  }
%		\end{tablenotes}
%	\end{threeparttable}
%	\vspace{-12pt}
\end{table}

\begin{table}[htb!]
		\caption{Candidate model classes used and falsification results for Example III(b).} \label{tab:exIII_mod_cls}
	\centering
	\begin{tabular}{ c  |l l l  | c c c c } 
		\hline
		\Tstrut  \multirow{2}{*}{Model Class} & \multicolumn{3}{c|}{Relationship} & \multirow{2}{*}{\% Unfalsified} \\
		& Rubber Bearings & Elastic Sliding Bearings & Steel Dampers & \Bstrut \\ 
		\hline
		\Tstrut $\C_1$ &  Linear & Linear & Linear & ~~0.0\\
		$\C_2$	& Linear & Hysteretic & Linear & ~~0.0\\
		$\C_3$ & Linear & Linear & Hysteretic & ~~0.0\\
		$\C_4$ & Linear & Hysteretic & Hysteretic & 75.0\Bstrut\\
		\hline
	\end{tabular}
	
\end{table}

\noindent\textit{Results:}\\
\indent Assuming target identification probability $\phi=0.90$, model falsification is implemented with response measurement data from Test 012 and $\Ns=10000$ models from each class. As the intensity of the excitation is higher than Test 010 used in case (a), the nonlinear behavior of the isolation devices is prominent in the measurement data. Note that, as the behavior of the models is nonlinear in this case, a larger number of candidate models are used for falsification. 
%{{In Test~012, all isolation-layer devices were assumed to be linear and random excitations from all table DOF's were used as the inputs, just like Test~010. The only difference is that the random excitaions in Test~012 have much more intense magnitudes compared with Test~010. Thus, if the validation results turn out to be desirable, some confidence could be given to the estimated parameters from model falsification as least when the magnitudes of the inputs are within those of Test~010 and Test~012.}}.
The likelihood function is assumed as zero-mean Gaussian with a standard deviation {{30$\%$ of the response RMS}}. Among four candidate model classes, only models from $\C_4$ remain unfalsified (see Table \ref{tab:exIII_mod_cls}). The estimated parameters using \eqref{eq:param_est2} for the isolator devices are shown in Table \ref{tab:ex3b_param}. The unfalsified models are then used for response prediction for Test 014 using \eqref{eq:rob_pred2}. \fref{fig:ex3b_resppred} shows the measured and predicted response for the sensor near 1-A.  These figures show that the proposed method performs well in response prediction for a full-scale structure in the presence of noises encountered during experimentation. The force-displacement characteristics of different isolator devices are also compared and shown in \fref{fig:ex3b_force}. These plots confirm that the behavior of the rubber bearing is predominantly linear whereas the other two types of devices show hysteretic behavior.
The RMS errors in acceleration responses are computed next and shown in Table \ref{tab:ex3b_errorM4} for model class $\C_4$. Note that the RMS errors in prediction are reduced compared to case (a).

% \begin{table}[htb!]
% 		\caption{Falsification results for Example III(b).} \label{tab:ex3b_fals}
% 	\centering
% 	\begin{tabular}{ l  |c c c  c c c c} 
% 		\hline
% 		\Tstrut  Model Class & \% Unfalsified\Bstrut \\ 
% 		%& \multicolumn{2}{|c|}{8 Hz sampling rate} & \multicolumn{2}{|c|}{20 Hz sampling rate} \Tstrut\Bstrut\\ \cline{2-5} \Tstrut
% 		%	\Tstrut	& \multicolumn{3}{c|}{Error-bound} & \multicolumn{4}{c}{Likelihood-bound}\Bstrut\\
% 		%		& Bonferroni & \v{S}id\'ak & BH &  Bonferroni & \v{S}id\'ak & BH & CPM \Tstrut\Bstrut\\
% 		\hline
% 		\Tstrut $\C_1$ &  ~~0.0\\
% 		%		Cubic Polynomial &  96.25  \\
% 		%		CALTRANS   & 0 \\
% 		%		 mod.~AASHTO  & 0 \\
% 		%		Bouc-Wen & 32.2& 32& 29.75& 100&100 & 89.45 & 16.7 \\
% 		$\C_2$	& ~~0.0\\
% 		$\C_3$ & ~~0.0\\
% 		$\C_4$ & {75.0}\Bstrut\\
% 		%\\
% 		%modified AASHTO & ~~0.0\\
% 		%Caltrans & ~~0.0
% 		%		Bilinear  & 5.1\Bstrut\\
% 		\hline
% 	\end{tabular}
	
% \end{table}

\begin{table}[htb!]
			\caption{ Estimated parameters of model class $\C_4$ for Example III(b). }%\vspace{-15pt}
	\label{tab:ex3b_param}
	\begin{center}
		
		\begin{tabular}{l |l c l c c c} 
			\hline
			\Tstrut	Device  & {Parameter} & $\widehat{\thetaa}$ \Bstrut\\
			\hline
			\Tstrut	{Rubber Bearing} & $k_\mathrm{RB}$ & 1077.60 kN/m   \Bstrut\\ \hline
			\Tstrut	\multirow{2}{*}{Elastic Sliding Bearing} & $\beta$ &  0.4426~cm$^{-1}$ \Bstrut\\
			& $\gamma$ & {{0.1492~cm$^{-1}$}}   \Bstrut\\ \hline
			\Tstrut	\multirow{5}{*}{Steel Damper} & $k_\mathrm{SD}$ &  {3056~kN/m} \Bstrut\\
			& $k_{xy}$ & {{0.4868~kN/cm}}   \Bstrut\\ 
			& $\alpha$ & {0.0663}   \Bstrut\\ 
			& $\beta$ & 0.0673~cm$^{-2}$  \Bstrut\\ 
			& $\gamma$ & $-0.0151$~cm$^{-2}$   \Bstrut\\ 
			\hline
		\end{tabular}
	\end{center}
	%\vspace{-20pt}
\end{table}

\begin{figure}[htb!]
    \centering
    \begin{subfigure}[t]{0.5\textwidth}
        \centering
        \includegraphics[scale=0.3]{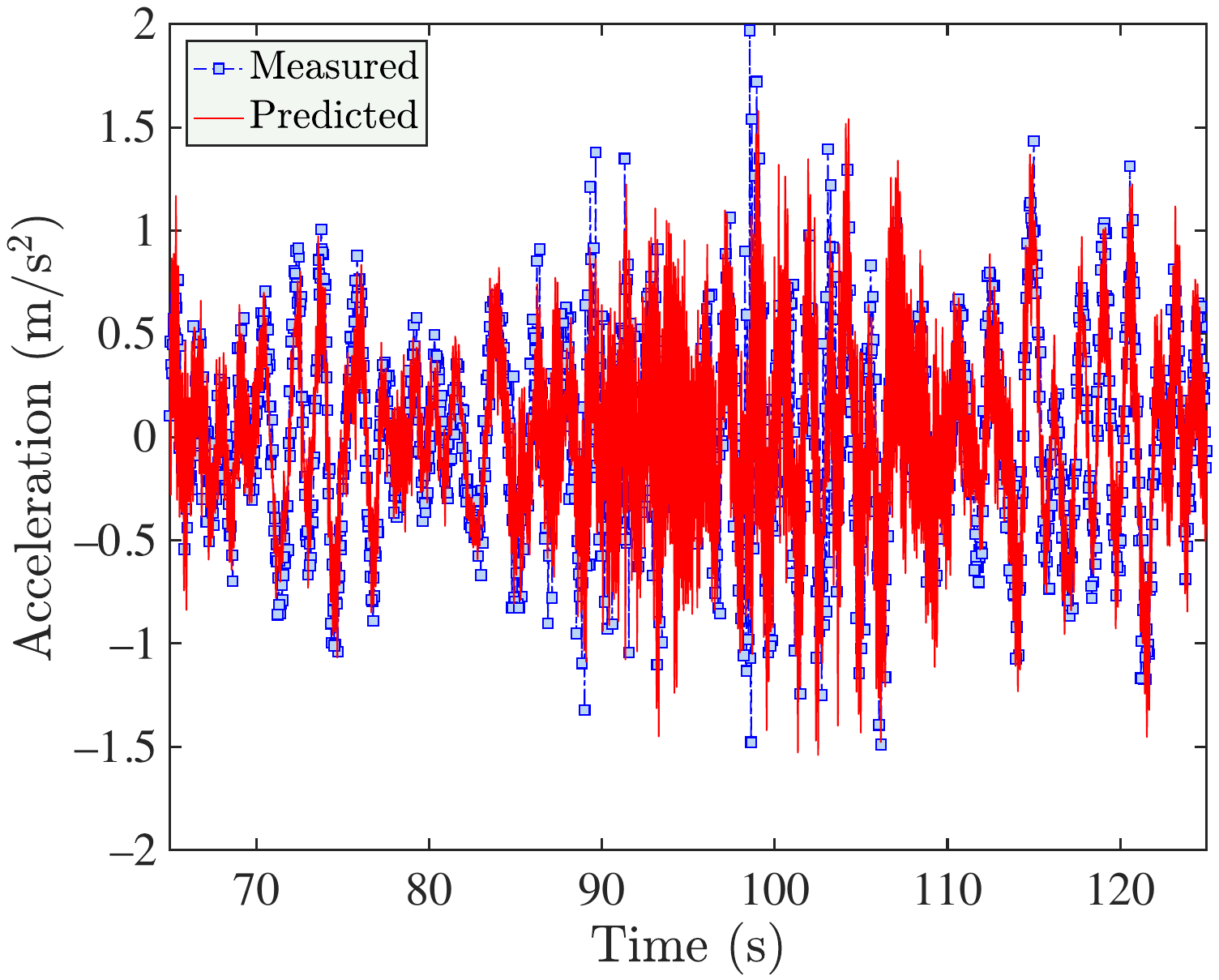}
        \caption{Base acceleration in $x$ direction.}
    \end{subfigure}%
    ~ 
    \begin{subfigure}[t]{0.5\textwidth}
        \centering
        \includegraphics[scale=0.3]{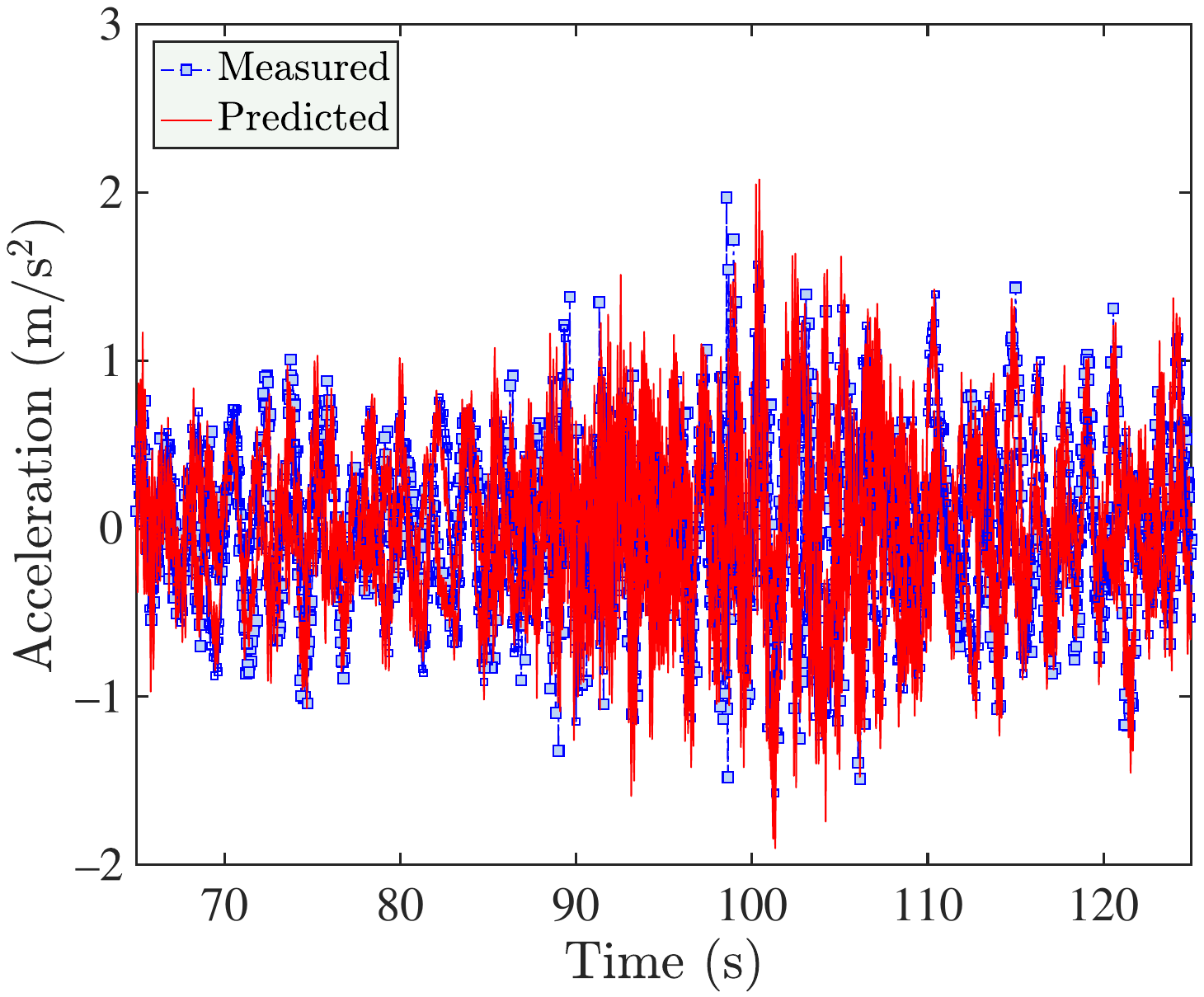}
        \caption{Base acceleration in $y$ direction.}
    \end{subfigure}
    \caption{Predicted base response is compared with measured during Test 014 using the sensor near 1-A (see \fref{fig:sensors}). For other floors, see \fref{fig:ex3b_resppred_allfloors} in Appendix \ref{sec:appendix1}.} \label{fig:ex3b_resppred}
\end{figure}

\begin{figure}[!htb]
    \centering
    \begin{subfigure}[t]{0.5\textwidth}
        \centering
        \includegraphics[scale=0.3]{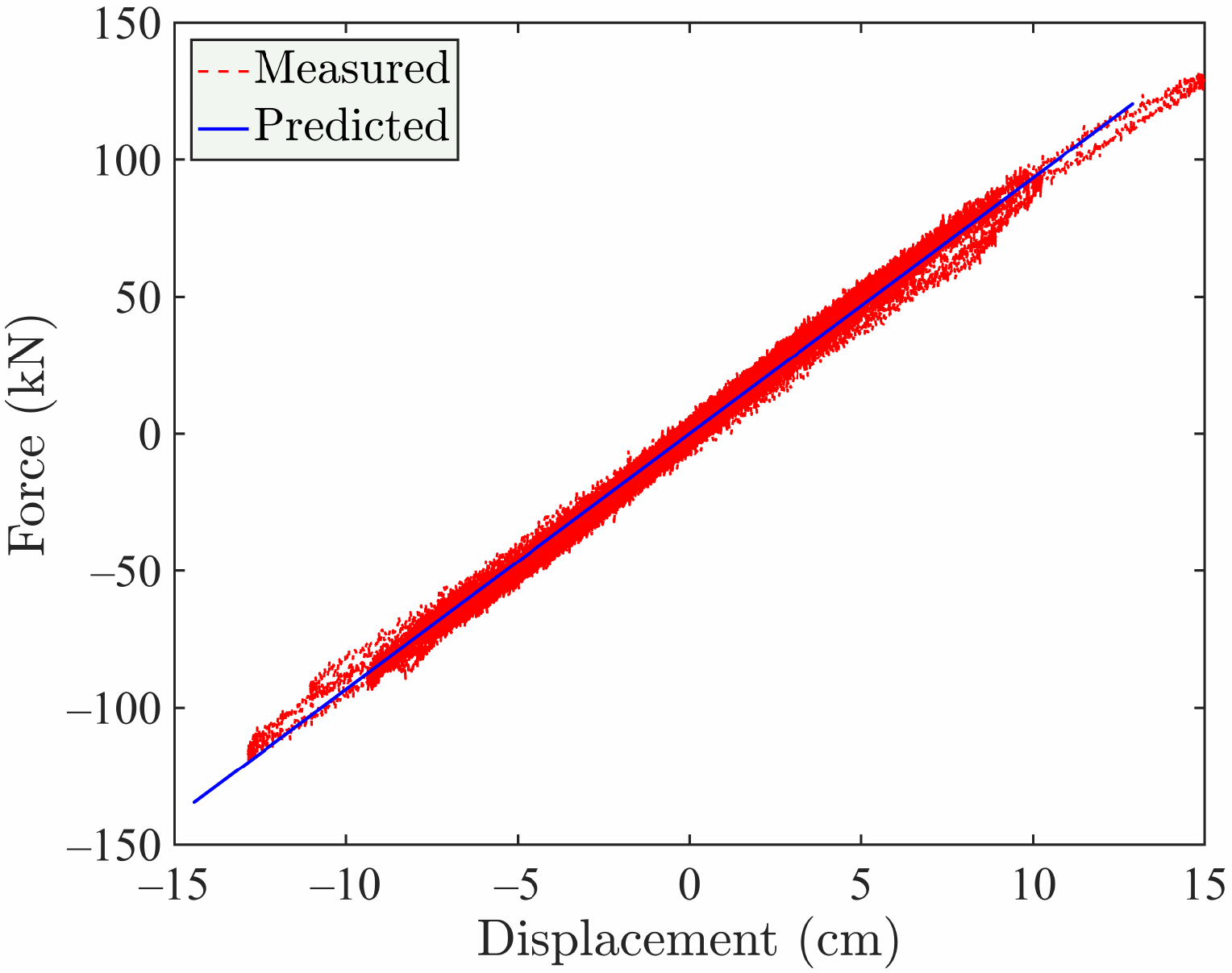}
        \caption{Rubber bearing at 1--A in $x$ direction.}
    \end{subfigure}%
    ~ 
    \begin{subfigure}[t]{0.5\textwidth}
        \centering
        \includegraphics[scale=0.3]{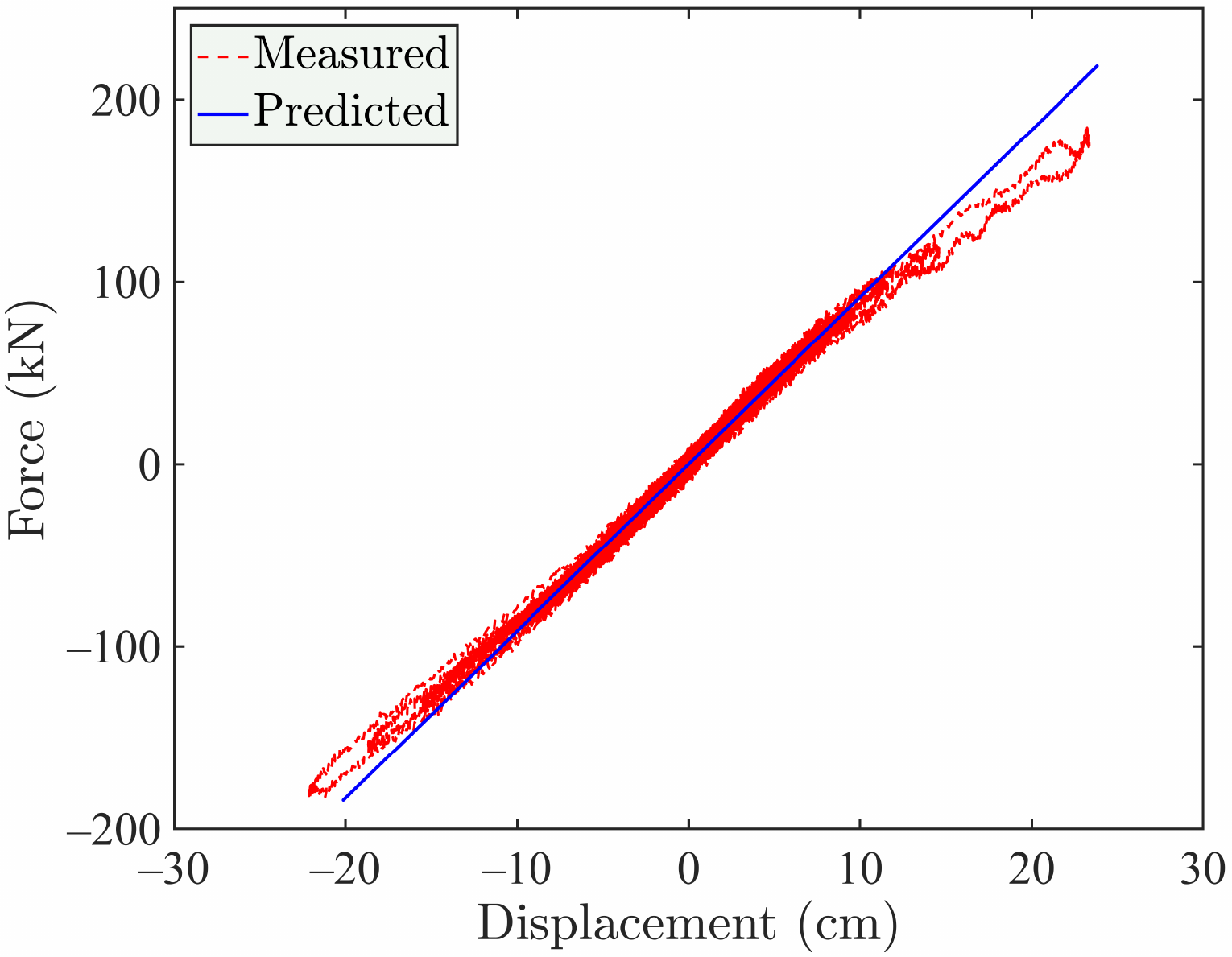}
        \caption{Rubber bearing at 1--A in $y$ direction.}
    \end{subfigure}
    \begin{subfigure}[t]{0.5\textwidth}
        \centering
        \includegraphics[scale=0.3]{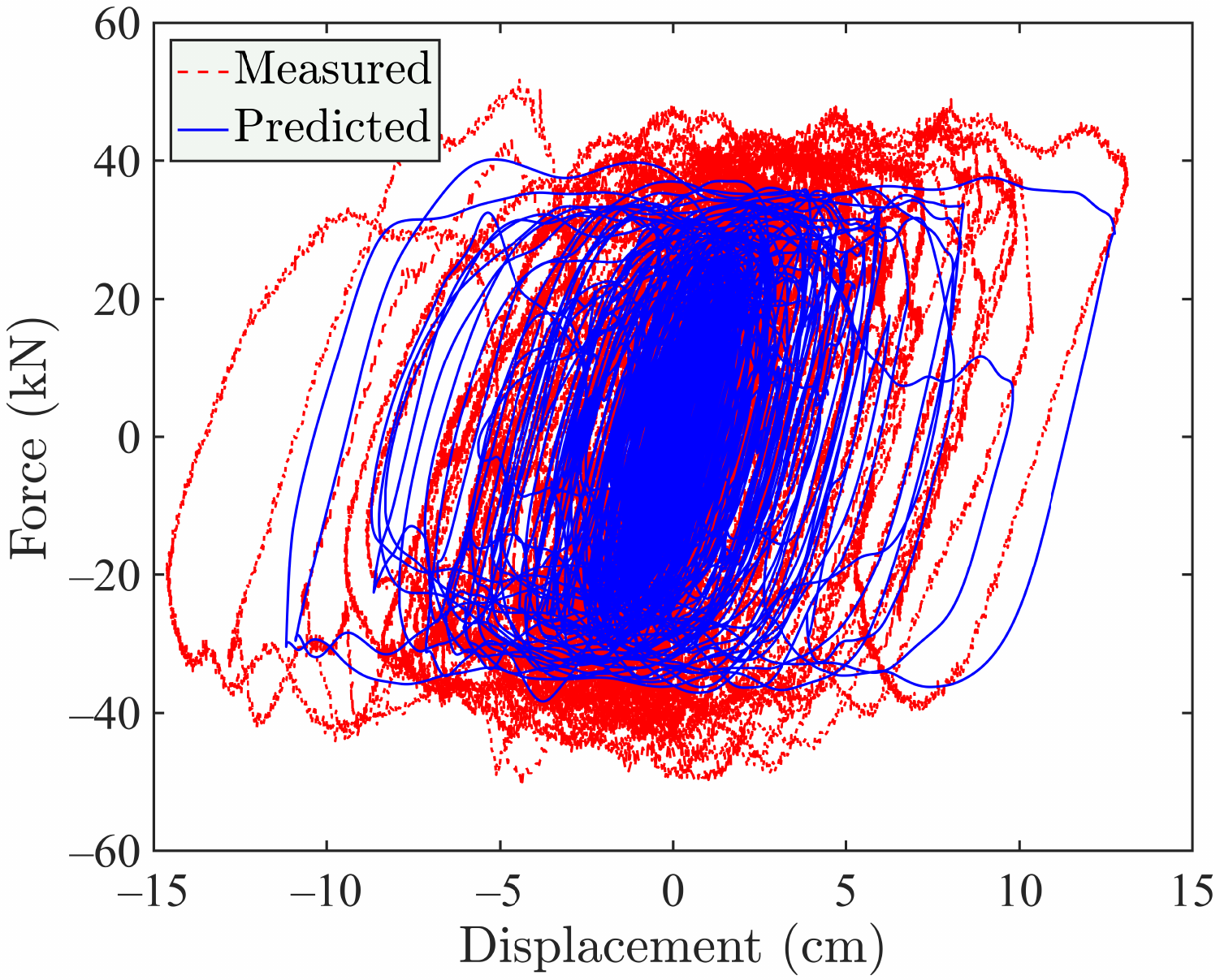}
        \caption{Elastic sliding bearing at 1--C in $x$ direction.}
    \end{subfigure}%
    ~ 
    \begin{subfigure}[t]{0.5\textwidth}
        \centering
        \includegraphics[scale=0.3]{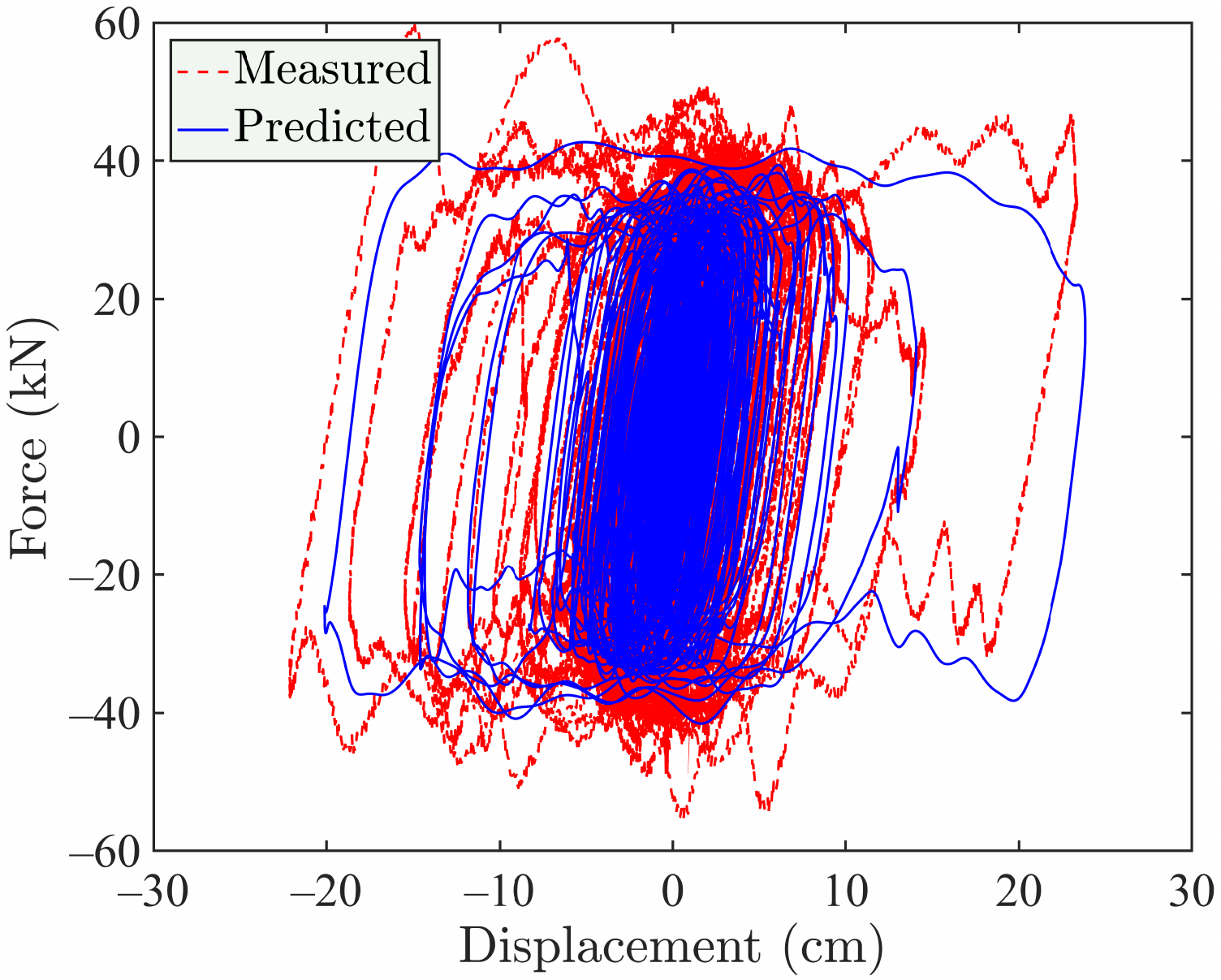}
        \caption{Elastic sliding bearing at 1--C in $y$ direction.}
    \end{subfigure}
    \begin{subfigure}[t]{0.5\textwidth}
        \centering
        \includegraphics[scale=0.3]{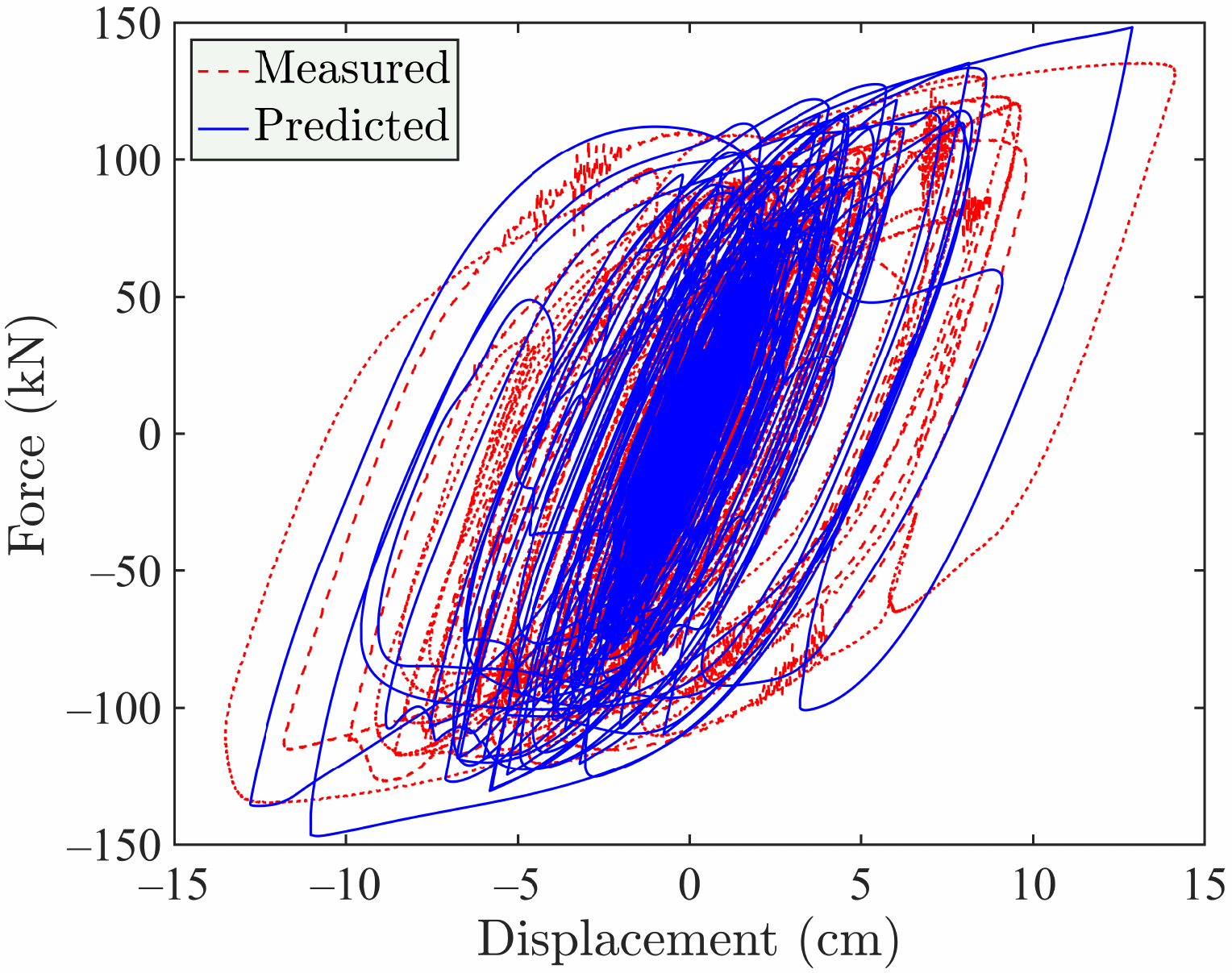}
        \caption{Steel damper at 1--B in $x$ direction.}
    \end{subfigure}%
    ~ 
    \begin{subfigure}[t]{0.5\textwidth}
        \centering
        \includegraphics[scale=0.3]{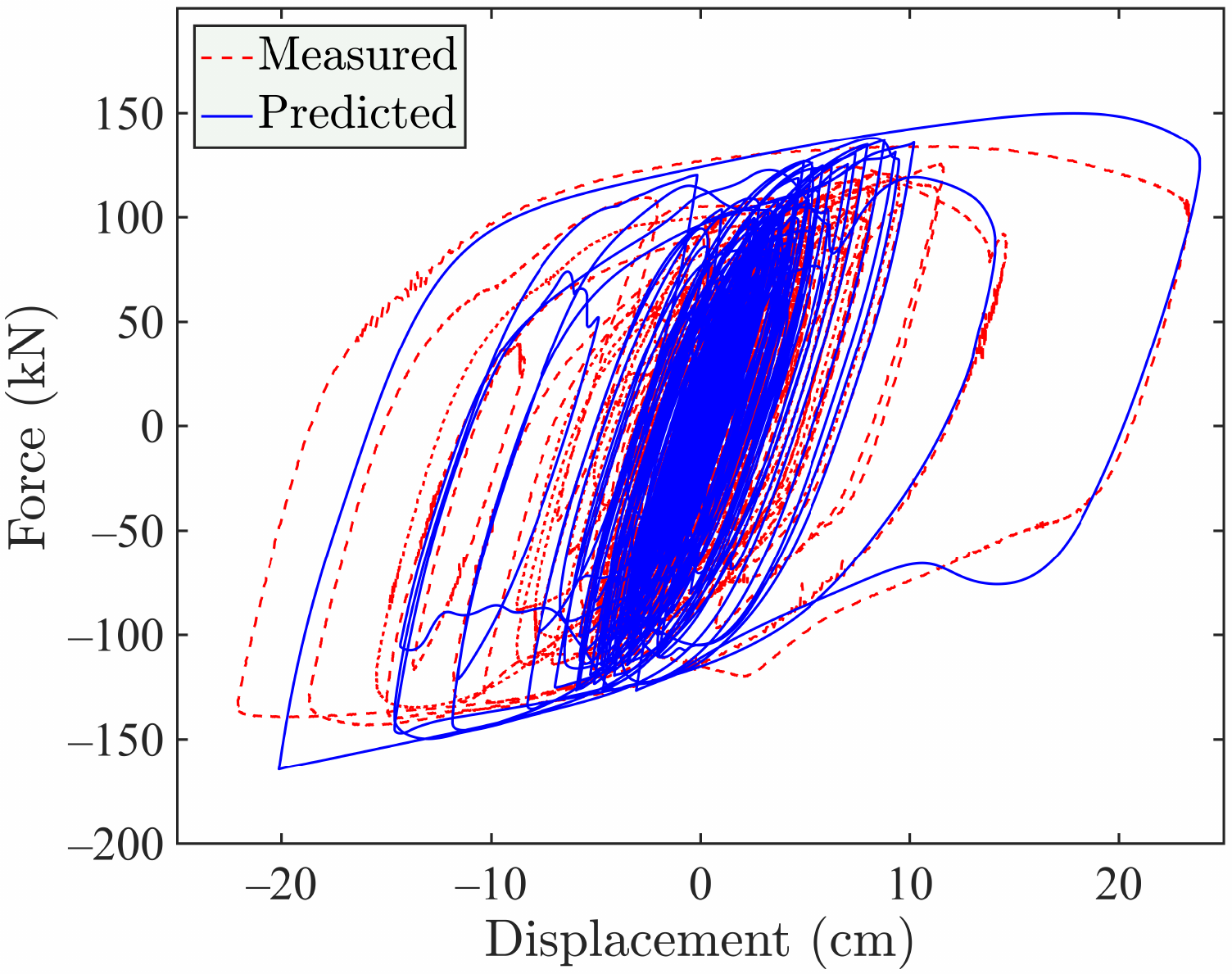}
        \caption{Steel damper at 1--B in $y$ direction.}
    \end{subfigure}
    \caption{Predicted and measured force displacement characteristics of different isolator devices during Test 014.} \label{fig:ex3b_force}
    \end{figure}

    \begin{table}[htb!]
				\caption{Prediction errors using unfalsified models from model class $\C_4$ for Example III(b).}%\vspace{-15pt}
	\label{tab:ex3b_errorM4}
	\centering
\begin{tabular}{c|c c c c c c} 
			\hline
			\Tstrut	 \multirow{2}{*}{Sensor}  & \multirow{2}{*}{Direction} & \multicolumn{5}{c}{Prediction error} \Bstrut \\ \cline{3-7}
            \Tstrut & & Base & 1st floor & 2nd floor & 3rd floor & 4th floor \Bstrut\\
			\hline
			\Tstrut	 \multirow{2}{*}{1-A} & $x$ & 0.0790 & 0.1224 & 0.1057 & 0.0209 & 0.0747 \Bstrut\\
			& $y$ & 0.0743 & $0.0163$ & 0.0417 & 0.1057 & 0.1267 \Bstrut\\ \cline{2-7}
\Tstrut		\multirow{2}{*}{1-C} & $x$ & 0.0553 & 0.0097 & 0.0912 & 0.0771 & 0.0913 \Bstrut\\
			& $y$ & 0.0851 & $0.0249$ & 0.0374 & 0.0843 & 0.0092 \Bstrut\\\cline{2-7}
\Tstrut		\multirow{2}{*}{3-C} & $x$ & 0.0395 & 0.0072 & 0.0933 & 0.0741 & --- \Bstrut\\
			& $y$ & 0.0062 & $0.0654$ & 0.0463 & 0.0031 & --- \Bstrut\\
			\hline
		\end{tabular} 
\end{table}

    \FloatBarrier

\section{Conclusions}
Identifying a valid model (or models) for a physical system is essential for response prediction to provide further insights into the system behavior, which ultimately assists design, analysis, and maintenance decisions.
%, designing control strategy, lifetime prognosis, and so on.
A  likelihood bound, determined using the false discovery rate (FDR), is used herein for falsifying invalid models of a physical system from measured responses to one input scenario. Weights are assigned to the unfalsified models according to Bayes' theorem for subsequent use in
predicting system response to other input scenario(s). % in a manner that is robust to the modeling uncertainty. %the robust prediction of system response.
{\color{black}The proposed approach is implemented to predict the dynamical responses of structures with uncertain passive control devices, where different model classes are assumed to describe the structure or the devices.} %Uncertainty is assumed to be present in structure as well as in the passive control devices. 
% Two types of passive control devices, namely, base isolation and tuned mass dampers are investigated herein.
First, a 4-DOF base-isolated building with six model classes to represent its hysteretic isolation layer is used to illustrate the proposed approach. With measurements from the building subjected to the 1940 El Centro earthquake, likelihood-bound model falsification is applied, resulting in four falsified linear model classes. The unfalsified models from the two nonlinear classes are then used to predict the responses to a different earthquake excitation. The second example employs a 1623-DOF wind-excited building model with three tuned mass dampers attached to its roof. The candidate model classes do not include the true model class used to generate the measurements. Again, the method falsifies multiple model classes and only retains the near-truth model classes, which are subsequently used for response prediction of the roof acceleration from a different wind excitation. The third example uses an experimental setup for testing a full-scale four-story structure with a base isolation layer on the world's largest shake table in Japan's EDefense laboratory. Multiple model classes are used first for the superstructure with a linear base isolation layer and then for a nonlinear base isolation layer. In the first case, uncertainty is introduced in the superstructure but, for the second case, uncertainty is assumed in the base isolation layer. The response predictions using unfalsified models are performed and compared with different test results. 
These examples all show that the proposed method provides accurate response predictions, and do so with a computational cost that is a fraction of the cost without falsification. 
%The predictions' robustness to the modeling uncertainty, induced by the presence of measurement noise and a randomly-chosen set of candidate models, is also demonstrated for the two examples.
%Future directions include a further analysis step to distinguish between unfalsified models from more than one model classes.

\section{Acknowledgment}
The authors gratefully acknowledge the partial support of this work by the National Science Foundation through awards CMMI 14-36018/14-36058 and 16-63667/16-62992. Any opinions, findings, and conclusions or recommendations expressed in this material are those of the authors and do not necessarily reflect the views of the National Science Foundation. Dr.~De, Dr.~Yu, and Dr.~Brewick acknowledge the support of a Viterbi Ph.D.~Fellowship, a Provost Fellowship, and a Viterbi Postdoctoral Fellowship, respectively, from University of Southern California. The authors thank Dr.~Eiji Sato (NIED) and Dr.~Tomohiro Sasaki (Obayashi Corp.) for their extensive efforts on the base isolation experiments and their collaboration studying the resulting data that is also used in the third example. 
 
%
% Here's the list of references:
%
% \label{section:references}
% \bibliography{ascexmpl-new}
%
\bibliography{Proposal_Refs}%

\appendix 
\section{Supplementary Results from Example III} \label{sec:appendix1}

This appendix presents additional figures supporting the results discussed in Example III. \fref{fig:ex3a_resppredM3_allfloors} and \fref{fig:ex3a_resppredM4_allfloors} compare predicted acceleration responses of floors 1st-4th obtained using unfalsified models from model classes $\C_3$ and $\C_4$, respectively, are compared with experimentally measured responses from Test 012. \fref{fig:ex3b_resppred_allfloors} shows a comparison of predicted acceleration responses of floors 1st-4th obtained using unfalsified models from model class $\C_4$ with experimentally measured responses from Test 014. 
\begin{figure}[htb!]
    \centering
    % \begin{subfigure}[t]{0.5\textwidth}
    %     \centering
    %     \includegraphics[scale=0.3]{figures/floor1_x_linear_class4.pdf}
    %     \caption{Base acceleration in $x$ direction.}
    % \end{subfigure}%
    % ~ 
    % \begin{subfigure}[t]{0.5\textwidth}
    %     \centering
    %     \includegraphics[scale=0.3]{figures/floor1_y_linear_class4.pdf}
    %     \caption{Base acceleration in $y$ direction. {\color{purple!70!blue}(EAJ: Zoom in. SD: I can zoom in here. But, in case (b), I don't have the fig files.)}}
    % \end{subfigure}
    \begin{subfigure}[t]{0.5\textwidth}
        \centering
        \includegraphics[scale=0.3]{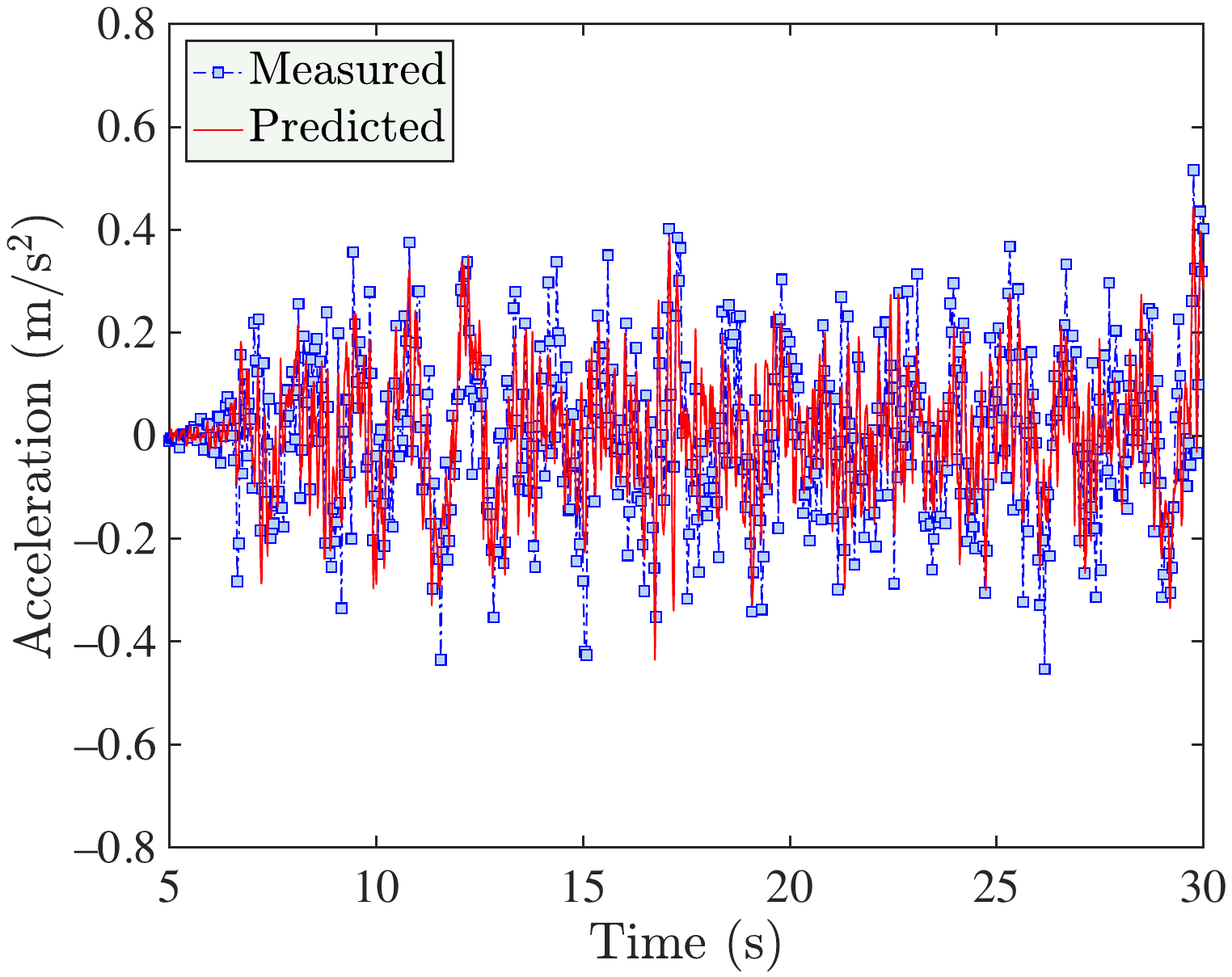}
        \caption{1st floor acceleration in $x$ direction.}
    \end{subfigure}%
    ~ 
    \begin{subfigure}[t]{0.5\textwidth}
        \centering
        \includegraphics[scale=0.3]{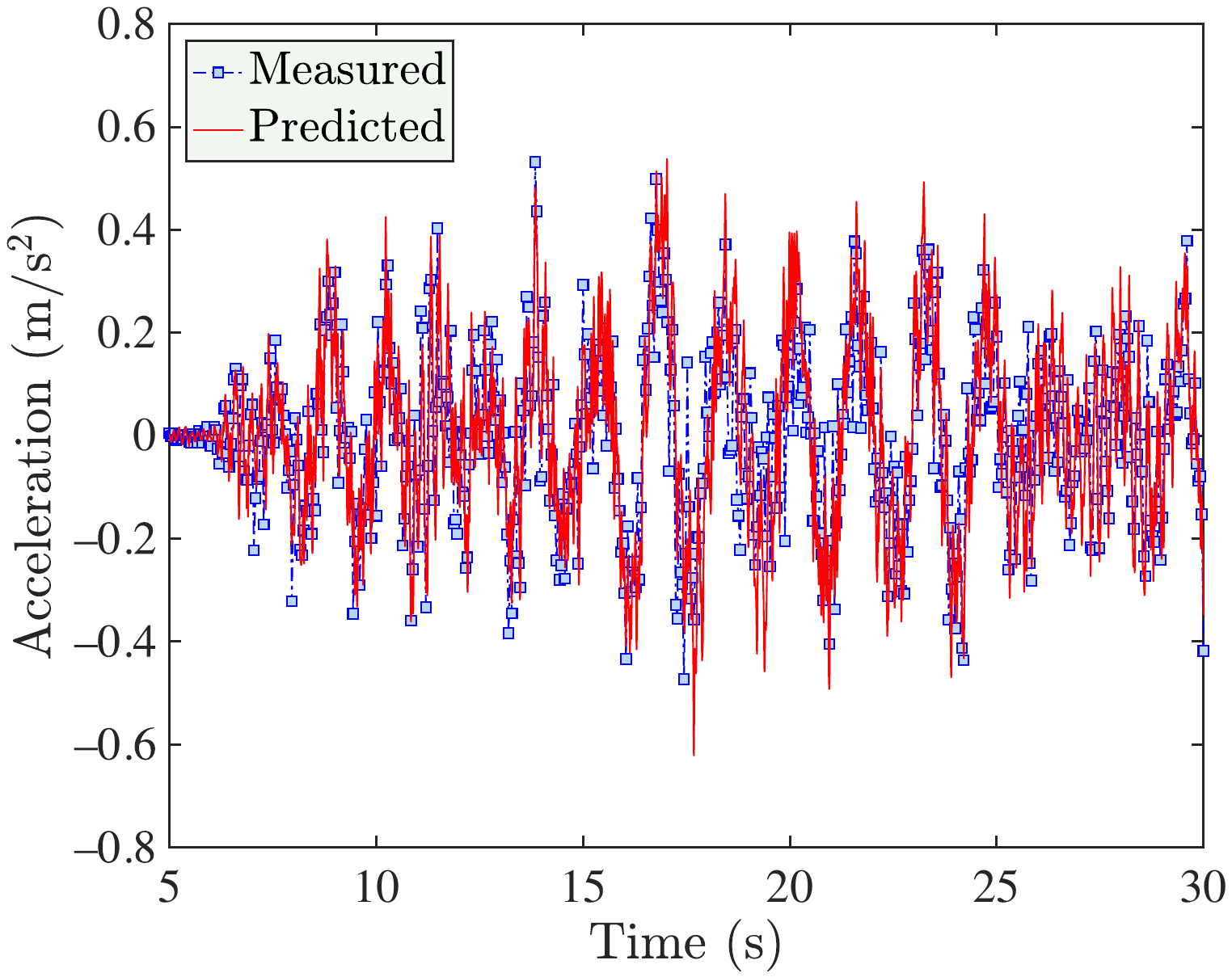}
        \caption{1st floor acceleration in $y$ direction.}
    \end{subfigure}
    \begin{subfigure}[t]{0.5\textwidth}
        \centering
        \includegraphics[scale=0.3]{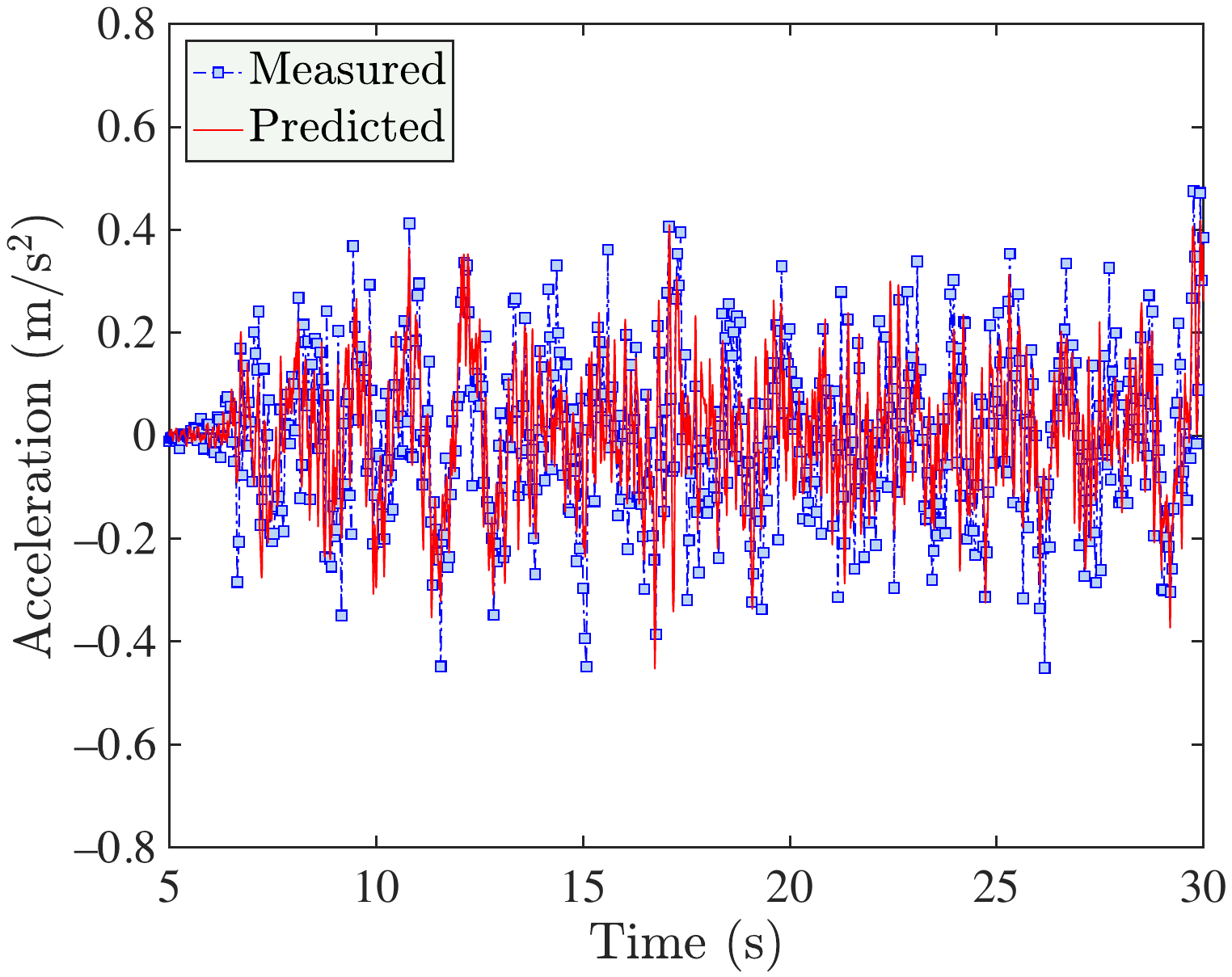}
        \caption{2nd floor acceleration in $x$ direction.}
    \end{subfigure}%
    ~ 
    \begin{subfigure}[t]{0.5\textwidth}
        \centering
        \includegraphics[scale=0.3]{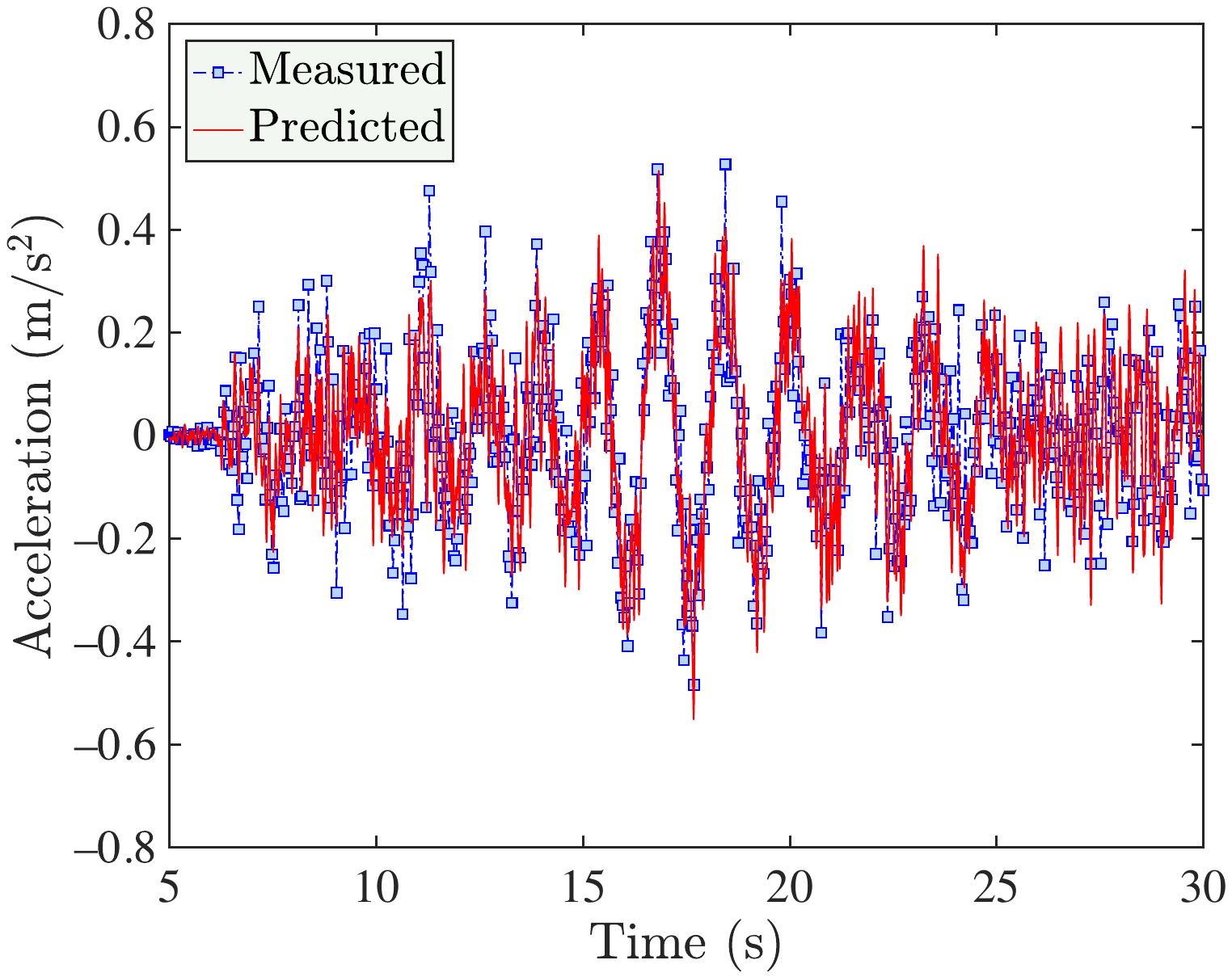}
        \caption{2nd floor acceleration in $y$ direction.}
    \end{subfigure}
    \end{figure}
\begin{figure}[htb!]\ContinuedFloat
    \begin{subfigure}[t]{0.5\textwidth}
        \centering
        \includegraphics[scale=0.3]{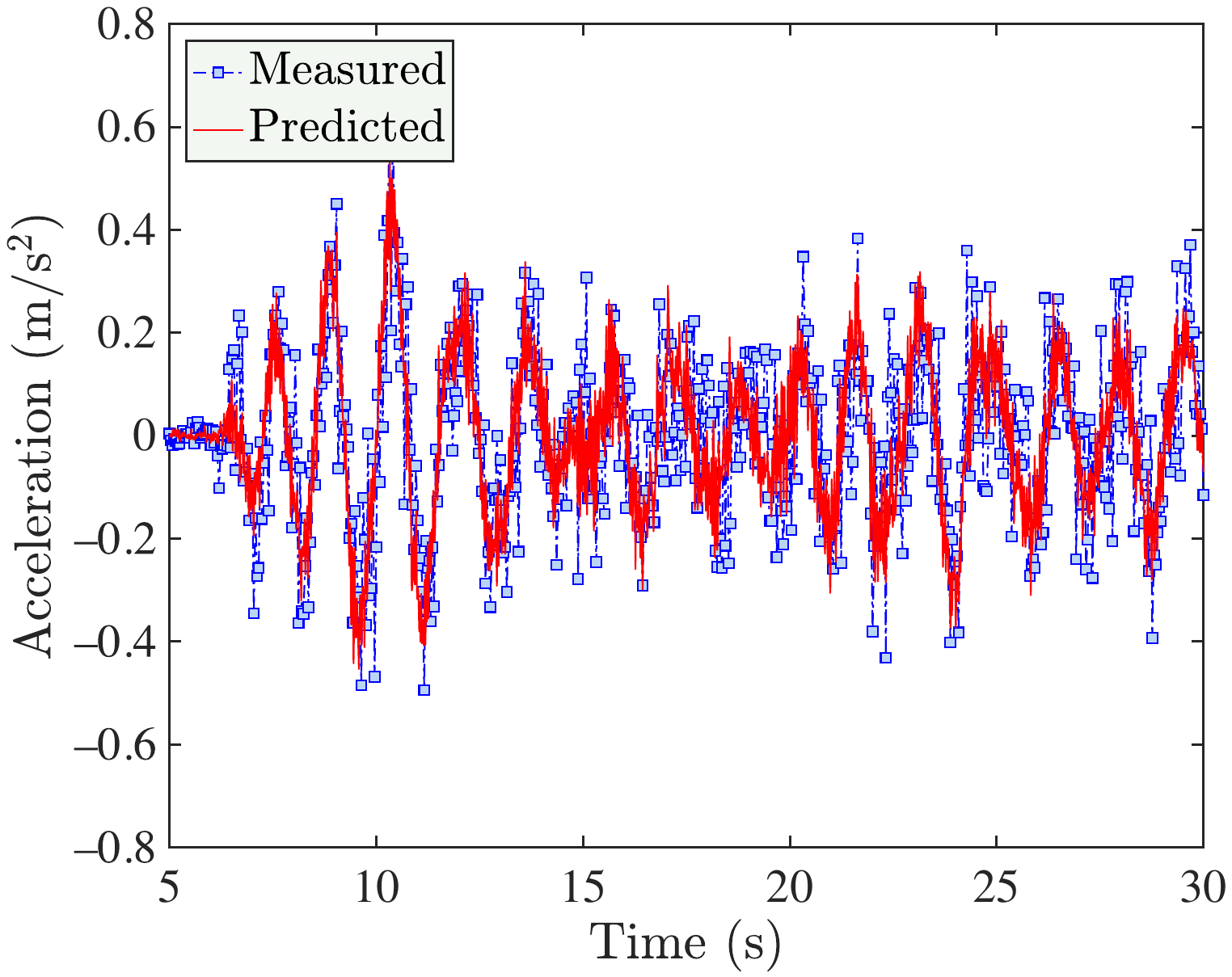}
        \caption{3rd floor acceleration in $x$ direction.}
    \end{subfigure}%
    ~ 
    \begin{subfigure}[t]{0.5\textwidth}
        \centering
        \includegraphics[scale=0.3]{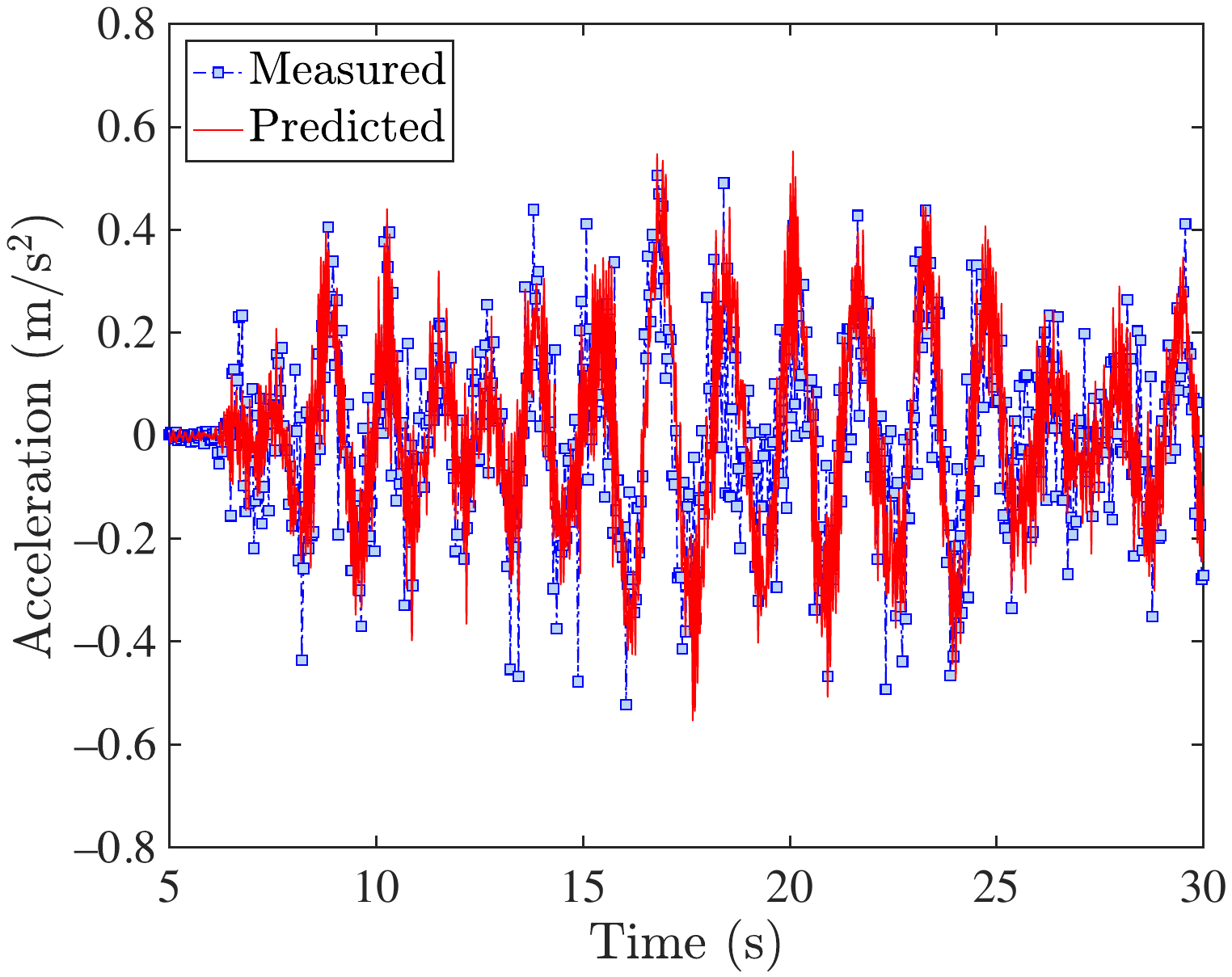}
        \caption{3rd floor acceleration in $y$ direction.}
    \end{subfigure}
    \begin{subfigure}[t]{0.5\textwidth}
        \centering
        \includegraphics[scale=0.3]{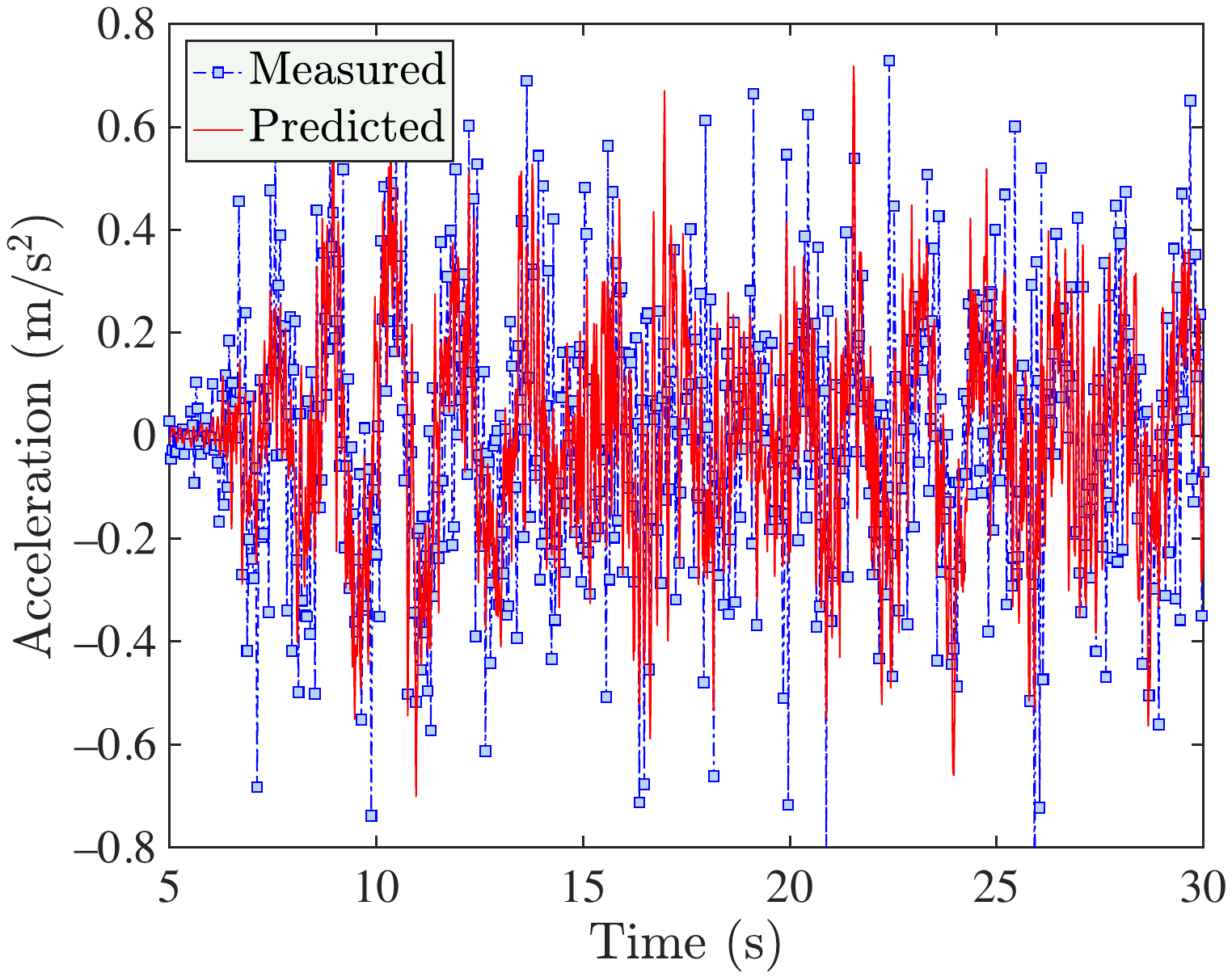}
        \caption{4th floor acceleration in $x$ direction.}
    \end{subfigure}%
    ~ 
    \begin{subfigure}[t]{0.5\textwidth}
        \centering
        \includegraphics[scale=0.3]{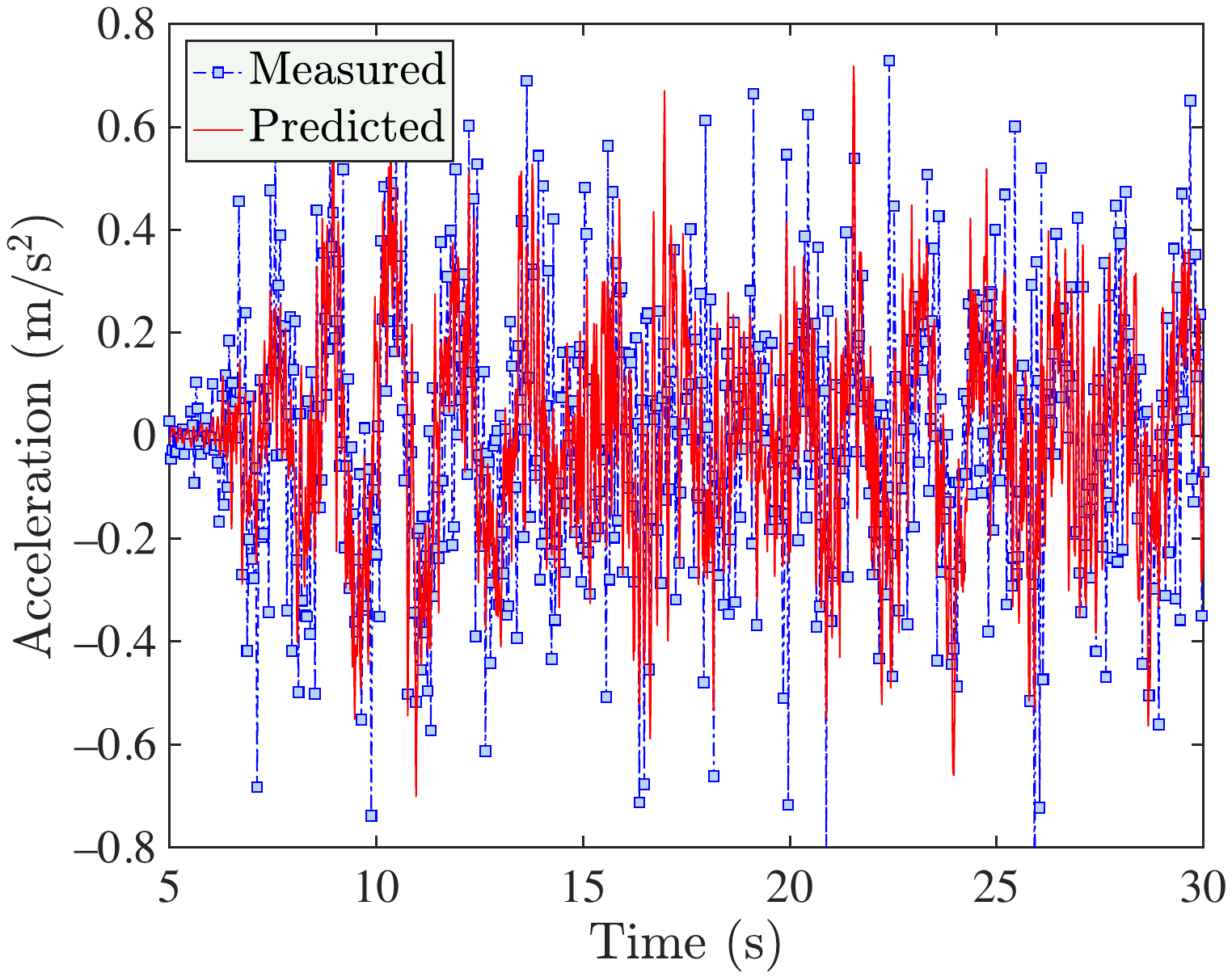}
        \caption{4th floor acceleration in $y$ direction.}
    \end{subfigure}
    \caption{Predicted responses using unfalsified models from $\C_3$ are compared with measured during Test 012 using the sensor near 1-A (see \fref{fig:sensors}).} \label{fig:ex3a_resppredM3_allfloors}
\end{figure} 

\begin{figure}[htb!]
    \centering
    % \begin{subfigure}[t]{0.5\textwidth}
    %     \centering
    %     \includegraphics[scale=0.3]{figures/floor1_x_linear_class5.pdf}
    %     \caption{1st floor acceleration in $x$ direction.}
    % \end{subfigure}%
    % ~ 
    % \begin{subfigure}[t]{0.5\textwidth}
    %     \centering
    %     \includegraphics[scale=0.3]{figures/floor1_y_linear_class5.pdf}
    %     \caption{1st floor acceleration in $y$ direction.}
    % \end{subfigure}
    \begin{subfigure}[t]{0.5\textwidth}
        \centering
        \includegraphics[scale=0.3]{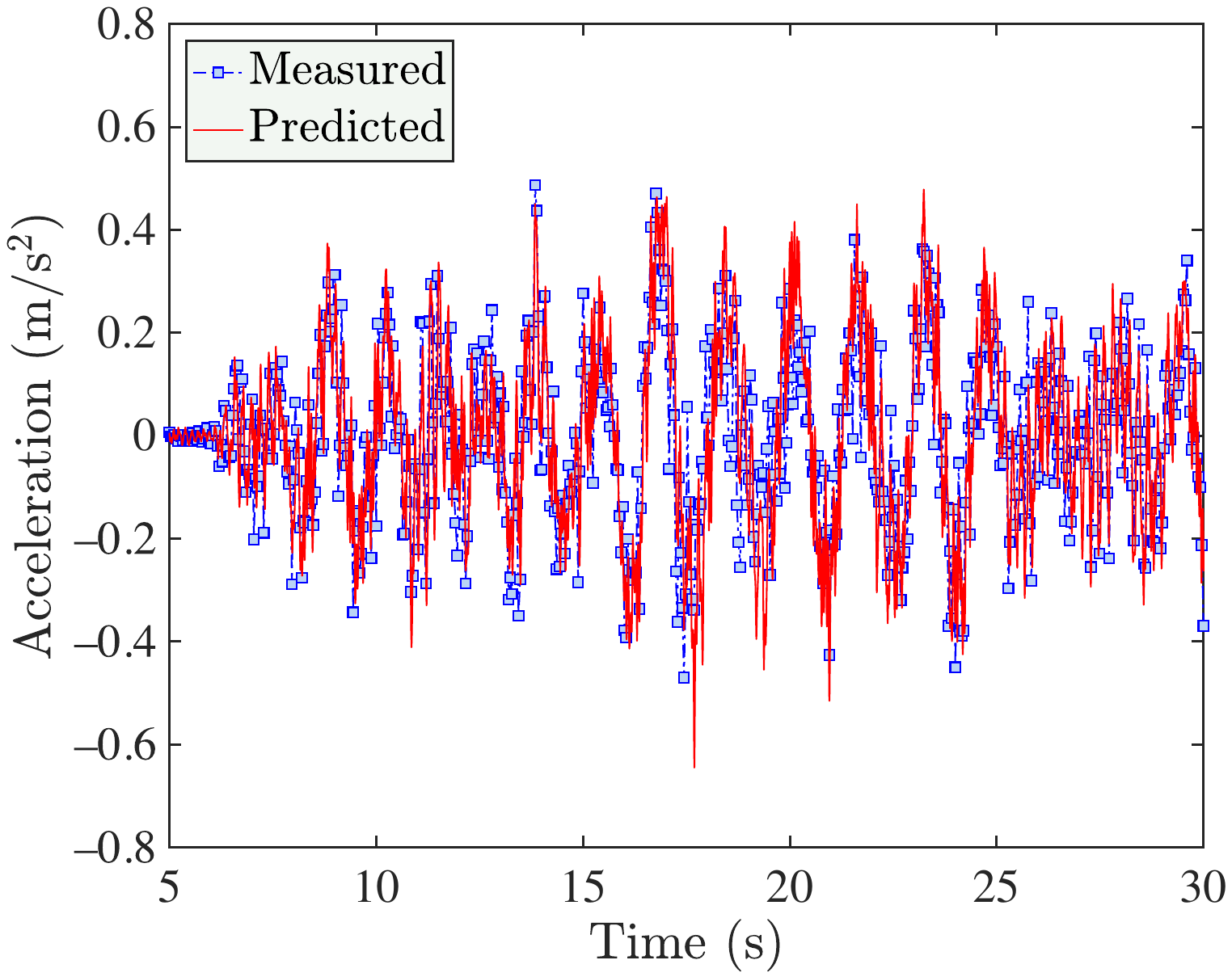}
        \caption{1st floor acceleration in $x$ direction.}
    \end{subfigure}%
    ~ 
    \begin{subfigure}[t]{0.5\textwidth}
        \centering
        \includegraphics[scale=0.3]{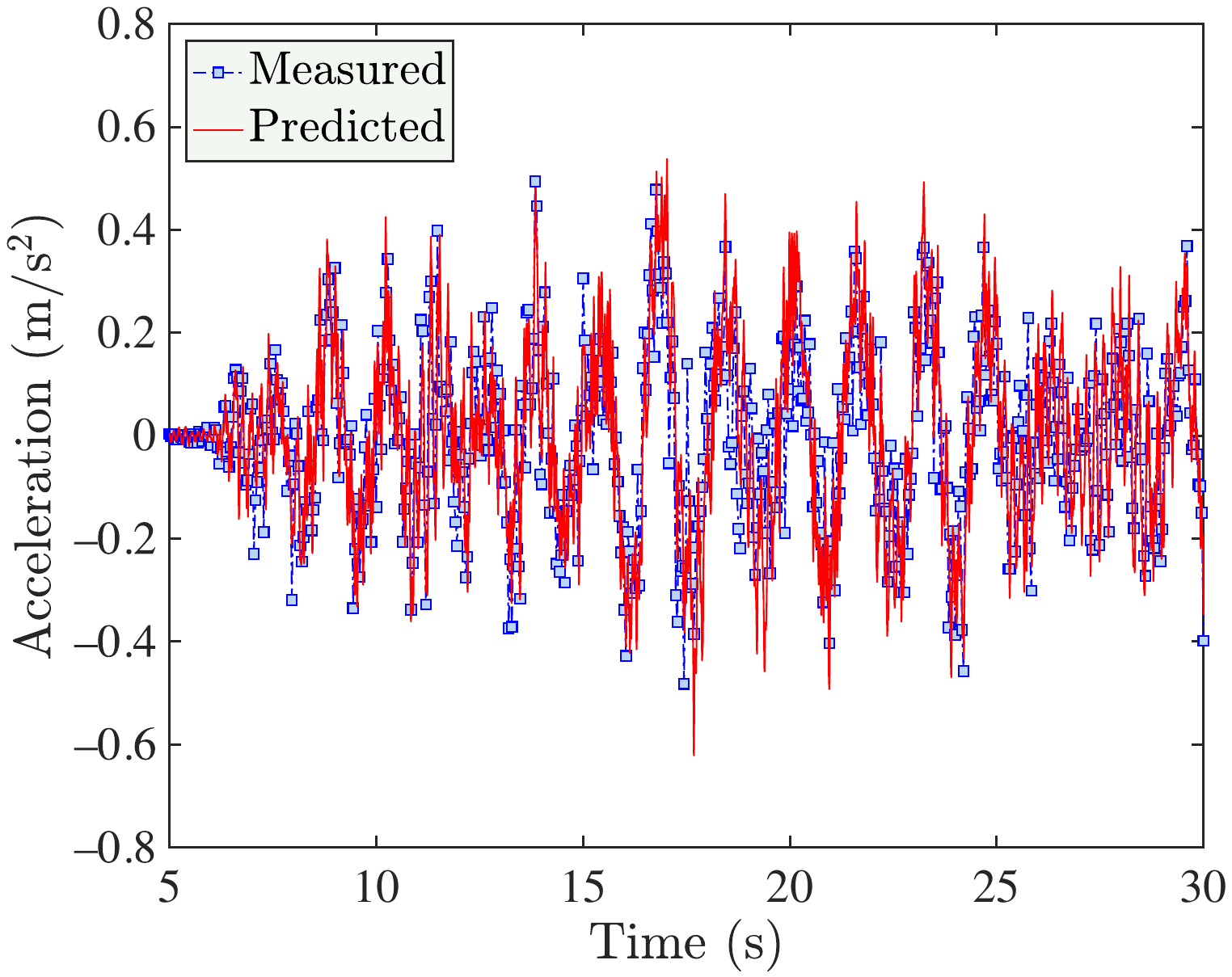}
        \caption{1st floor acceleration in $y$ direction.}
    \end{subfigure}
    \begin{subfigure}[t]{0.5\textwidth}
        \centering
        \includegraphics[scale=0.3]{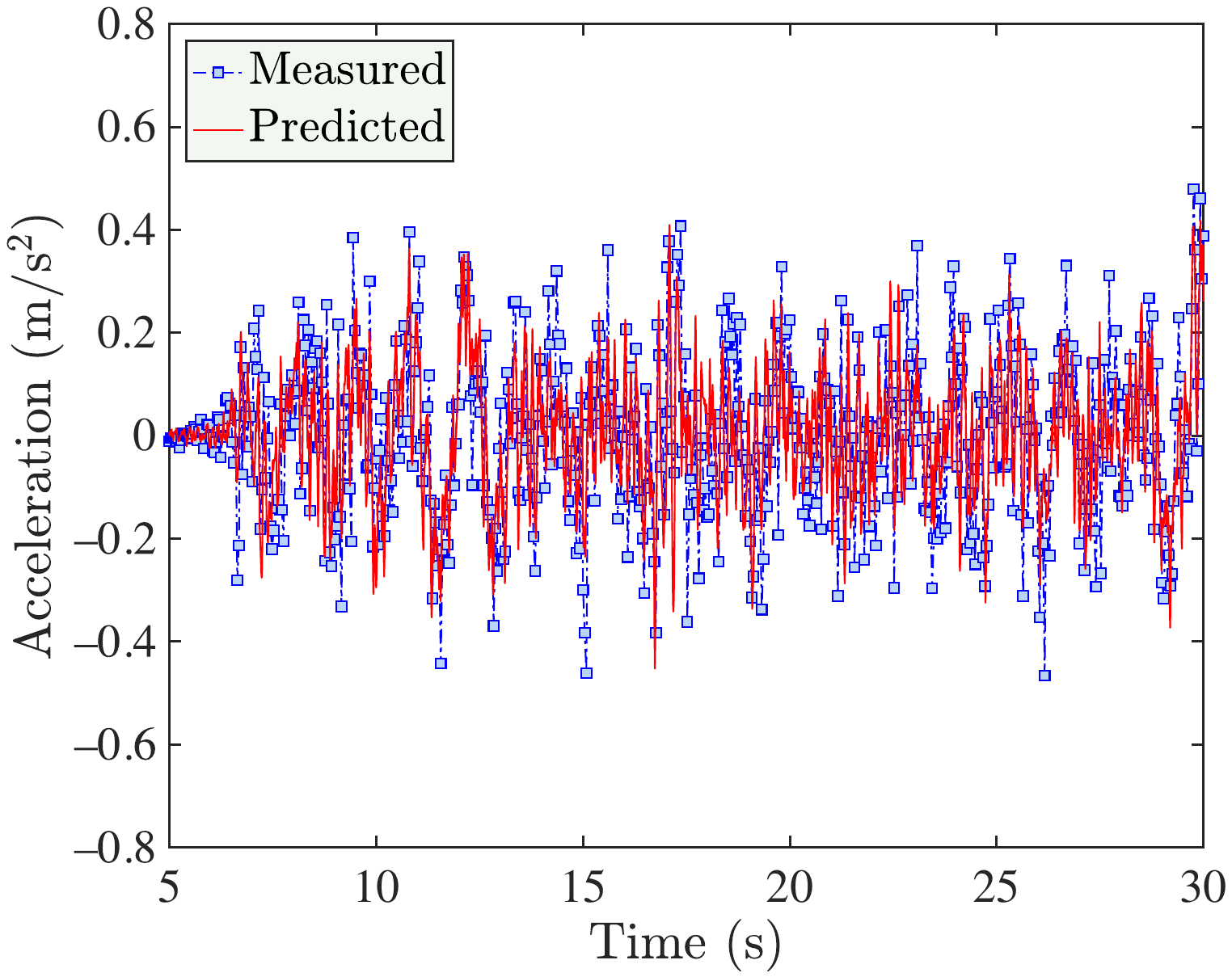}
        \caption{2nd floor acceleration in $x$ direction.}
    \end{subfigure}%
    ~ 
    \begin{subfigure}[t]{0.5\textwidth}
        \centering
        \includegraphics[scale=0.3]{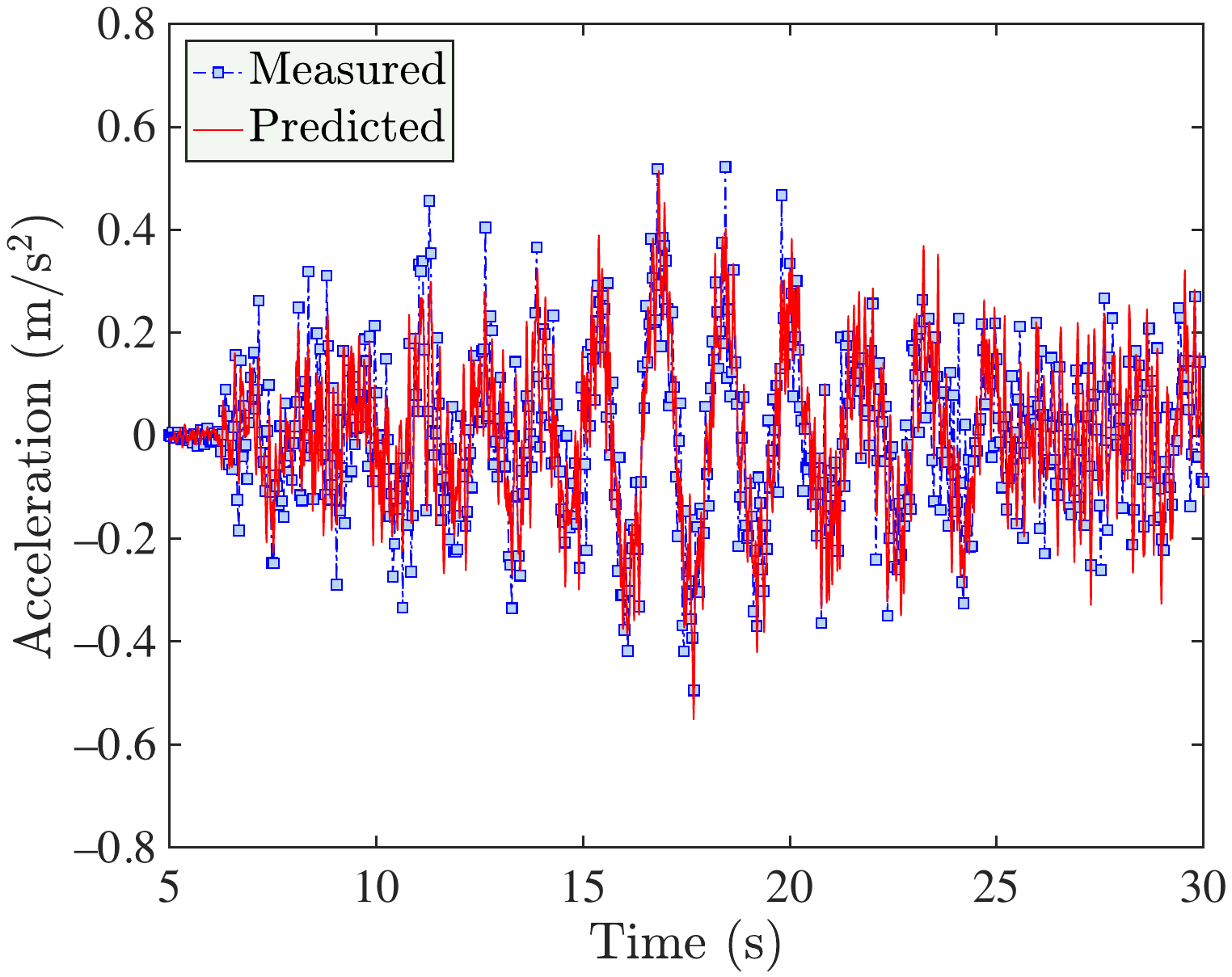}
        \caption{2nd floor acceleration in $y$ direction.}
    \end{subfigure}
    \end{figure}
\begin{figure}[htb!]\ContinuedFloat
    \begin{subfigure}[t]{0.5\textwidth}
        \centering
        \includegraphics[scale=0.3]{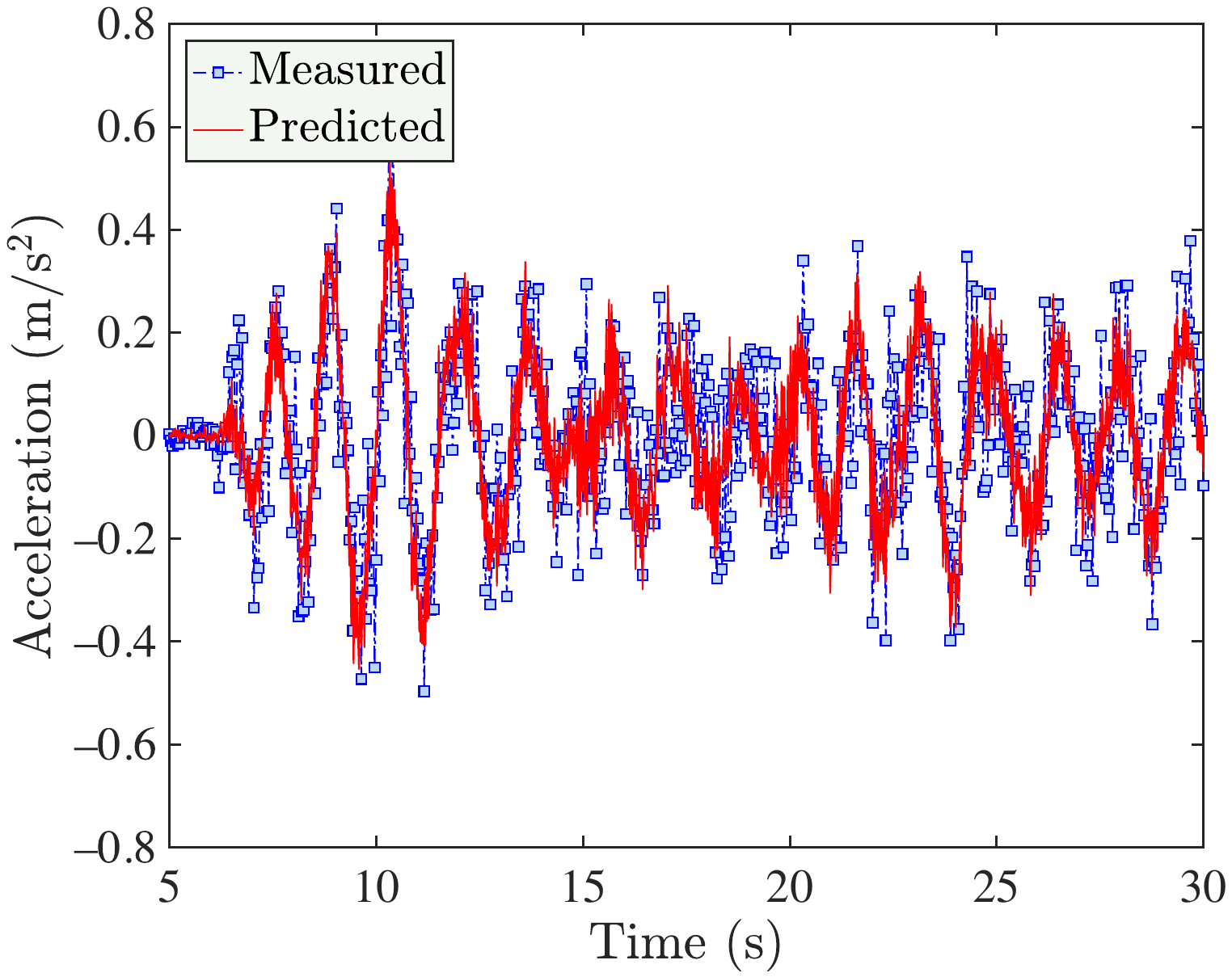}
        \caption{3rd floor acceleration in $x$ direction.}
    \end{subfigure}%
    ~ 
    \begin{subfigure}[t]{0.5\textwidth}
        \centering
        \includegraphics[scale=0.3]{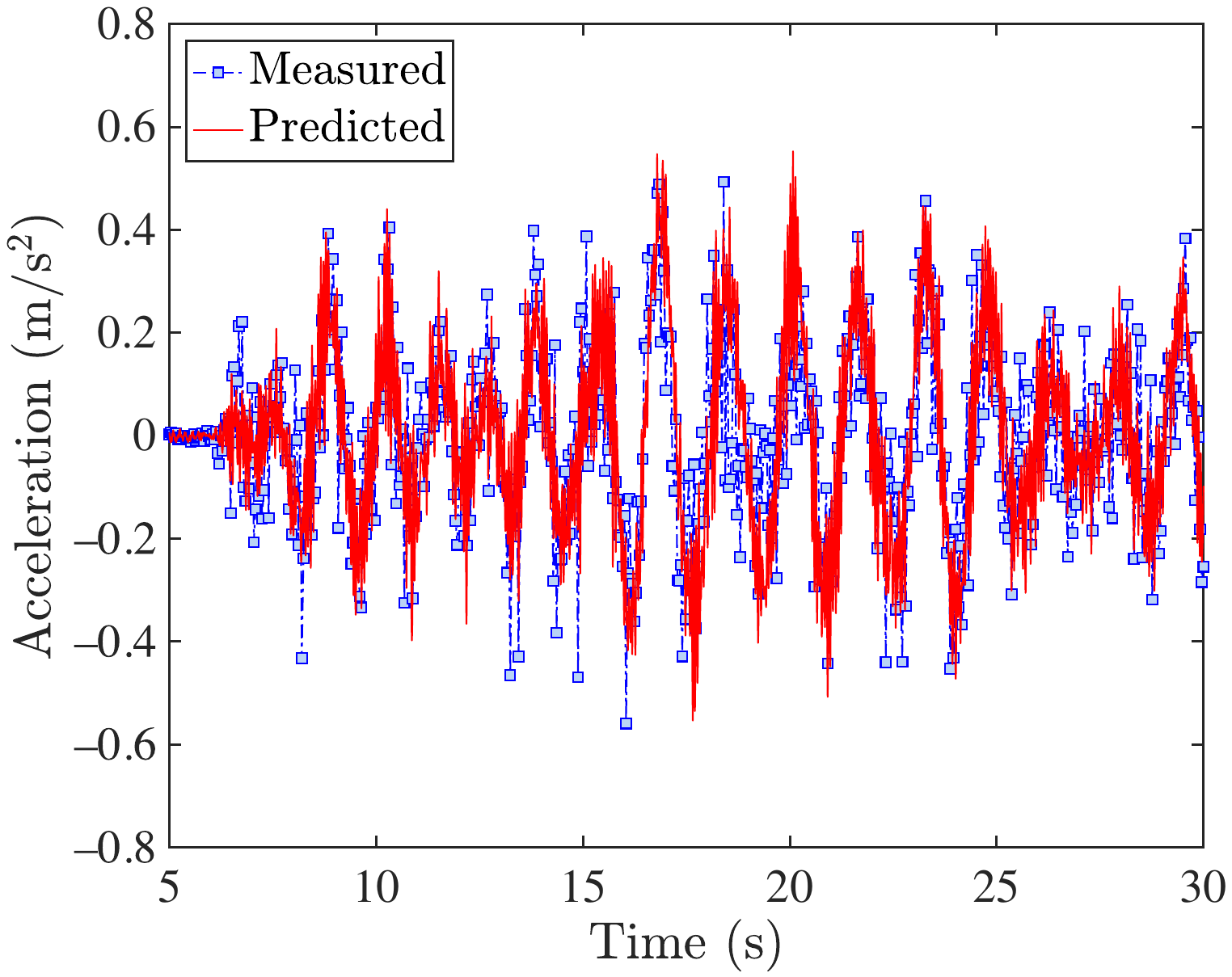}
        \caption{3rd floor acceleration in $y$ direction.}
    \end{subfigure}
    \begin{subfigure}[t]{0.5\textwidth}
        \centering
        \includegraphics[scale=0.3]{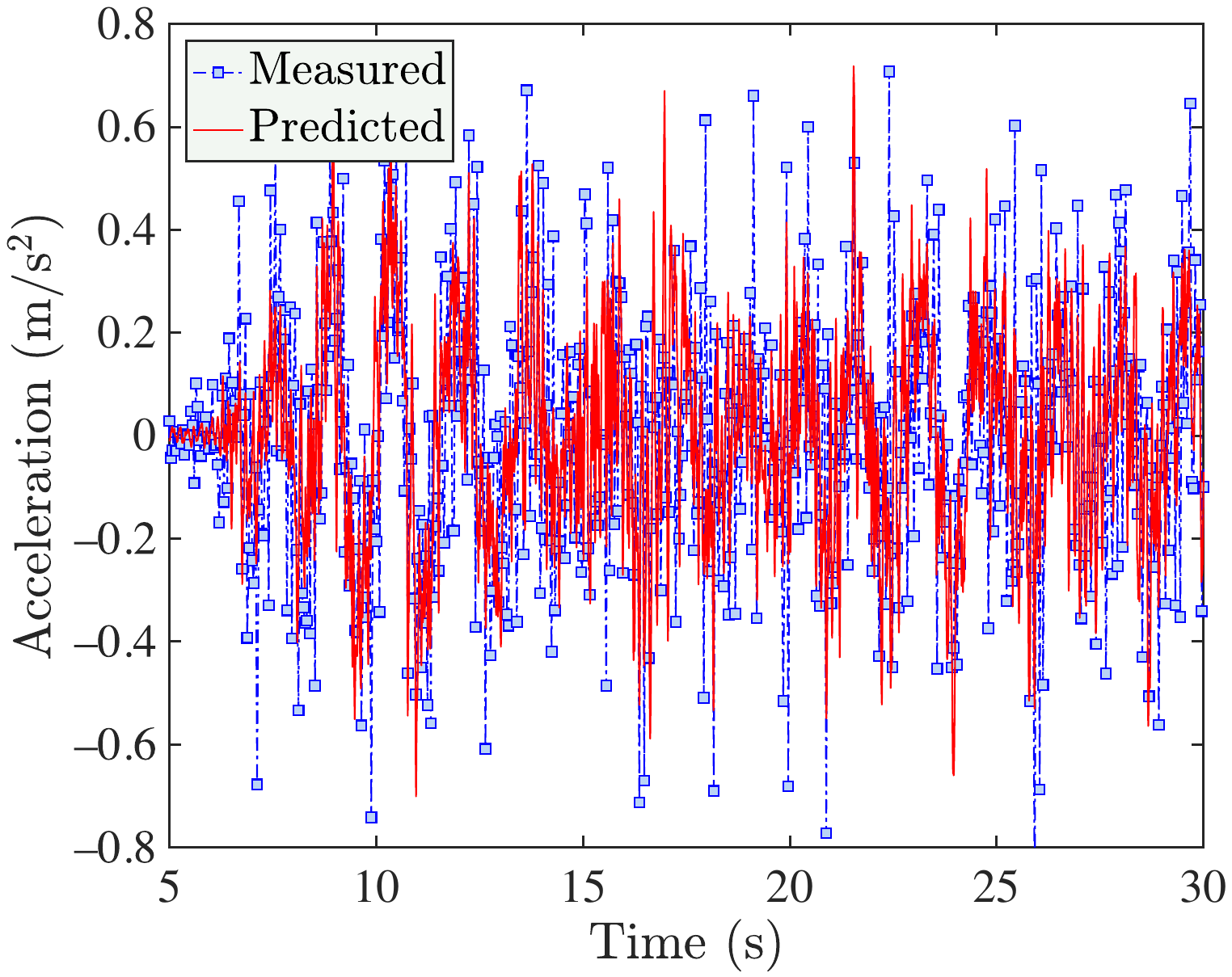}
        \caption{4th floor acceleration in $x$ direction.}
    \end{subfigure}%
    ~ 
    \begin{subfigure}[t]{0.5\textwidth}
        \centering
        \includegraphics[scale=0.3]{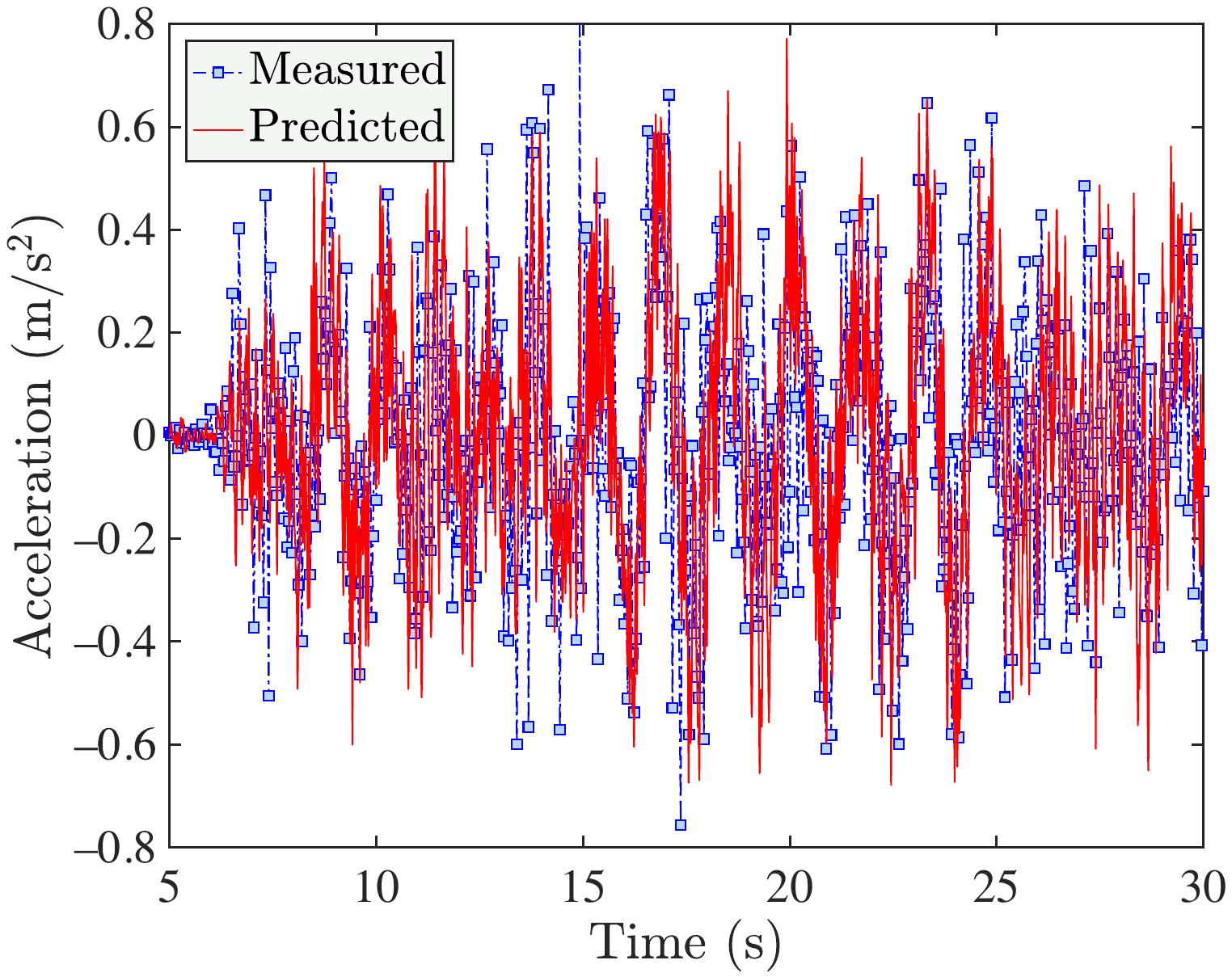}
        \caption{4th floor acceleration in $y$ direction.}
    \end{subfigure}
    \caption{Predicted responses using unfalsified models from $\C_4$ are compared with measured during Test 012 using the sensor near 1-A (see \fref{fig:sensors}).} \label{fig:ex3a_resppredM4_allfloors}
\end{figure} 

\begin{figure}[htb!]
    \centering
    % \begin{subfigure}[t]{0.5\textwidth}
    %     \centering
    %     \includegraphics[scale=0.3]{figures/floor1_x.pdf}
    %     \caption{Base acceleration in $x$ direction.}
    % \end{subfigure}%
    % ~ 
    % \begin{subfigure}[t]{0.5\textwidth}
    %     \centering
    %     \includegraphics[scale=0.3]{figures/floor1_y.pdf}
    %     \caption{Base acceleration in $y$ direction.}
    % \end{subfigure}
    \begin{subfigure}[t]{0.5\textwidth}
        \centering
        \includegraphics[scale=0.3]{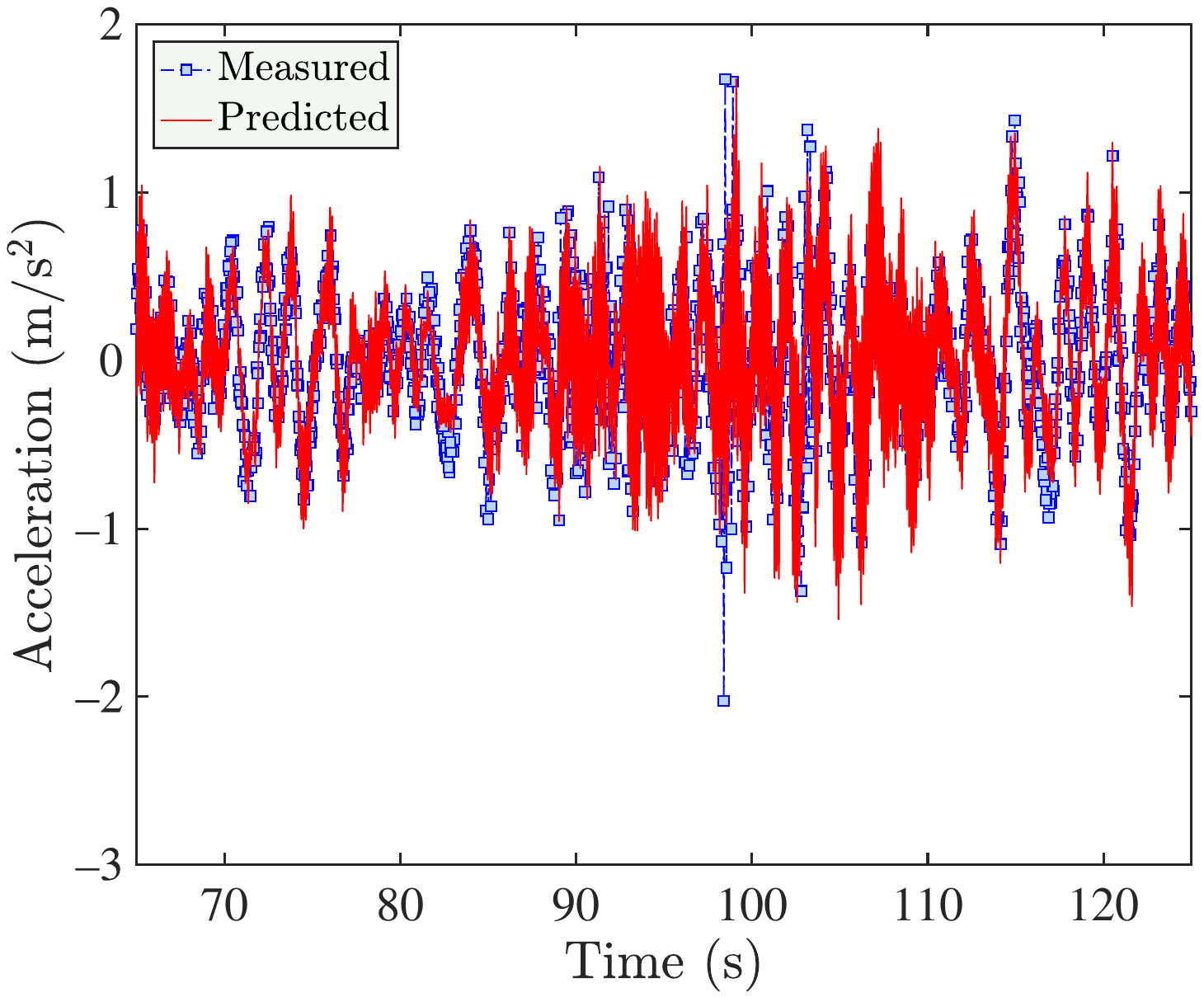}
        \caption{1st floor acceleration in $x$ direction.}
    \end{subfigure}%
    ~ 
    \begin{subfigure}[t]{0.5\textwidth}
        \centering
        \includegraphics[scale=0.3]{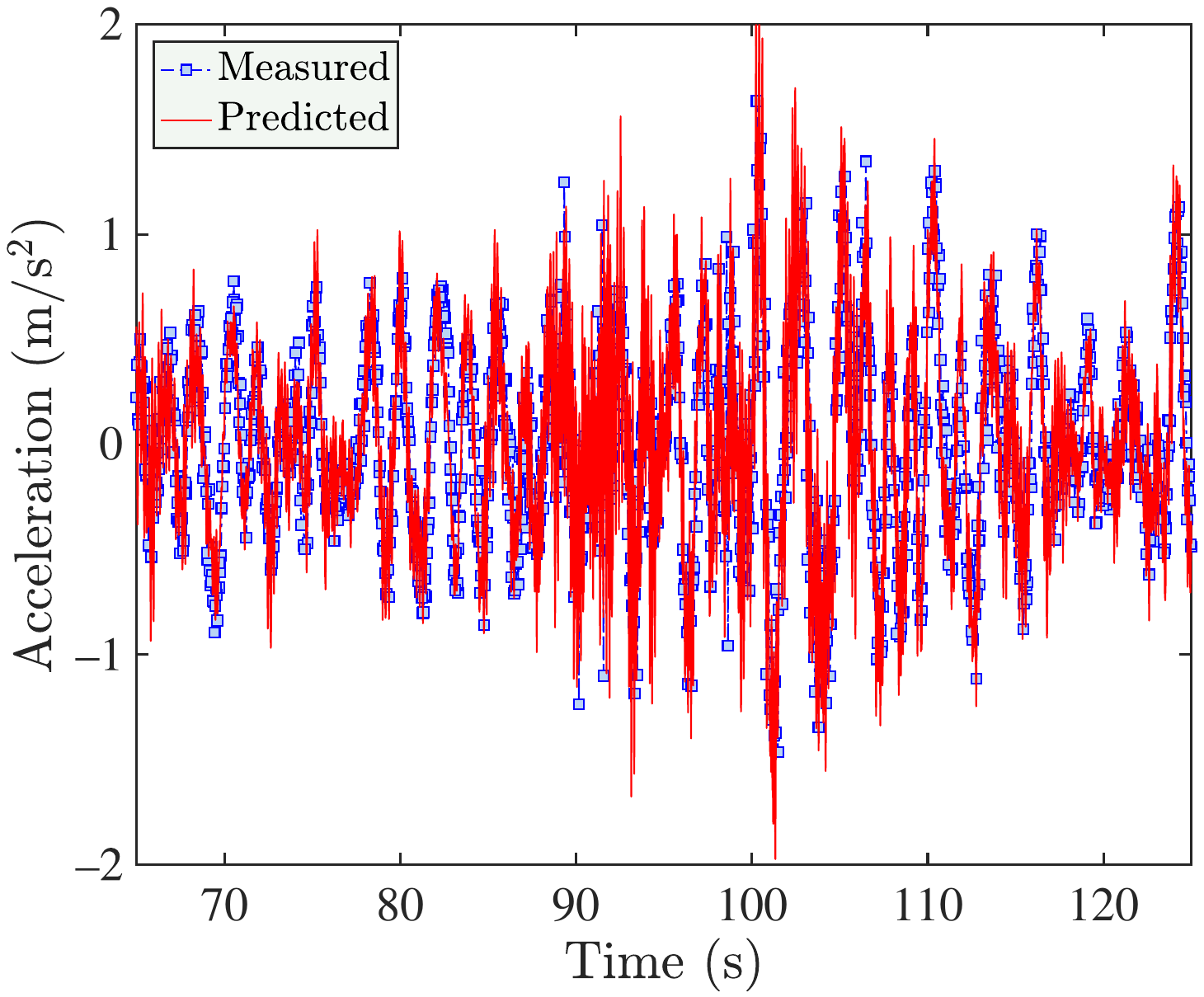}
        \caption{1st floor acceleration in $y$ direction.}
    \end{subfigure}
    \begin{subfigure}[t]{0.5\textwidth}
        \centering
        \includegraphics[scale=0.3]{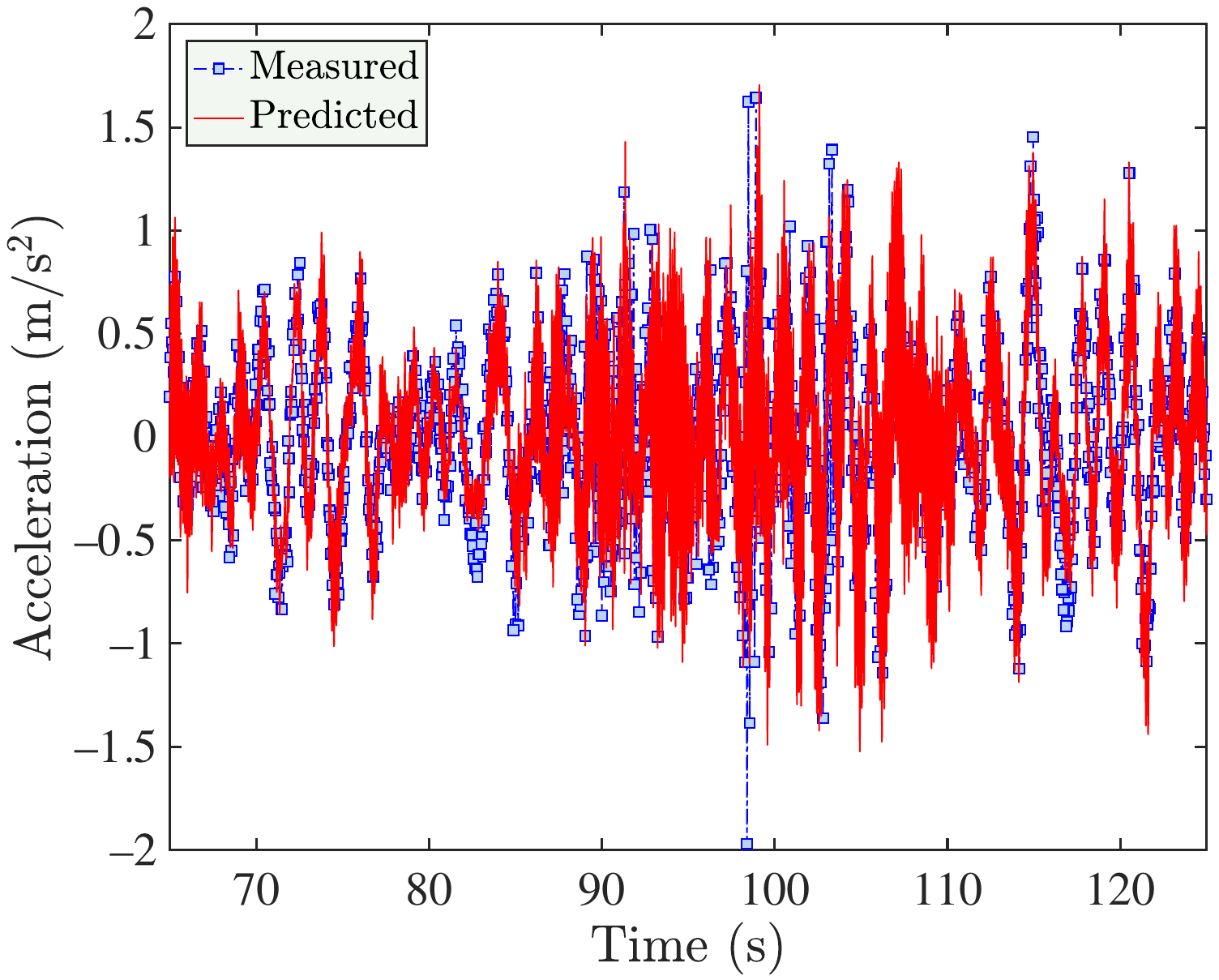}
        \caption{2nd floor acceleration in $x$ direction.}
    \end{subfigure}%
    ~ 
    \begin{subfigure}[t]{0.5\textwidth}
        \centering
        \includegraphics[scale=0.3]{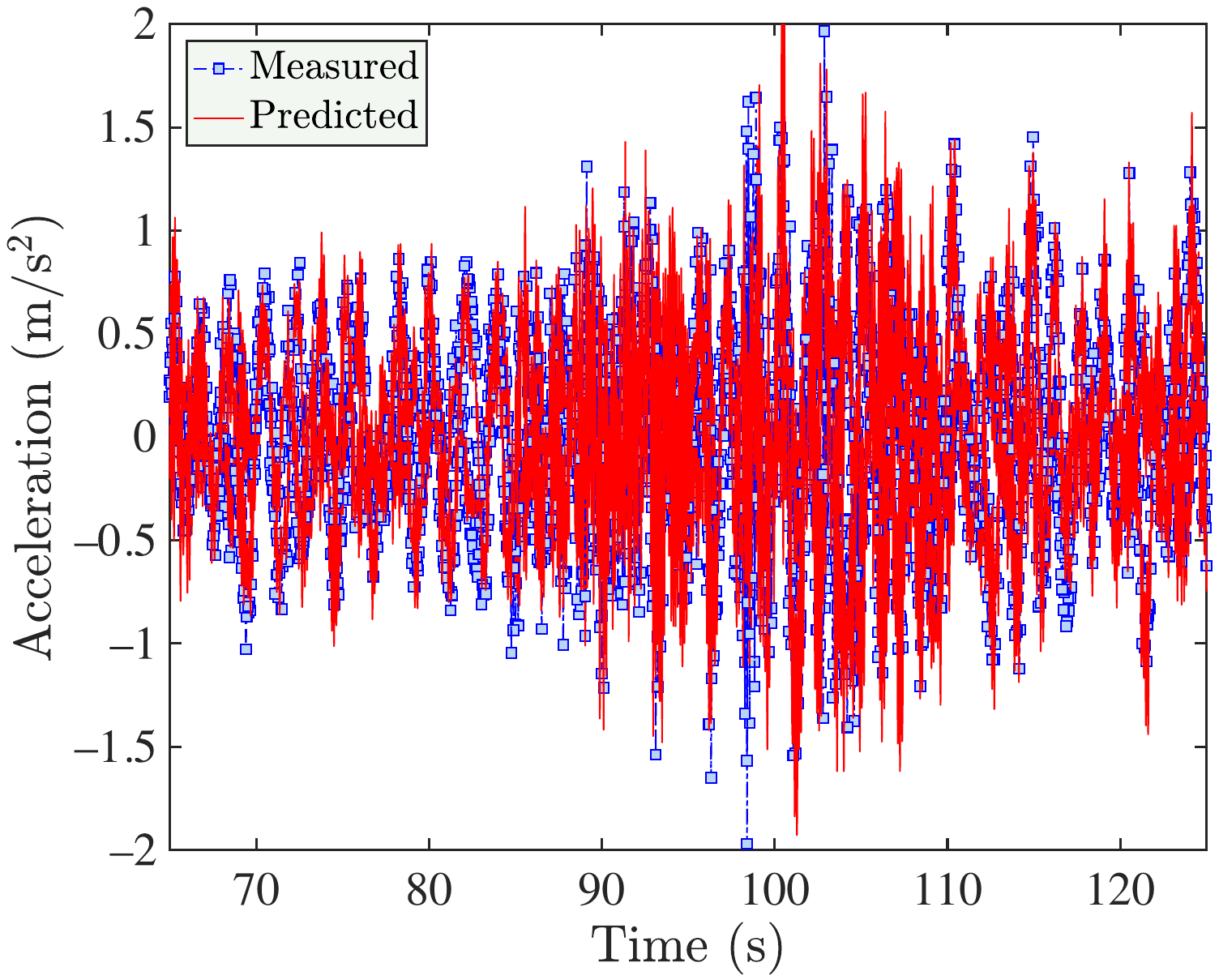}
        \caption{2nd floor acceleration in $y$ direction.}
    \end{subfigure}
    \end{figure}
\begin{figure}[htb!]\ContinuedFloat
    \begin{subfigure}[t]{0.5\textwidth}
        \centering
        \includegraphics[scale=0.3]{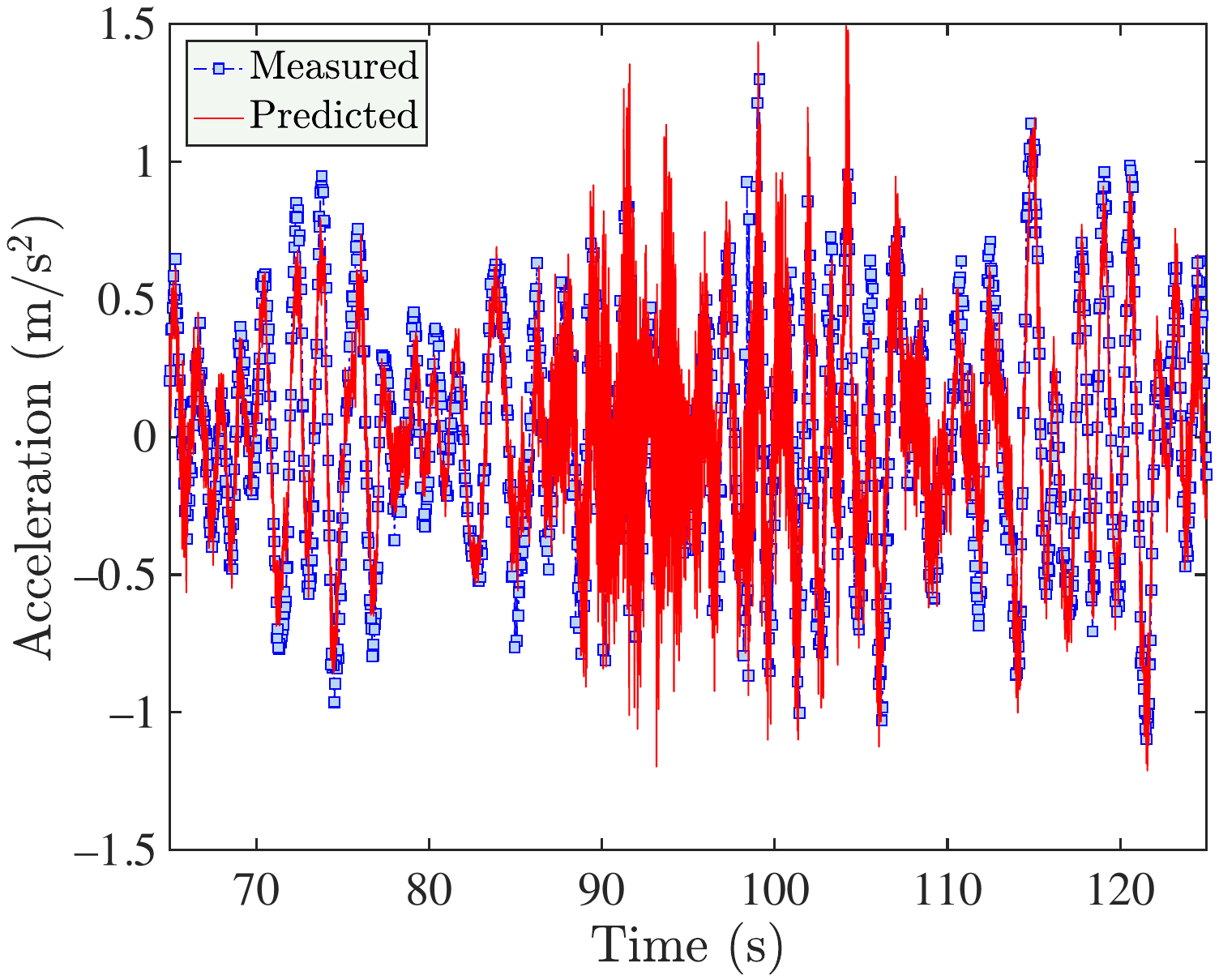}
        \caption{3rd floor acceleration in $x$ direction.}
    \end{subfigure}%
    ~ 
    \begin{subfigure}[t]{0.5\textwidth}
        \centering
        \includegraphics[scale=0.3]{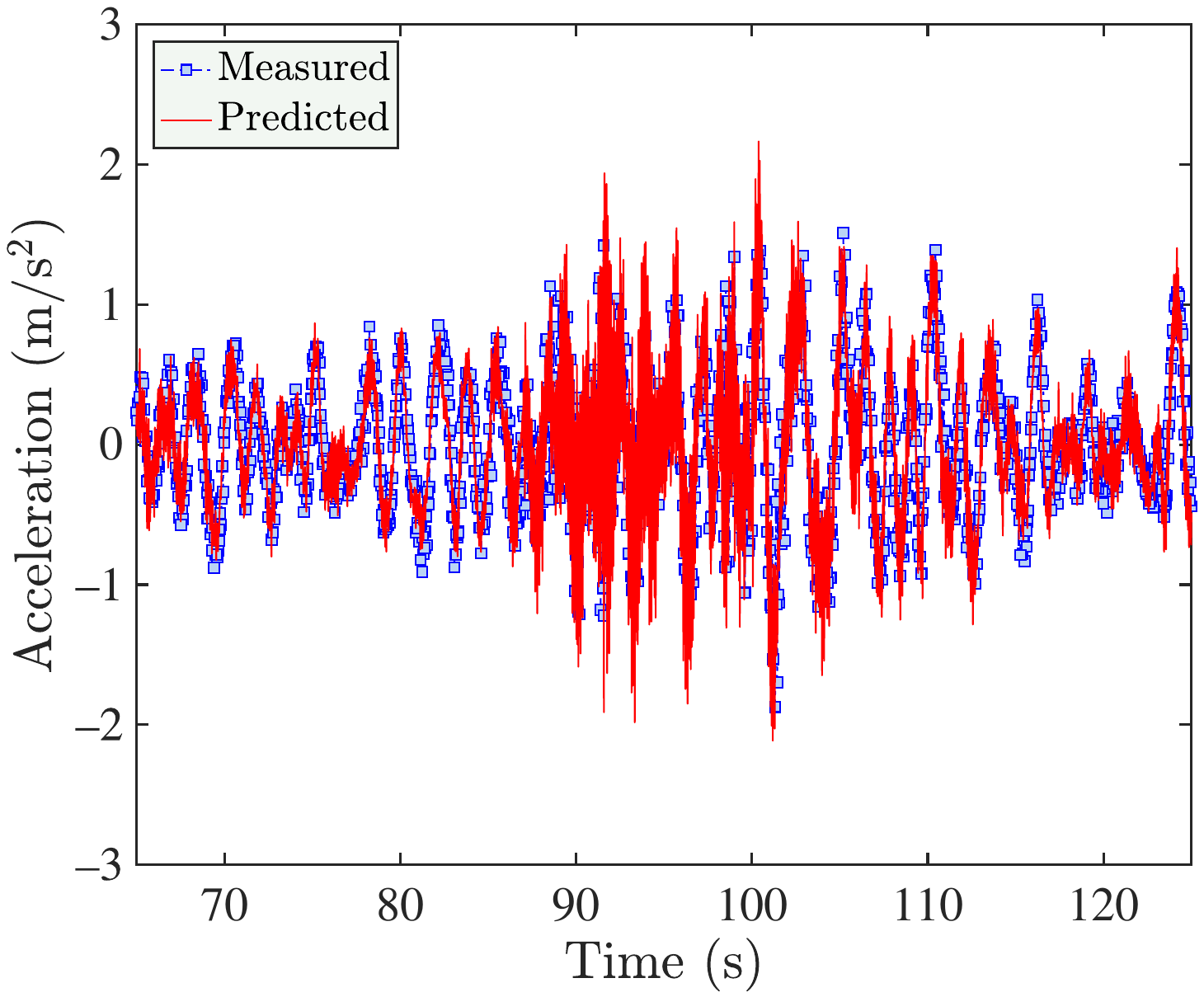}
        \caption{3rd floor acceleration in $y$ direction.}
    \end{subfigure}
    \begin{subfigure}[t]{0.5\textwidth}
        \centering
        \includegraphics[scale=0.3]{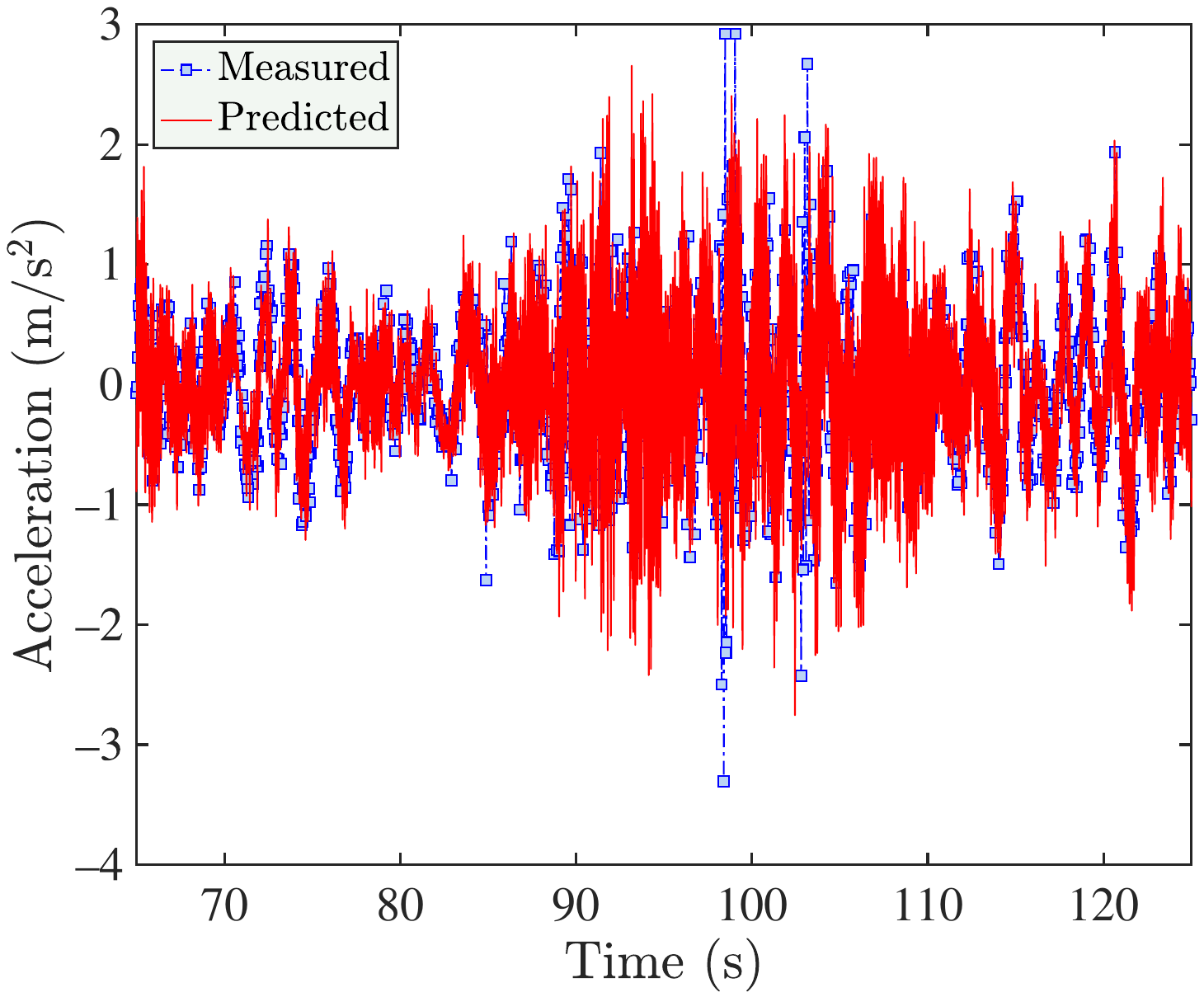}
        \caption{4th floor acceleration in $x$ direction.}
    \end{subfigure}%
    ~ 
    \begin{subfigure}[t]{0.5\textwidth}
        \centering
        \includegraphics[scale=0.3]{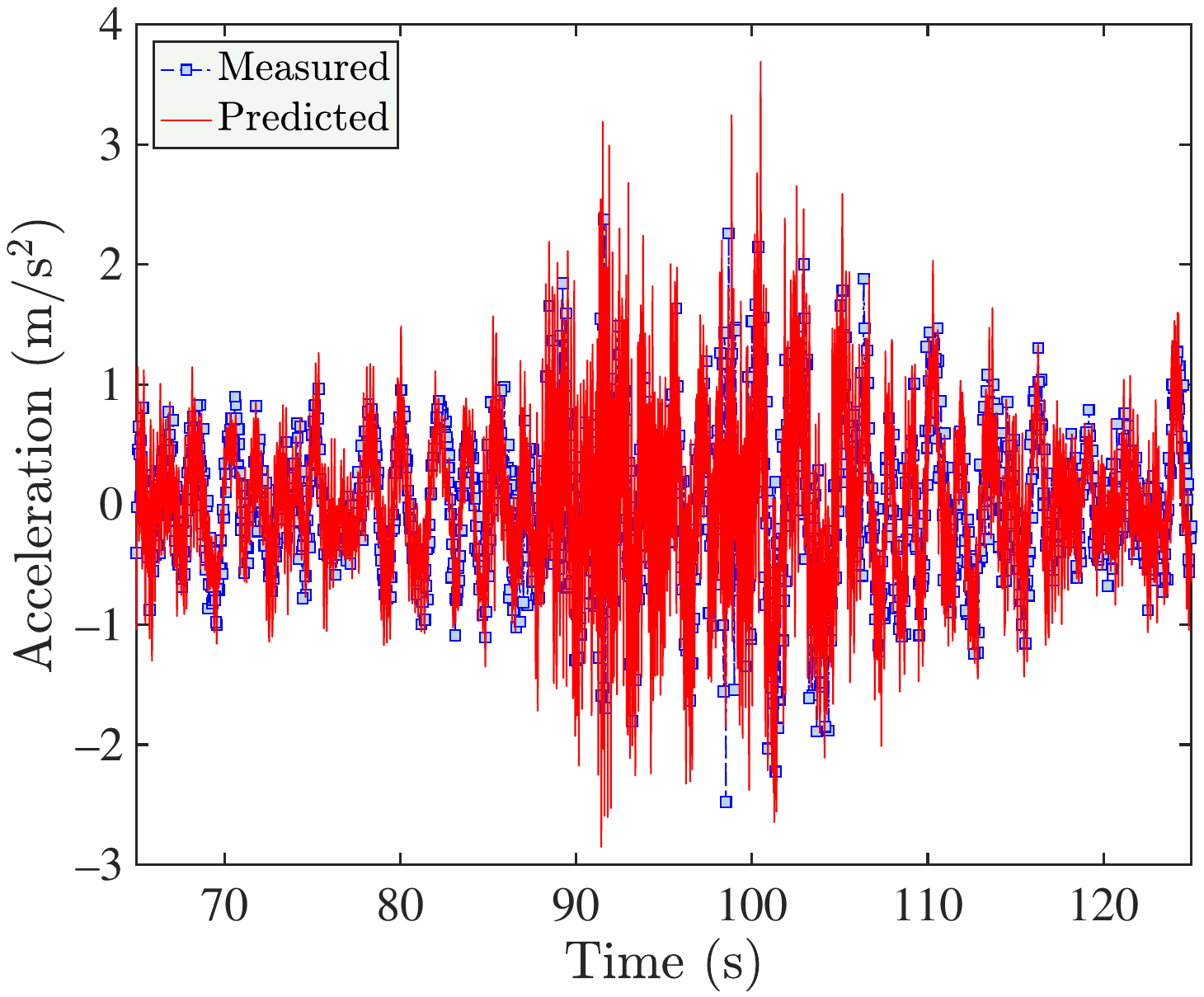}
        \caption{4th floor acceleration in $y$ direction.} 
    \end{subfigure}
    \caption{Predicted response is compared with measured during Test 014 using the sensor near 1-A (see \fref{fig:sensors}).} \label{fig:ex3b_resppred_allfloors}
\end{figure} 

\end{document}